\pdfoutput=1
\documentclass{JINST}
\usepackage{times}
\usepackage{xspace}
\usepackage{array}
\usepackage{graphicx}
\usepackage{amsmath}
\usepackage{upgreek}
\usepackage{cite}

\newcommand{\GeV}{\ensuremath{\mbox{GeV}}\xspace}
\newcommand{\TeV}{\ensuremath{\mbox{TeV}}\xspace}

\newcommand{\micron}{\ensuremath{\upmu \mbox{m}}\xspace}
\newcommand{\ns}{\ensuremath{\mbox{ns}}\xspace}

\newcommand{\mrad}{\ensuremath{\mbox{mrad}}\xspace}
\newcommand{\micrad}{\ensuremath{\upmu\mbox{rad}}\xspace}

\newcommand{\PZ}[1][]{\ensuremath{{P_0^{\mathrm{#1}}}}\xspace}
\newcommand{\sigeff}[2][]{\ensuremath{{\sigma^{#1}_{\mathrm{#2}}}}\xspace}
\newcommand{\mueff}[2][]{\ensuremath{{\mu^{#1}_{\mathrm{#2}}}}\xspace}
\newcommand{\eff}{\ensuremath{\mathrm{vis}}}
\newcommand{\thetac}{\ensuremath{\alpha}\xspace}
\newcommand{\thetab}{\ensuremath{\alpha}\xspace}
\newcommand{\bx}[1]{\ensuremath{{\mathsf{#1}}}\xspace}
\newcommand{\nb}[1]{\ensuremath{n_{\mathrm{#1}}}\xspace}
\newcommand{\ntot}{\nb{tot}}
\newcommand{\ncol}{\nb{coll}}
\newcommand{\ntrks}[1]{\ensuremath{N_{\mathrm{Tr}_{#1}}}\xspace}
\newcommand{\ntrk}{\ensuremath{N_{\mathrm{Tr}}}\xspace}
\newcommand{\ndof}{\ensuremath{\mathrm{ndf}}\xspace}
\newcommand{\cres}{\ensuremath{F_{z}}\xspace}
\newcommand{\mpos}{\ensuremath{\xi}\xspace}
\newcommand{\sigxp}[2][]{\ensuremath{\sigma_{\otimes\mathrm{#2}}^{#1}}\xspace}
\newcommand{\posxp}[2][]{\ensuremath{\mpos_{\otimes\mathrm{#2}}^{#1}}\xspace}
\newcommand{\xp}{\otimes}
\newcommand{\tms}{\times}

\newcommand{\Deltax}{\ensuremath{\Delta_x}\xspace}
\newcommand{\Deltay}{\ensuremath{\Delta_y}\xspace}
\newcommand{\Deltaxz}{\ensuremath{\Delta_{x_0}}\xspace}
\newcommand{\Deltayz}{\ensuremath{\Delta_{y_0}}\xspace}

\title{
 \vskip -1.4cm
 { \normalsize \normalfont
 \rightline{LHCb-PAPER-2011-015}
 \rightline{CERN-PH-EP-2011-157}
 \rightline {15 December 2011}
 }
 \vglue 1.8cm
 {
 Absolute luminosity measurements with the LHCb detector at the LHC
 }
}
\author{The LHCb Collaboration\footnote{Authors are listed on the
following pages}}
\abstract{
 Absolute luminosity measurements are of general interest for
 colliding-beam experiments at storage rings. 
 These measurements are necessary to determine the absolute
 cross-sections of reaction processes and are valuable to quantify the
 performance of the accelerator. 
 Using data taken in 2010,
 LHCb has applied two methods to determine the absolute scale of its
 luminosity measurements for proton-proton collisions at the LHC with
 a centre-of-mass energy of 7~\TeV.
 In addition to the classic ``van der Meer scan'' method  a novel technique has
 been developed which makes use of direct imaging of the individual
 beams using beam-gas and beam-beam interactions.
 This beam imaging method is made possible by the high resolution of the
 LHCb vertex detector and the close proximity of the detector to the beams,
 and allows beam parameters such as positions, angles and widths to be
 determined.  
 The results of the two methods have comparable precision and are in good
 agreement. 
 Combining the two methods, an overal precision of 3.5\% in the absolute
 luminosity determination is reached.
 The techniques used to transport the absolute
 luminosity calibration to the full 2010 data-taking period are
 presented.
 }

\begin{document}

\newpage
\noindent {\Large LHCb Collaboration}\\[2ex]
R.~Aaij$^{23}$, 
B.~Adeva$^{36}$, 
M.~Adinolfi$^{42}$, 
C.~Adrover$^{6}$, 
A.~Affolder$^{48}$, 
Z.~Ajaltouni$^{5}$, 
J.~Albrecht$^{37}$, 
F.~Alessio$^{37}$, 
M.~Alexander$^{47}$, 
G.~Alkhazov$^{29}$, 
P.~Alvarez~Cartelle$^{36}$, 
A.A.~Alves~Jr$^{22}$, 
S.~Amato$^{2}$, 
Y.~Amhis$^{38}$, 
J.~Anderson$^{39}$, 
R.B.~Appleby$^{50}$, 
O.~Aquines~Gutierrez$^{10}$, 
F.~Archilli$^{18,37}$, 
L.~Arrabito$^{53}$, 
A.~Artamonov$^{34}$, 
M.~Artuso$^{52,37}$, 
E.~Aslanides$^{6}$, 
G.~Auriemma$^{22,m}$, 
S.~Bachmann$^{11}$, 
J.J.~Back$^{44}$, 
D.S.~Bailey$^{50}$, 
V.~Balagura$^{30,37}$, 
W.~Baldini$^{16}$, 
R.J.~Barlow$^{50}$, 
C.~Barschel$^{37}$, 
S.~Barsuk$^{7}$, 
W.~Barter$^{43}$, 
A.~Bates$^{47}$, 
C.~Bauer$^{10}$, 
Th.~Bauer$^{23}$, 
A.~Bay$^{38}$, 
I.~Bediaga$^{1}$, 
K.~Belous$^{34}$, 
I.~Belyaev$^{30,37}$, 
E.~Ben-Haim$^{8}$, 
M.~Benayoun$^{8}$, 
G.~Bencivenni$^{18}$, 
S.~Benson$^{46}$, 
J.~Benton$^{42}$, 
R.~Bernet$^{39}$, 
M.-O.~Bettler$^{17}$, 
M.~van~Beuzekom$^{23}$, 
A.~Bien$^{11}$, 
S.~Bifani$^{12}$, 
A.~Bizzeti$^{17,h}$, 
P.M.~Bj\o rnstad$^{50}$, 
T.~Blake$^{49}$, 
F.~Blanc$^{38}$, 
C.~Blanks$^{49}$, 
J.~Blouw$^{11}$, 
S.~Blusk$^{52}$, 
A.~Bobrov$^{33}$, 
V.~Bocci$^{22}$, 
A.~Bondar$^{33}$, 
N.~Bondar$^{29}$, 
W.~Bonivento$^{15}$, 
S.~Borghi$^{47}$, 
A.~Borgia$^{52}$, 
T.J.V.~Bowcock$^{48}$, 
C.~Bozzi$^{16}$, 
T.~Brambach$^{9}$, 
J.~van~den~Brand$^{24}$, 
J.~Bressieux$^{38}$, 
D.~Brett$^{50}$, 
S.~Brisbane$^{51}$, 
M.~Britsch$^{10}$, 
T.~Britton$^{52}$, 
N.H.~Brook$^{42}$, 
H.~Brown$^{48}$, 
A.~B\"{u}chler-Germann$^{39}$, 
I.~Burducea$^{28}$, 
A.~Bursche$^{39}$, 
J.~Buytaert$^{37}$, 
S.~Cadeddu$^{15}$, 
J.M.~Caicedo~Carvajal$^{37}$, 
O.~Callot$^{7}$, 
M.~Calvi$^{20,j}$, 
M.~Calvo~Gomez$^{35,n}$, 
A.~Camboni$^{35}$, 
P.~Campana$^{18,37}$, 
A.~Carbone$^{14}$, 
G.~Carboni$^{21,k}$, 
R.~Cardinale$^{19,i,37}$, 
A.~Cardini$^{15}$, 
L.~Carson$^{36}$, 
K.~Carvalho~Akiba$^{23}$, 
G.~Casse$^{48}$, 
M.~Cattaneo$^{37}$, 
M.~Charles$^{51}$, 
Ph.~Charpentier$^{37}$, 
N.~Chiapolini$^{39}$, 
K.~Ciba$^{37}$, 
X.~Cid~Vidal$^{36}$, 
G.~Ciezarek$^{49}$, 
P.E.L.~Clarke$^{46,37}$, 
M.~Clemencic$^{37}$, 
H.V.~Cliff$^{43}$, 
J.~Closier$^{37}$, 
C.~Coca$^{28}$, 
V.~Coco$^{23}$, 
J.~Cogan$^{6}$, 
P.~Collins$^{37}$, 
F.~Constantin$^{28}$, 
G.~Conti$^{38}$, 
A.~Contu$^{51}$, 
A.~Cook$^{42}$, 
M.~Coombes$^{42}$, 
G.~Corti$^{37}$, 
G.A.~Cowan$^{38}$, 
R.~Currie$^{46}$, 
B.~D'Almagne$^{7}$, 
C.~D'Ambrosio$^{37}$, 
P.~David$^{8}$, 
I.~De~Bonis$^{4}$, 
S.~De~Capua$^{21,k}$, 
M.~De~Cian$^{39}$, 
F.~De~Lorenzi$^{12}$, 
J.M.~De~Miranda$^{1}$, 
L.~De~Paula$^{2}$, 
P.~De~Simone$^{18}$, 
D.~Decamp$^{4}$, 
M.~Deckenhoff$^{9}$, 
H.~Degaudenzi$^{38,37}$, 
M.~Deissenroth$^{11}$, 
L.~Del~Buono$^{8}$, 
C.~Deplano$^{15}$, 
O.~Deschamps$^{5}$, 
F.~Dettori$^{15,d}$, 
J.~Dickens$^{43}$, 
H.~Dijkstra$^{37}$, 
P.~Diniz~Batista$^{1}$, 
S.~Donleavy$^{48}$, 
F.~Dordei$^{11}$, 
A.~Dosil~Su\'{a}rez$^{36}$, 
D.~Dossett$^{44}$, 
A.~Dovbnya$^{40}$, 
F.~Dupertuis$^{38}$, 
R.~Dzhelyadin$^{34}$, 
C.~Eames$^{49}$, 
S.~Easo$^{45}$, 
U.~Egede$^{49}$, 
V.~Egorychev$^{30}$, 
S.~Eidelman$^{33}$, 
D.~van~Eijk$^{23}$, 
F.~Eisele$^{11}$, 
S.~Eisenhardt$^{46}$, 
R.~Ekelhof$^{9}$, 
L.~Eklund$^{47}$, 
Ch.~Elsasser$^{39}$, 
D.G.~d'Enterria$^{35,o}$, 
D.~Esperante~Pereira$^{36}$, 
L.~Est\`{e}ve$^{43}$, 
A.~Falabella$^{16,e}$, 
E.~Fanchini$^{20,j}$, 
C.~F\"{a}rber$^{11}$, 
G.~Fardell$^{46}$, 
C.~Farinelli$^{23}$, 
S.~Farry$^{12}$, 
V.~Fave$^{38}$, 
V.~Fernandez~Albor$^{36}$, 
M.~Ferro-Luzzi$^{37}$, 
S.~Filippov$^{32}$, 
C.~Fitzpatrick$^{46}$, 
M.~Fontana$^{10}$, 
F.~Fontanelli$^{19,i}$, 
R.~Forty$^{37}$, 
M.~Frank$^{37}$, 
C.~Frei$^{37}$, 
M.~Frosini$^{17,f,37}$, 
S.~Furcas$^{20}$, 
A.~Gallas~Torreira$^{36}$, 
D.~Galli$^{14,c}$, 
M.~Gandelman$^{2}$, 
P.~Gandini$^{51}$, 
Y.~Gao$^{3}$, 
J-C.~Garnier$^{37}$, 
J.~Garofoli$^{52}$, 
J.~Garra~Tico$^{43}$, 
L.~Garrido$^{35}$, 
C.~Gaspar$^{37}$, 
N.~Gauvin$^{38}$, 
M.~Gersabeck$^{37}$, 
T.~Gershon$^{44,37}$, 
Ph.~Ghez$^{4}$, 
V.~Gibson$^{43}$, 
V.V.~Gligorov$^{37}$, 
C.~G\"{o}bel$^{54}$, 
D.~Golubkov$^{30}$, 
A.~Golutvin$^{49,30,37}$, 
A.~Gomes$^{2}$, 
H.~Gordon$^{51}$, 
M.~Grabalosa~G\'{a}ndara$^{35}$, 
R.~Graciani~Diaz$^{35}$, 
L.A.~Granado~Cardoso$^{37}$, 
E.~Graug\'{e}s$^{35}$, 
G.~Graziani$^{17}$, 
A.~Grecu$^{28}$, 
S.~Gregson$^{43}$, 
B.~Gui$^{52}$, 
E.~Gushchin$^{32}$, 
Yu.~Guz$^{34}$, 
T.~Gys$^{37}$, 
G.~Haefeli$^{38}$, 
C.~Haen$^{37}$, 
S.C.~Haines$^{43}$, 
T.~Hampson$^{42}$, 
S.~Hansmann-Menzemer$^{11}$, 
R.~Harji$^{49}$, 
N.~Harnew$^{51}$, 
J.~Harrison$^{50}$, 
P.F.~Harrison$^{44}$, 
J.~He$^{7}$, 
V.~Heijne$^{23}$, 
K.~Hennessy$^{48}$, 
P.~Henrard$^{5}$, 
J.A.~Hernando~Morata$^{36}$, 
E.~van~Herwijnen$^{37}$, 
E.~Hicks$^{48}$, 
W.~Hofmann$^{10}$, 
K.~Holubyev$^{11}$, 
P.~Hopchev$^{4}$, 
W.~Hulsbergen$^{23}$, 
P.~Hunt$^{51}$, 
T.~Huse$^{48}$, 
R.S.~Huston$^{12}$, 
D.~Hutchcroft$^{48}$, 
D.~Hynds$^{47}$, 
V.~Iakovenko$^{41}$, 
P.~Ilten$^{12}$, 
J.~Imong$^{42}$, 
R.~Jacobsson$^{37}$, 
A.~Jaeger$^{11}$, 
M.~Jahjah~Hussein$^{5}$, 
E.~Jans$^{23}$, 
F.~Jansen$^{23}$, 
P.~Jaton$^{38}$, 
B.~Jean-Marie$^{7}$, 
F.~Jing$^{3}$, 
M.~John$^{51}$, 
D.~Johnson$^{51}$, 
C.R.~Jones$^{43}$, 
B.~Jost$^{37}$, 
S.~Kandybei$^{40}$, 
M.~Karacson$^{37}$, 
T.M.~Karbach$^{9}$, 
J.~Keaveney$^{12}$, 
U.~Kerzel$^{37}$, 
T.~Ketel$^{24}$, 
A.~Keune$^{38}$, 
B.~Khanji$^{6}$, 
Y.M.~Kim$^{46}$, 
M.~Knecht$^{38}$, 
S.~Koblitz$^{37}$, 
P.~Koppenburg$^{23}$, 
A.~Kozlinskiy$^{23}$, 
L.~Kravchuk$^{32}$, 
K.~Kreplin$^{11}$, 
M.~Kreps$^{44}$, 
G.~Krocker$^{11}$, 
P.~Krokovny$^{11}$, 
F.~Kruse$^{9}$, 
K.~Kruzelecki$^{37}$, 
M.~Kucharczyk$^{20,25,37}$, 
S.~Kukulak$^{25}$, 
R.~Kumar$^{14,37}$, 
T.~Kvaratskheliya$^{30,37}$, 
V.N.~La~Thi$^{38}$, 
D.~Lacarrere$^{37}$, 
G.~Lafferty$^{50}$, 
A.~Lai$^{15}$, 
D.~Lambert$^{46}$, 
R.W.~Lambert$^{37}$, 
E.~Lanciotti$^{37}$, 
G.~Lanfranchi$^{18}$, 
C.~Langenbruch$^{11}$, 
T.~Latham$^{44}$, 
R.~Le~Gac$^{6}$, 
J.~van~Leerdam$^{23}$, 
J.-P.~Lees$^{4}$, 
R.~Lef\`{e}vre$^{5}$, 
A.~Leflat$^{31,37}$, 
J.~Lefran\c{c}ois$^{7}$, 
O.~Leroy$^{6}$, 
T.~Lesiak$^{25}$, 
L.~Li$^{3}$, 
L.~Li~Gioi$^{5}$, 
M.~Lieng$^{9}$, 
M.~Liles$^{48}$, 
R.~Lindner$^{37}$, 
C.~Linn$^{11}$, 
B.~Liu$^{3}$, 
G.~Liu$^{37}$, 
J.H.~Lopes$^{2}$, 
E.~Lopez~Asamar$^{35}$, 
N.~Lopez-March$^{38}$, 
J.~Luisier$^{38}$, 
F.~Machefert$^{7}$, 
I.V.~Machikhiliyan$^{4,30}$, 
F.~Maciuc$^{10}$, 
O.~Maev$^{29,37}$, 
J.~Magnin$^{1}$, 
S.~Malde$^{51}$, 
R.M.D.~Mamunur$^{37}$, 
G.~Manca$^{15,d}$, 
G.~Mancinelli$^{6}$, 
N.~Mangiafave$^{43}$, 
U.~Marconi$^{14}$, 
R.~M\"{a}rki$^{38}$, 
J.~Marks$^{11}$, 
G.~Martellotti$^{22}$, 
A.~Martens$^{7}$, 
L.~Martin$^{51}$, 
A.~Mart\'{i}n~S\'{a}nchez$^{7}$, 
D.~Martinez~Santos$^{37}$, 
A.~Massafferri$^{1}$, 
R.~Matev$^{37,p}$, 
Z.~Mathe$^{12}$, 
C.~Matteuzzi$^{20}$, 
M.~Matveev$^{29}$, 
E.~Maurice$^{6}$, 
B.~Maynard$^{52}$, 
A.~Mazurov$^{16,32,37}$, 
G.~McGregor$^{50}$, 
R.~McNulty$^{12}$, 
C.~Mclean$^{14}$, 
M.~Meissner$^{11}$, 
M.~Merk$^{23}$, 
J.~Merkel$^{9}$, 
R.~Messi$^{21,k}$, 
S.~Miglioranzi$^{37}$, 
D.A.~Milanes$^{13,37}$, 
M.-N.~Minard$^{4}$, 
S.~Monteil$^{5}$, 
D.~Moran$^{12}$, 
P.~Morawski$^{25}$, 
R.~Mountain$^{52}$, 
I.~Mous$^{23}$, 
F.~Muheim$^{46}$, 
K.~M\"{u}ller$^{39}$, 
R.~Muresan$^{28,38}$, 
B.~Muryn$^{26}$, 
M.~Musy$^{35}$, 
J.~Mylroie-Smith$^{48}$, 
P.~Naik$^{42}$, 
T.~Nakada$^{38}$, 
R.~Nandakumar$^{45}$, 
J.~Nardulli$^{45}$, 
I.~Nasteva$^{1}$, 
M.~Nedos$^{9}$, 
M.~Needham$^{46}$, 
N.~Neufeld$^{37}$, 
C.~Nguyen-Mau$^{38,q}$, 
M.~Nicol$^{7}$, 
S.~Nies$^{9}$, 
V.~Niess$^{5}$, 
N.~Nikitin$^{31}$, 
A.~Oblakowska-Mucha$^{26}$, 
V.~Obraztsov$^{34}$, 
S.~Oggero$^{23}$, 
S.~Ogilvy$^{47}$, 
O.~Okhrimenko$^{41}$, 
R.~Oldeman$^{15,d}$, 
M.~Orlandea$^{28}$, 
J.M.~Otalora~Goicochea$^{2}$, 
P.~Owen$^{49}$, 
B.~Pal$^{52}$, 
J.~Palacios$^{39}$, 
M.~Palutan$^{18}$, 
J.~Panman$^{37}$, 
A.~Papanestis$^{45}$, 
M.~Pappagallo$^{13,b}$, 
C.~Parkes$^{47,37}$, 
C.J.~Parkinson$^{49}$, 
G.~Passaleva$^{17}$, 
G.D.~Patel$^{48}$, 
M.~Patel$^{49}$, 
S.K.~Paterson$^{49}$, 
G.N.~Patrick$^{45}$, 
C.~Patrignani$^{19,i}$, 
C.~Pavel-Nicorescu$^{28}$, 
A.~Pazos~Alvarez$^{36}$, 
A.~Pellegrino$^{23}$, 
G.~Penso$^{22,l}$, 
M.~Pepe~Altarelli$^{37}$, 
S.~Perazzini$^{14,c}$, 
D.L.~Perego$^{20,j}$, 
E.~Perez~Trigo$^{36}$, 
A.~P\'{e}rez-Calero~Yzquierdo$^{35}$, 
P.~Perret$^{5}$, 
M.~Perrin-Terrin$^{6}$, 
G.~Pessina$^{20}$, 
A.~Petrella$^{16,37}$, 
A.~Petrolini$^{19,i}$, 
B.~Pie~Valls$^{35}$, 
B.~Pietrzyk$^{4}$, 
T.~Pilar$^{44}$, 
D.~Pinci$^{22}$, 
R.~Plackett$^{47}$, 
S.~Playfer$^{46}$, 
M.~Plo~Casasus$^{36}$, 
G.~Polok$^{25}$, 
A.~Poluektov$^{44,33}$, 
E.~Polycarpo$^{2}$, 
D.~Popov$^{10}$, 
B.~Popovici$^{28}$, 
C.~Potterat$^{35}$, 
A.~Powell$^{51}$, 
T.~du~Pree$^{23}$, 
J.~Prisciandaro$^{38}$, 
V.~Pugatch$^{41}$, 
A.~Puig~Navarro$^{35}$, 
W.~Qian$^{52}$, 
J.H.~Rademacker$^{42}$, 
B.~Rakotomiaramanana$^{38}$, 
M.S.~Rangel$^{2}$, 
I.~Raniuk$^{40}$, 
G.~Raven$^{24}$, 
S.~Redford$^{51}$, 
M.M.~Reid$^{44}$, 
A.C.~dos~Reis$^{1}$, 
S.~Ricciardi$^{45}$, 
K.~Rinnert$^{48}$, 
D.A.~Roa~Romero$^{5}$, 
P.~Robbe$^{7}$, 
E.~Rodrigues$^{47}$, 
F.~Rodrigues$^{2}$, 
P.~Rodriguez~Perez$^{36}$, 
G.J.~Rogers$^{43}$, 
S.~Roiser$^{37}$, 
V.~Romanovsky$^{34}$, 
J.~Rouvinet$^{38}$, 
T.~Ruf$^{37}$, 
H.~Ruiz$^{35}$, 
G.~Sabatino$^{21,k}$, 
J.J.~Saborido~Silva$^{36}$, 
N.~Sagidova$^{29}$, 
P.~Sail$^{47}$, 
B.~Saitta$^{15,d}$, 
C.~Salzmann$^{39}$, 
M.~Sannino$^{19,i}$, 
R.~Santacesaria$^{22}$, 
C.~Santamarina~Rios$^{36}$, 
R.~Santinelli$^{37}$, 
E.~Santovetti$^{21,k}$, 
M.~Sapunov$^{6}$, 
A.~Sarti$^{18,l}$, 
C.~Satriano$^{22,m}$, 
A.~Satta$^{21}$, 
M.~Savrie$^{16,e}$, 
D.~Savrina$^{30}$, 
P.~Schaack$^{49}$, 
M.~Schiller$^{11}$, 
S.~Schleich$^{9}$, 
M.~Schmelling$^{10}$, 
B.~Schmidt$^{37}$, 
O.~Schneider$^{38}$, 
A.~Schopper$^{37}$, 
M.-H.~Schune$^{7}$, 
R.~Schwemmer$^{37}$, 
A.~Sciubba$^{18,l}$, 
M.~Seco$^{36}$, 
A.~Semennikov$^{30}$, 
K.~Senderowska$^{26}$, 
I.~Sepp$^{49}$, 
N.~Serra$^{39}$, 
J.~Serrano$^{6}$, 
P.~Seyfert$^{11}$, 
B.~Shao$^{3}$, 
M.~Shapkin$^{34}$, 
I.~Shapoval$^{40,37}$, 
P.~Shatalov$^{30}$, 
Y.~Shcheglov$^{29}$, 
T.~Shears$^{48}$, 
L.~Shekhtman$^{33}$, 
O.~Shevchenko$^{40}$, 
V.~Shevchenko$^{30}$, 
A.~Shires$^{49}$, 
R.~Silva~Coutinho$^{54}$, 
H.P.~Skottowe$^{43}$, 
T.~Skwarnicki$^{52}$, 
A.C.~Smith$^{37}$, 
N.A.~Smith$^{48}$, 
K.~Sobczak$^{5}$, 
F.J.P.~Soler$^{47}$, 
A.~Solomin$^{42}$, 
F.~Soomro$^{49}$, 
B.~Souza~De~Paula$^{2}$, 
B.~Spaan$^{9}$, 
A.~Sparkes$^{46}$, 
P.~Spradlin$^{47}$, 
F.~Stagni$^{37}$, 
S.~Stahl$^{11}$, 
O.~Steinkamp$^{39}$, 
S.~Stoica$^{28}$, 
S.~Stone$^{52,37}$, 
B.~Storaci$^{23}$, 
M.~Straticiuc$^{28}$, 
U.~Straumann$^{39}$, 
N.~Styles$^{46}$, 
V.K.~Subbiah$^{37}$, 
S.~Swientek$^{9}$, 
M.~Szczekowski$^{27}$, 
P.~Szczypka$^{38}$, 
T.~Szumlak$^{26}$, 
S.~T'Jampens$^{4}$, 
E.~Teodorescu$^{28}$, 
F.~Teubert$^{37}$, 
C.~Thomas$^{51,45}$, 
E.~Thomas$^{37}$, 
J.~van~Tilburg$^{11}$, 
V.~Tisserand$^{4}$, 
M.~Tobin$^{39}$, 
S.~Topp-Joergensen$^{51}$, 
M.T.~Tran$^{38}$, 
A.~Tsaregorodtsev$^{6}$, 
N.~Tuning$^{23}$, 
M.~Ubeda~Garcia$^{37}$, 
A.~Ukleja$^{27}$, 
P.~Urquijo$^{52}$, 
U.~Uwer$^{11}$, 
V.~Vagnoni$^{14}$, 
G.~Valenti$^{14}$, 
R.~Vazquez~Gomez$^{35}$, 
P.~Vazquez~Regueiro$^{36}$, 
S.~Vecchi$^{16}$, 
J.J.~Velthuis$^{42}$, 
M.~Veltri$^{17,g}$, 
K.~Vervink$^{37}$, 
B.~Viaud$^{7}$, 
I.~Videau$^{7}$, 
X.~Vilasis-Cardona$^{35,n}$, 
J.~Visniakov$^{36}$, 
A.~Vollhardt$^{39}$, 
D.~Voong$^{42}$, 
A.~Vorobyev$^{29}$, 
H.~Voss$^{10}$, 
K.~Wacker$^{9}$, 
S.~Wandernoth$^{11}$, 
J.~Wang$^{52}$, 
D.R.~Ward$^{43}$, 
A.D.~Webber$^{50}$, 
D.~Websdale$^{49}$, 
M.~Whitehead$^{44}$, 
D.~Wiedner$^{11}$, 
L.~Wiggers$^{23}$, 
G.~Wilkinson$^{51}$, 
M.P.~Williams$^{44,45}$, 
M.~Williams$^{49}$, 
F.F.~Wilson$^{45}$, 
J.~Wishahi$^{9}$, 
M.~Witek$^{25,37}$, 
W.~Witzeling$^{37}$, 
S.A.~Wotton$^{43}$, 
K.~Wyllie$^{37}$, 
Y.~Xie$^{46}$, 
F.~Xing$^{51}$, 
Z.~Yang$^{3}$, 
R.~Young$^{46}$, 
O.~Yushchenko$^{34}$, 
M.~Zavertyaev$^{10,a}$, 
F.~Zhang$^{3}$, 
L.~Zhang$^{52}$, 
W.C.~Zhang$^{12}$, 
Y.~Zhang$^{3}$, 
A.~Zhelezov$^{11}$, 
L.~Zhong$^{3}$, 
E.~Zverev$^{31}$, 
A.~Zvyagin~$^{37}$.\\

\noindent
{\it \footnotesize
$ ^{1}$Centro Brasileiro de Pesquisas F\'{i}sicas (CBPF), Rio de Janeiro, Brazil\\
$ ^{2}$Universidade Federal do Rio de Janeiro (UFRJ), Rio de Janeiro, Brazil\\
$ ^{3}$Center for High Energy Physics, Tsinghua University, Beijing, China\\
$ ^{4}$LAPP, Universit\'{e} de Savoie, CNRS/IN2P3, Annecy-Le-Vieux, France\\
$ ^{5}$Clermont Universit\'{e}, Universit\'{e} Blaise Pascal, CNRS/IN2P3, LPC, Clermont-Ferrand, France\\
$ ^{6}$CPPM, Aix-Marseille Universit\'{e}, CNRS/IN2P3, Marseille, France\\
$ ^{7}$LAL, Universit\'{e} Paris-Sud, CNRS/IN2P3, Orsay, France\\
$ ^{8}$LPNHE, Universit\'{e} Pierre et Marie Curie, Universit\'{e} Paris Diderot, CNRS/IN2P3, Paris, France\\
$ ^{9}$Fakult\"{a}t Physik, Technische Universit\"{a}t Dortmund, Dortmund, Germany\\
$ ^{10}$Max-Planck-Institut f\"{u}r Kernphysik (MPIK), Heidelberg, Germany\\
$ ^{11}$Physikalisches Institut, Ruprecht-Karls-Universit\"{a}t Heidelberg, Heidelberg, Germany\\
$ ^{12}$School of Physics, University College Dublin, Dublin, Ireland\\
$ ^{13}$Sezione INFN di Bari, Bari, Italy\\
$ ^{14}$Sezione INFN di Bologna, Bologna, Italy\\
$ ^{15}$Sezione INFN di Cagliari, Cagliari, Italy\\
$ ^{16}$Sezione INFN di Ferrara, Ferrara, Italy\\
$ ^{17}$Sezione INFN di Firenze, Firenze, Italy\\
$ ^{18}$Laboratori Nazionali dell'INFN di Frascati, Frascati, Italy\\
$ ^{19}$Sezione INFN di Genova, Genova, Italy\\
$ ^{20}$Sezione INFN di Milano Bicocca, Milano, Italy\\
$ ^{21}$Sezione INFN di Roma Tor Vergata, Roma, Italy\\
$ ^{22}$Sezione INFN di Roma La Sapienza, Roma, Italy\\
$ ^{23}$Nikhef National Institute for Subatomic Physics, Amsterdam, Netherlands\\
$ ^{24}$Nikhef National Institute for Subatomic Physics and Vrije Universiteit, Amsterdam, Netherlands\\
$ ^{25}$Henryk Niewodniczanski Institute of Nuclear Physics  Polish Academy of Sciences, Cracow, Poland\\
$ ^{26}$Faculty of Physics \& Applied Computer Science, Cracow, Poland\\
$ ^{27}$Soltan Institute for Nuclear Studies, Warsaw, Poland\\
$ ^{28}$Horia Hulubei National Institute of Physics and Nuclear Engineering, Bucharest-Magurele, Romania\\
$ ^{29}$Petersburg Nuclear Physics Institute (PNPI), Gatchina, Russia\\
$ ^{30}$Institute of Theoretical and Experimental Physics (ITEP), Moscow, Russia\\
$ ^{31}$Institute of Nuclear Physics, Moscow State University (SINP MSU), Moscow, Russia\\
$ ^{32}$Institute for Nuclear Research of the Russian Academy of Sciences (INR RAN), Moscow, Russia\\
$ ^{33}$Budker Institute of Nuclear Physics (SB RAS) and Novosibirsk State University, Novosibirsk, Russia\\
$ ^{34}$Institute for High Energy Physics (IHEP), Protvino, Russia\\
$ ^{35}$Universitat de Barcelona, Barcelona, Spain\\
$ ^{36}$Universidad de Santiago de Compostela, Santiago de Compostela, Spain\\
$ ^{37}$European Organization for Nuclear Research (CERN), Geneva, Switzerland\\
$ ^{38}$Ecole Polytechnique F\'{e}d\'{e}rale de Lausanne (EPFL), Lausanne, Switzerland\\
$ ^{39}$Physik-Institut, Universit\"{a}t Z\"{u}rich, Z\"{u}rich, Switzerland\\
$ ^{40}$NSC Kharkiv Institute of Physics and Technology (NSC KIPT), Kharkiv, Ukraine\\
$ ^{41}$Institute for Nuclear Research of the National Academy of Sciences (KINR), Kyiv, Ukraine\\
$ ^{42}$H.H. Wills Physics Laboratory, University of Bristol, Bristol, United Kingdom\\
$ ^{43}$Cavendish Laboratory, University of Cambridge, Cambridge, United Kingdom\\
$ ^{44}$Department of Physics, University of Warwick, Coventry, United Kingdom\\
$ ^{45}$STFC Rutherford Appleton Laboratory, Didcot, United Kingdom\\
$ ^{46}$School of Physics and Astronomy, University of Edinburgh, Edinburgh, United Kingdom\\
$ ^{47}$School of Physics and Astronomy, University of Glasgow, Glasgow, United Kingdom\\
$ ^{48}$Oliver Lodge Laboratory, University of Liverpool, Liverpool, United Kingdom\\
$ ^{49}$Imperial College London, London, United Kingdom\\
$ ^{50}$School of Physics and Astronomy, University of Manchester, Manchester, United Kingdom\\
$ ^{51}$Department of Physics, University of Oxford, Oxford, United Kingdom\\
$ ^{52}$Syracuse University, Syracuse, NY, United States\\
$ ^{53}$CC-IN2P3, CNRS/IN2P3, Lyon-Villeurbanne, France, associated member\\
$ ^{54}$Pontif\'{i}cia Universidade Cat\'{o}lica do Rio de Janeiro (PUC-Rio), Rio de Janeiro, Brazil, associated to $^2 $\\

\noindent
$ ^{a}$P.N. Lebedev Physical Institute, Russian Academy of Science (LPI RAS), Moscow, Russia\\
$ ^{b}$Universit\`{a} di Bari, Bari, Italy\\
$ ^{c}$Universit\`{a} di Bologna, Bologna, Italy\\
$ ^{d}$Universit\`{a} di Cagliari, Cagliari, Italy\\
$ ^{e}$Universit\`{a} di Ferrara, Ferrara, Italy\\
$ ^{f}$Universit\`{a} di Firenze, Firenze, Italy\\
$ ^{g}$Universit\`{a} di Urbino, Urbino, Italy\\
$ ^{h}$Universit\`{a} di Modena e Reggio Emilia, Modena, Italy\\
$ ^{i}$Universit\`{a} di Genova, Genova, Italy\\
$ ^{j}$Universit\`{a} di Milano Bicocca, Milano, Italy\\
$ ^{k}$Universit\`{a} di Roma Tor Vergata, Roma, Italy\\
$ ^{l}$Universit\`{a} di Roma La Sapienza, Roma, Italy\\
$ ^{m}$Universit\`{a} della Basilicata, Potenza, Italy\\
$ ^{n}$LIFAELS, La Salle, Universitat Ramon Llull, Barcelona, Spain\\
$ ^{o}$Instituci\'{o} Catalana de Recerca i Estudis Avan\c{c}ats (ICREA), Barcelona, Spain\\
$ ^{p}$University of Sofia, Sofia, Bulgaria\\
$ ^{q}$Hanoi University of Science, Hanoi, Viet Nam\\
}
\newpage

\section{Introduction}
\label{sec:intro}

Absolute luminosity measurements are of general interest to
colliding-beam experiments at storage rings. 
Such measurements are
necessary to determine the absolute cross-sections of reaction processes
and to quantify the performance of the accelerator. 
The required accuracy
on the value of the cross-section depends on both the process
of interest and the precision of the theoretical predictions. 
At the LHC, the required precision on the cross-section is expected to
be of order 1--2\%. 
This estimate is motivated by the accuracy of theoretical
predictions for the production of vector bosons and for the two-photon
production of muon 
pairs~\cite{ref:ws-precision-mangano,ref:ws-precision-anderson,ref:thorne,ref:delorenzi}. 

In a cyclical collider, such as the LHC, the average instantaneous
luminosity of one pair of colliding bunches can be expressed 
as~\cite{ref:moller}
\begin{equation}
 \label{eq:luminosity}
  L =  N_1 \, N_2 \, f  \sqrt{(\vec{v}_1-\vec{v}_2)^2-\frac{(\vec{v}_1\times \vec{v}_2)^2}{c^2}}\int{\rho_1(x,y,z,t)\rho_2(x,y,z,t) \
  {\rm d}x\,{\rm d}y\,{\rm d}z\,{\rm d}t} \ ,
\end{equation}
where we have introduced the revolution frequency $f$ (11245 Hz at the LHC),
the numbers of protons $N_1$ and $N_2$ in the two bunches,
the corresponding velocities $\vec{v}_1$ and $\vec{v}_2$ of the 
particles,\footnote{In the approximation of zero emittance the velocities
are the same within one bunch.} 
and the 
particle densities for beam~1 and beam~2, $\rho_{1,2}(x,y,z,t)$.
The particle densities are normalized such that their individual
integrals over all space are unity. 
For highly relativistic beams colliding with a very small 
half crossing-angle \thetac,
the M{\o}ller factor 
$\sqrt{(\vec{v}_1-\vec{v}_2)^2-{(\vec{v}_1\times \vec{v}_2)^2}/{c^2}}$
reduces to $2c \cos^2{\thetac} \simeq 2c$.
The integral in Eq.~\ref{eq:luminosity} is known as the beam overlap
integral.

Methods for absolute luminosity determination are generally classified 
as either direct or indirect.
Indirect methods are {\em e.g.} the use of the optical theorem 
 to make a simultaneous measurement of the elastic and total
 cross-sections~\cite{ref:totem,ref:alfa}, or the comparison to a
 process of 
which the absolute cross-section is known, either from theory or by a
previous direct measurement. 
Direct measurements make use of Eq.~\ref{eq:luminosity} and employ several
strategies to measure the various parameters in the equation.

The analysis described in this paper relies on two direct methods to
determine the absolute luminosity calibration:  
the ``van der Meer scan'' method (VDM)~\cite{ref:vdm-method,ref:vdm-LHC} and 
the ``beam-gas imaging'' method (BGI)~\cite{ref:massi-bgmethod}.
The BGI method is based on reconstructing beam-gas interaction
vertices to measure the beam angles, positions and shapes.
It was applied for the first time in LHCb (see
Refs.~\cite{ref:lhcb-kshort,ref:balagura-moriond,ref:plamen-moriond}) using  
the first LHC data collected at the end of 2009 at
$\sqrt{s} = 900~\GeV$. 
The BGI method relies on the high precision of the measurement
of interaction vertices obtained with the LHCb vertex detector. 
The VDM method exploits the ability to move the beams in both
transverse coordinates with high precision and to thus scan the
colliding beams with respect to each other. 
This method is also being used by other LHC experiments~\cite{ref:otherlhc}. 
The method was first applied at the CERN ISR~\cite{ref:vdm-method}.
Recently it was demonstrated that additional information can be
extracted when the two beams probe each other such as during a VDM scan,
allowing the individual beam profiles to be determined by using vertex
measurements of $pp$ interactions in beam-beam
collisions (beam-beam imaging)~\cite{ref:vdmimaging}. 

In principle, beam profiles can also be obtained by scanning wires
across the beams~\cite{ref:wire-method} or by inferring the beam
properties by theoretical calculation from the beam optics.
Both methods lack precision, however, as they both rely on detailed
knowledge of the beam optics. 
The wire-scan method is limited by the achievable proximity of the wire
to the interaction region which introduces the dependence on the beam
optics model.

The LHC operated with a $pp$ centre-of-mass energy of 7~TeV 
(3.5~TeV per beam).  
Typical values observed for the transverse beam sizes are close to 
50~\micron and 55~mm for the bunch length.
The half-crossing angle was typically 0.2~\mrad.
  
Data taken with the LHCb detector, located at interaction point (IP) 8,
are used in conjunction with data from the LHC beam instrumentation. 
The measurements obtained with the VDM and BGI methods are found to be
consistent, and an average is made for the final result. 
The limiting systematics in both measurements come from the knowledge of
the bunch populations $N_1$ and $N_2$. 
All other sources of systematics are specific to the analysis
method. 
Therefore, the comparison of both methods provides an important cross
check of the results.
The beam-beam imaging method is applied to the data taken during the VDM
scan as an overall cross check of the absolute luminosity measurement.

Since the absolute calibration can only be performed during specific
running periods, a relative normalization method 
is needed to transport the results of the absolute calibration of the
luminosity to the complete data-taking period. To this end we defined a
class of visible interactions.
The cross-section for these interactions is determined using the
measurements of the absolute luminosity during specific data-taking
periods. 
Once this visible cross-section is determined, the integrated
luminosity for a period of data-taking is obtained by accumulating the
count rate of the corresponding visible interactions over this period. 
Thus, the calibration of the absolute luminosity is translated into a
determination of a well defined visible cross-section.

In the present paper we first describe briefly the LHCb detector in
Sect.~\ref{sec:lhcb}, and in particular those aspects relevant to the
analysis presented here. 
In  Sect.~\ref{sec:relative} the methods used for the relative
normalization technique are given.
The determination of the number of protons in the LHC bunches is
detailed in Sect.~\ref{sec:bct}. 
The two methods which are used to determine the absolute scale are
described in Sect.~\ref{sec:vdm} and \ref{sec:beamgas}, respectively.
The cross checks made with the beam-beam imaging method are
shown in Sect.~\ref{sec:vdmimagingsigma}.
Finally, the results are combined in
Sect.~\ref{sec:results-conclusions}. 

\section{The LHCb detector}
\label{sec:lhcb}

The LHCb detector is a magnetic dipole spectrometer with a polar angular
coverage of approximately 10 to 300~mrad in the horizontal
(bending) plane, and 10 to 250~mrad in the vertical plane. 
It is described in detail elsewhere~\cite{ref:lhcb}.  
A right-handed coordinate system is defined with its origin at the 
nominal $pp$ interaction point, the $z$ axis along the average nominal beam
line and pointing towards the magnet, and the $y$ axis pointing upwards. 
Beam~1 (beam~2) travels in the direction of positive (negative) $z$.

The apparatus contains tracking detectors, ring-imaging Cherenkov
detectors, calorimeters, and a muon system. 
The tracking system comprises the vertex locator (VELO) 
surrounding the $pp$ interaction region, a tracking station upstream of
the dipole magnet and three tracking stations located
downstream of the magnet.  
Particles traversing the spectrometer experience a bending-field integral
of around 4~Tm.

\begin{figure}[tbp]
 \centering
 \includegraphics*[width=0.9\textwidth]{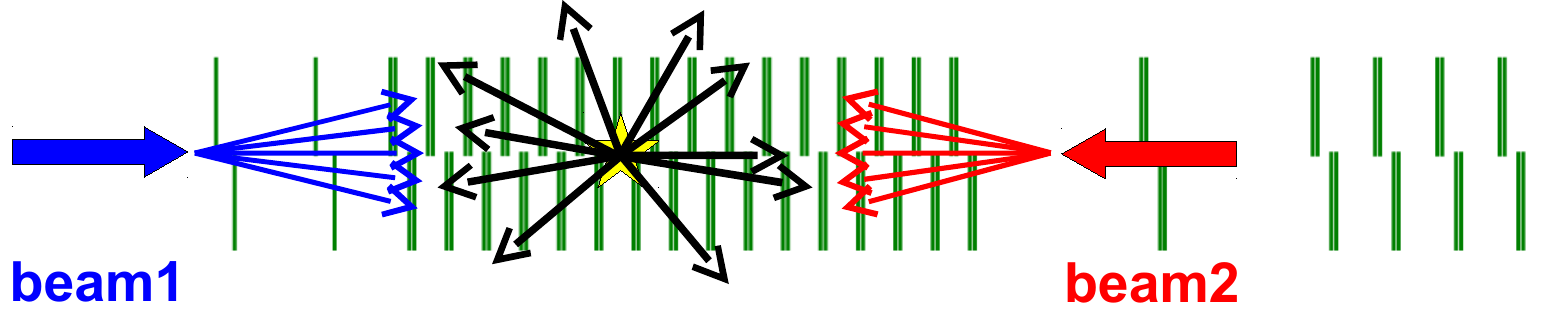}
 \caption{A sketch of the VELO, including the two Pile-Up stations on the
 left. 
 The VELO sensors are drawn as double lines while the PU sensors are
 indicated with single lines.
 The thick arrows 
 indicate the direction of the LHC beams (beam~1 going from left to right), while the 
 thin ones show example directions of flight of the products of the beam-gas and 
 beam-beam interactions.
 \label{fig:velo-sketch}
 }
\end{figure}

The VELO plays an essential role in the application of the
beam-gas imaging method at LHCb. 
It consists of two retractable halves, each
having 21 modules of radial and azimuthal silicon-strip sensors in a
half-circle shape, see Fig.~\ref{fig:velo-sketch}. 
Two additional stations ({\it Pile-Up System}, PU) upstream of the VELO
tracking stations are mainly used in the hardware trigger. 
The VELO has a large acceptance for beam-beam interactions 
owing to its many layers of silicon sensors and their close proximity to
the beam line. 
During nominal operation, the distance between sensor and beam is only
8~mm. 
During injection and beam adjustments, the two VELO halves are moved apart
in a retracted position away from the beams. 
They are brought to their
nominal position close to the beams during stable beam periods only. 

The LHCb trigger system consists of two separate
levels: a hardware trigger (L0), which is implemented in custom electronics, and a
software High Level Trigger (HLT), which is executed on a farm of
commercial processors.  
The L0 trigger system is designed to run at
1~MHz and uses information from the Pile-Up sensors of
the VELO, the calorimeters and the muon system.
They send information to the L0 decision
unit (L0DU) where selection algorithms are run synchronously with the
40~MHz LHC bunch-crossing signal. 
For every nominal bunch-crossing slot ({\em i.e.} each 25 ns) the L0DU
sends decisions to the LHCb readout supervisor.  
The full event information of all sub-detectors is available to the HLT
algorithms. 

A trigger strategy is adopted to select $pp$ inelastic interactions and
collisions of the beam with the residual gas in the vacuum chamber.
Events are collected for the four bunch-crossing types:
two colliding bunches (\bx{bb}), one beam~1 bunch with no beam~2 bunch (\bx{be}),
one beam~2 bunch with no beam~1 bunch (\bx{eb}) and nominally empty bunch slots
(\bx{ee}).  
Here ``\bx{b}'' stands for ``beam'' and ``\bx{e}'' stands for ``empty''.
The first two categories of crossings, which produce particles in the
forward direction, are triggered using calorimeter information.
An additional PU veto is applied for \bx{be} crossings.
Crossings of the type \bx{eb}, which produce particles 
in the backward direction, are triggered by demanding a minimal hit
multiplicity in the PU, and vetoed by calorimeter activity.
The trigger for \bx{ee} crossings is defined as the logical OR of the
conditions used for the \bx{be} and \bx{eb} crossings in order to be
sensitive to background from both beams.
During VDM scans specialized trigger conditions are defined which
optimize the data taking for these measurements (see Sect.~\ref{sec:vdm:conditions}).

The precise reconstruction of interaction vertices (``primary
vertices'', PV) is an essential ingredient in the analysis described in
this paper. 
The initial estimate of the PV position is based on an iterative
clustering of tracks (``seeding'').  
Only tracks with hits in the VELO are considered. 
For each track the distance of closest approach (DOCA) with all
other tracks is calculated and tracks are clustered into a seed if their 
DOCA is less than 1~mm. 
The position of the seed is then obtained using an iterative procedure. 
The point of closest approach between all track pairs is calculated and
its coordinates are used to discard outliers and to determine the
weighted average position. 
The final PV coordinates are determined by iteratively improving the
seed position with an adaptive, weighted, least-squares fit.
In each iteration a new PV position is evaluated. 
Participating tracks are extrapolated to the $z$ coordinate of the PV
and assigned weights depending on their impact parameter with respect to the PV. 
The procedure is repeated for all seeds, excluding tracks from
previously reconstructed primary vertices, retaining only PVs with at 
least five tracks. 
For this analysis only PVs with a larger number of tracks are used since they
have better resolution. 
For the study of beam-gas interactions
only PVs with at least ten tracks are used and at least 25 tracks are
required for the study of $pp$ interactions.

\section{Relative normalization method}
\label{sec:relative}

The absolute luminosity is obtained only for short periods of data-taking. 
To be able to perform cross-section measurements on any selected data sample,
the relative luminosity must be measured consistently during the full
period of data taking. 
The systematic relative normalization of all data-taking periods requires
specific procedures to be applied in the trigger, data-acquisition,
processing and final analysis. 
The basic principle is to acquire luminosity data together with the
physics data and to store it in the same files as the physics event data. 
During further processing of the physics data the relevant
luminosity data is kept together in the same storage entity.
In this way, it remains possible to select only part of the full
data-set for analysis and still keep the capability to determine
the corresponding integrated luminosity.

The luminosity is proportional to the average number of
visible proton-proton interactions in a beam-beam crossing, \mueff{\eff}. 
The subscript ``\eff'' is used to indicate that this holds for an
arbitrary definition of the visible cross-section.
Any stable interaction rate can be used as relative luminosity
monitor.
For a given period of data-taking, the integrated interaction rate can be
used to determine the integrated luminosity if the cross-section
for these visible interactions is known.
The determination of the cross-section corresponding to these visible
interactions is achieved by calibrating the absolute luminosity during
specific periods and simultaneously counting the visible interactions.  

Triggers which initiate the full
readout of the LHCb detector are created for random beam crossings.
These are called ``luminosity triggers''.
During normal physics data-taking, the overall rate is chosen to be
997 Hz, with 70\% assigned to \bx{bb}, 15\% to \bx{be}, 10\% to \bx{eb} and the
remaining 5\% to \bx{ee} crossings.
The events taken for crossing types other than \bx{bb} are used for
background subtraction and beam monitoring.
After a processing step in the HLT a small number of ``luminosity
counters'' are stored for each of these random luminosity triggers.
The set of luminosity counters comprise
the number of vertices and tracks in the VELO, the number of hits in the
PU and in the 
scintillator pad detector (SPD) in front of calorimeters, and the
transverse energy deposition in the calorimeters. 
Some of these counters are directly obtained from the L0, others are the
result of partial event-reconstruction in the HLT.

During the final analysis stage the event data and luminosity data are
available on the same files.
The luminosity counters are summed (when necessary after
time-dependent calibration) and an absolute calibration factor is
applied to obtain the absolute integrated luminosity. 
The absolute calibration factor is universal and is the result of the
luminosity calibration procedure described in this paper. 

The relative luminosity can be determined by summing the values of
any counter which is linear with the instantaneous luminosity.  
Alternatively, one may determine the relative luminosity from the
fraction of ``empty'' or invisible events in \bx{bb} crossings which we
denote by \PZ.  
An invisible event is defined by applying a counter-specific threshold 
below which it is considered that no $pp$ interaction
was seen in the corresponding bunch crossing.
Since the number of events per bunch crossing follows a Poisson
distribution with mean value proportional to the luminosity, the
luminosity is proportional to $-\ln \PZ$.  
This ``zero count'' method is both robust and easy to
implement~\cite{ref:zaitsev}.
In the absence of backgrounds, the average number of visible $pp$
interactions per crossing can be obtained from the fraction of empty
\bx{bb} crossings by $\mueff{\eff} = -\ln{\PZ[\bx{bb}]}$.
Backgrounds are subtracted using 
\begin{equation} 
\label{eq:mu} 
\mueff{\eff} = -\left(\ln{\PZ[\bx{bb}]} - \ln{\PZ[\bx{be}]} -
\ln{\PZ[\bx{eb}]} + \ln{\PZ[\bx{ee}]}\right) \, ,
\end{equation}
where $\PZ[i] (i=\bx{bb},\bx{ee},\bx{be},\bx{eb})$ are the
probabilities to find an empty event in a bunch-crossing slot for the
four different bunch-crossing types.
The $\PZ[\bx{ee}]$ contribution is added because it is also contained in
the $\PZ[\bx{be}]$ and $\PZ[\bx{eb}]$ terms. 
The purpose of the background subtraction, Eq.~\ref{eq:mu}, is to
correct the count-rate in the \bx{bb} crossings for the detector
response which is due to beam-gas interactions and detector noise.  
In principle, the noise background is measured during \bx{ee} crossings.
In the presence of parasitic beam protons in \bx{ee} bunch positions, as
will be discussed below, it is not correct to evaluate the noise 
from $\PZ[\bx{ee}]$.
In addition, the detector signals are not fully confined within one
25~ns bunch-crossing slot.
The empty (\bx{ee}) bunch-crossing slots immediately following a \bx{bb}, \bx{be}
or \bx{eb} crossing slot contain detector signals from interactions
occurring in the preceding slot (``spill-over'').
The spill-over background is not present in the \bx{bb}, \bx{be}
and \bx{eb} crossings.
Therefore, since the detector noise for the selected counters is
small ($< \, 10^{-5}$ relative to the typical values measured
during \bx{bb} crossings) the term $\ln{\PZ[\bx{ee}]}$ in 
Eq.~\ref{eq:mu} is neglected. 
Equation~\ref{eq:mu} assumes that the proton populations in the \bx{be}
and \bx{eb} crossings are the same as in the \bx{bb} crossings.
With a population spread of typically 10\% and a beam-gas background
fraction $< \, 10^{-4}$ compared to the $pp$ interactions the effect of
the spread is negligible, and is not taken into account.

\begin{figure}[tbp]
 \centering
 \includegraphics*[width=0.48\textwidth]{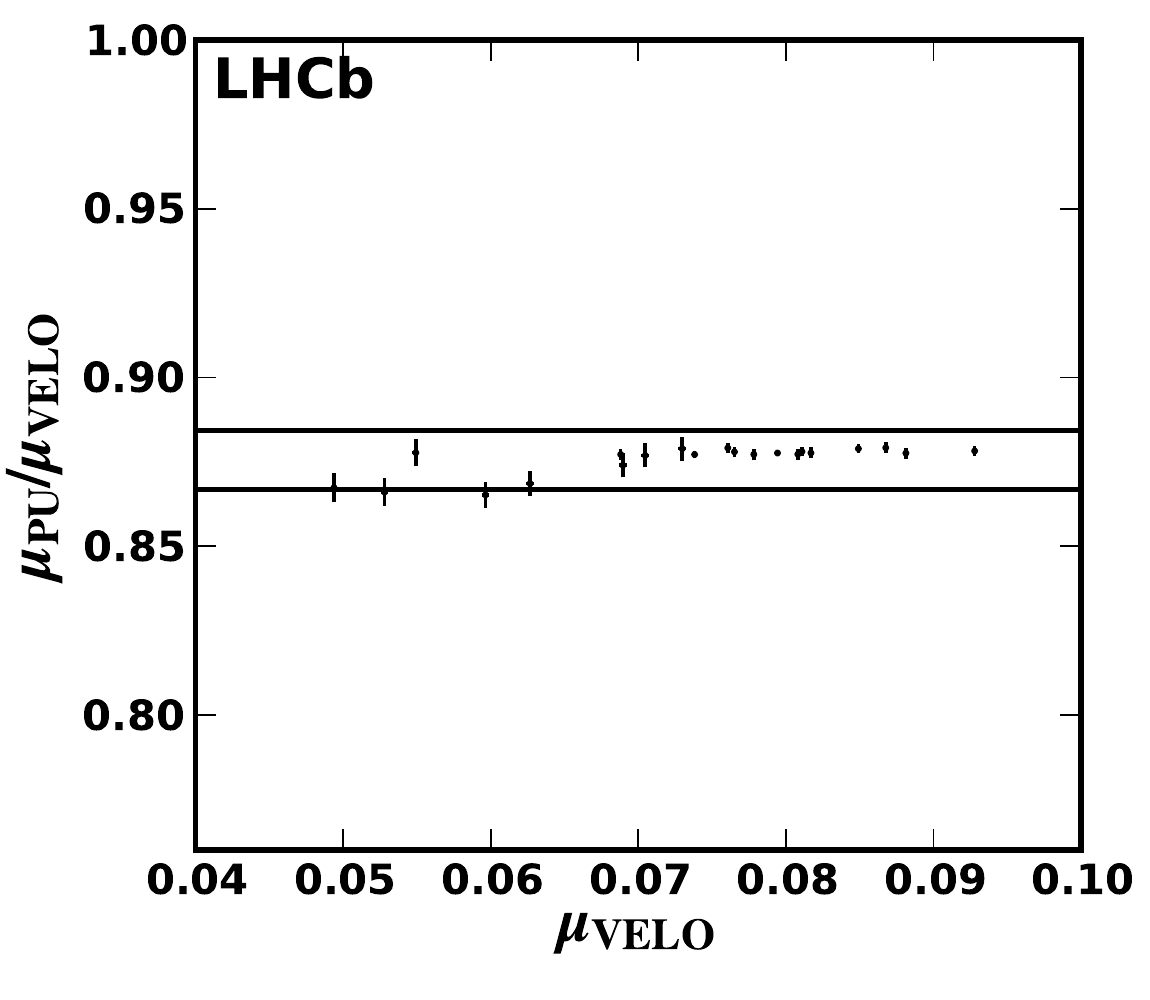}
 \includegraphics*[width=0.48\textwidth]{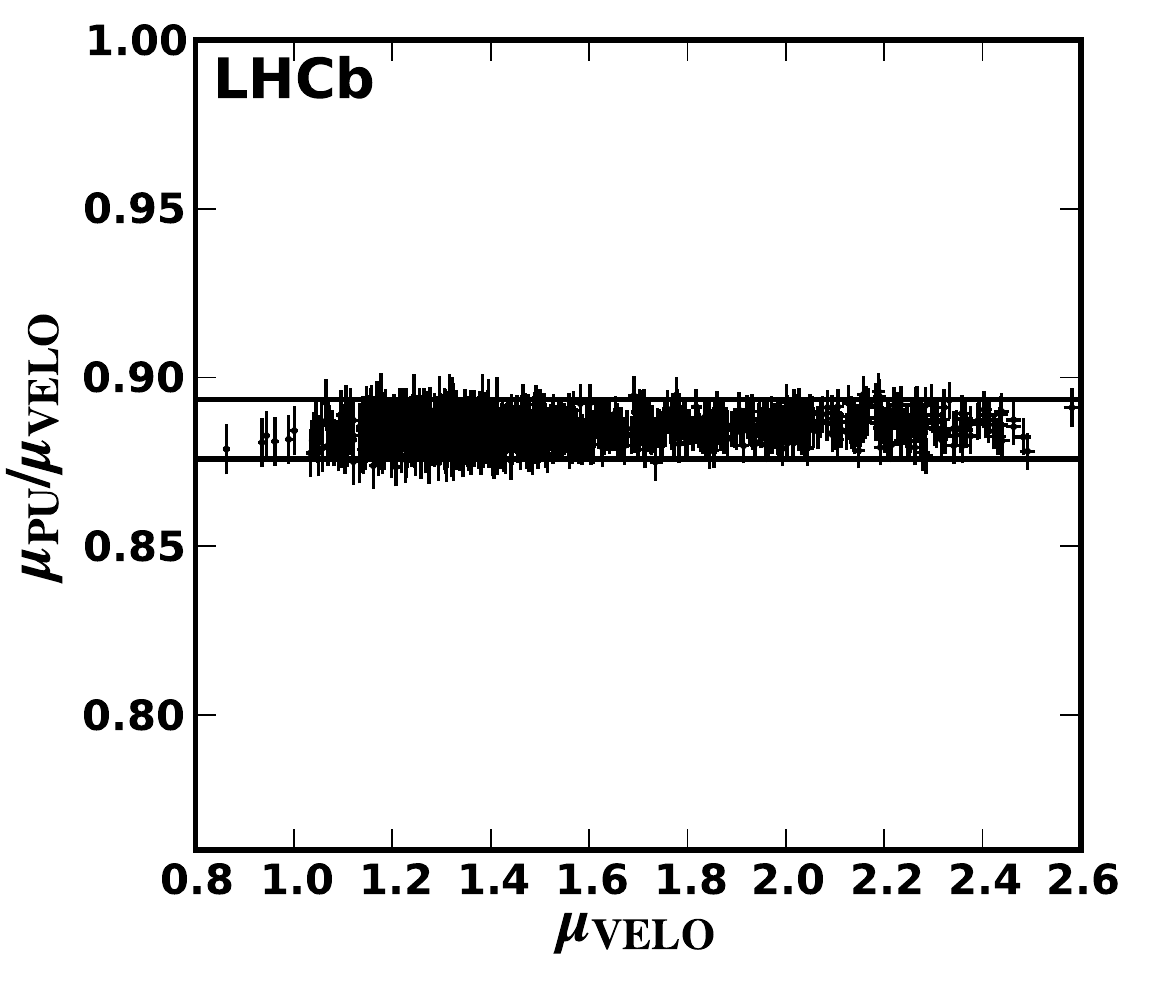}
 \caption{
 Ratio between \mueff{\eff} values
 obtained with the zero-count method using the number of hits in the PU
 and the track count in the VELO  versus $\mueff{VELO}$. 
 The deviation from unity is due to the difference in
 acceptance. The left (right) panel uses runs from the beginning (end) of
 the 2010 running period with lower (higher) values of \mueff{VELO}.
 The horizontal lines indicate a $\pm1$\% variation.
 \label{fig:muratio}
 }
\end{figure}

\begin{figure}[tbp]
 \centering
 \includegraphics[width=0.8\textwidth, clip, trim=0.cm 0.cm 0.3cm 0cm]{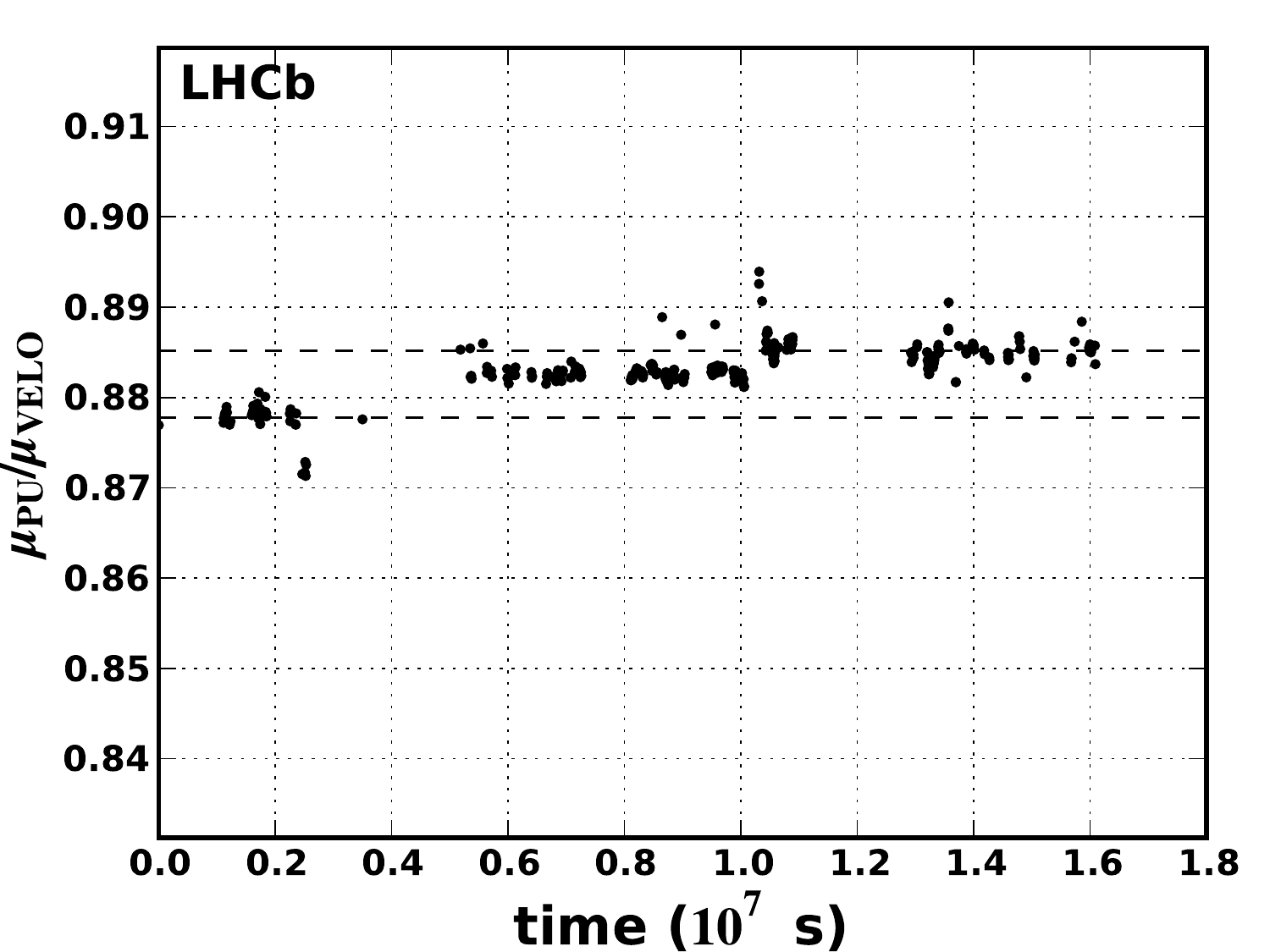}
 \caption{
 \label{fig:countersratio}
 Ratio between \mueff{\eff} values
 obtained with the zero-count method using the number of hits in the PU
 and the track count in the VELO as a function of time in seconds relative
 to the first run of LHCb in 2010. 
 The period spans the full 2010 data taking period (about half a year).  
 The dashed lines show the average  value of the starting and ending
 periods (the first and last 25 runs, respectively) and differ by
 $\approx1$\%.  
 The changes in the average values between the three main groups
 ($t < 0.4\tms 10^7$~s, $0.4\tms 10^7 < t < 1.2\tms 10^7$~s, $t > 1.2\tms 10^7$~s)
 coincide with known maintenance changes to the PU system.
 The upward excursion near $1.05 \tms 10^7$~s is due to background
 introduced by  parasitic collisions located at 37.5~m from the nominal
 IP present in the bunch filling scheme used for these
 fills to which the two counters have different sensitivity. 
 The downward excursion near $0.25 \tms 10^7$~s is due to known hardware
 failures in the PU (recovered after maintenance).
 The statistical errors are smaller than the symbol size of the data points. 
 }
\end{figure}

The results of the zero-count method based on the number of hits in the
PU and on the number of tracks in the VELO are found to be the most stable 
ones.
An empty event is defined to have $< 2$~hits when the PU
is considered or $< 2$~tracks when the VELO is considered.  A VELO track
is defined by at least three hits on a straight line in the radial
strips of the silicon detectors of the VELO.
The number of tracks reconstructed in the VELO is chosen as 
the most stable counter.  
In the following we will use the notation $\sigeff{\eff}$
($=\sigeff{VELO}$) for the visible cross-section measured using this
method, except when explicitly stated otherwise.
Modifications and alignment variations of the VELO also have
negligible impact on the method, since the efficiency for reconstructing
at least two tracks in an inelastic event is very stable against
detector effects.  
Therefore, the systematics associated with this choice of threshold is
negligible. 
The stability of the counter is demonstrated in Fig.~\ref{fig:muratio} which 
shows the ratio of the relative luminosities determined with the zero-count
method from the multiplicity of hits in the PU and from the number of
VELO tracks.  
Apart from a few threshold updates in the PU configuration, the PU was
also stable throughout LHCb 2010 running, and it was used as a cross check. 
Figure~\ref{fig:countersratio} covers the whole period of LHCb operation
in 2010, with both low and high number of interactions per crossing. 
Similar cross checks have been made with the counter based on the number
of reconstructed vertices.
These three counters have different systematics, and by comparing their
ratio as a function of time and instantaneous luminosity we
conclude that the relative luminosity measurement has a systematic error
of 0.5\%. 

The number of protons, beam sizes and transverse offsets at the
interaction point vary across bunches.  
Thus, the \mueff{\eff} value varies across \bx{bb} bunch crossings.
The spread in \mueff{\eff} is about 10\% of the mean value for typical runs. 
Due to the non-linearity of the logarithm function one first needs to
compute \mueff{\eff} values for different bunch crossings and then to
take the average. 
However, for short time intervals the statistics are insufficient to
distinguish between bunch-crossing IDs, while one cannot assume
\mueff{\eff} to be constant when the intervals are too long due to {\em
e.g.} loss of bunch population and emittance growth. 
If the spread in instantaneous \mueff{\eff} is known, the effect of
neglecting it in calculating an average value of \mueff{\eff} can be 
estimated.  
The difference between the naively computed \mueff{\eff} value and the
true one is then
\begin{equation}
\mueff[\mathrm{biased}]{\eff} - \mueff[{\mathrm{true}}]{\eff} = -\ln\langle \PZ[i]\rangle
 - (-\langle\, \ln \PZ[i]\,\rangle) =
 \langle\, \ln(\frac{\PZ[i]}{\,\langle \PZ[i]\,\rangle})\rangle \, ,
\end{equation}
where the average is taken over all beam-beam crossing slots $i$. 
Therefore, the biased \mueff{\eff} value can be calculated over short
time intervals and a correction for the spread of \mueff{\eff} can in
principle be applied by computing $\PZ[i]/\langle\, \PZ[i]\, \rangle $
over long time intervals. 
At the present level of accuracy, this correction is not 
required.\footnote{The relative luminosity increases by $0.5\%$
when the correction is applied.}
The effect is only weakly dependent on the luminosity counter used. 

\section{Bunch population measurements}
\label{sec:bct}

To measure the number of particles in the LHC beams
two types of beam current transformers are
installed in each ring~\cite{ref:bct}.
One type, the DCCT (DC Current Transformer), measures the total current
of the beams. 
The other type, the FBCT (Fast Beam Current Transformer),
is gated with 25~ns intervals and is used to measure the relative
charges of the individual bunches. 
The DCCT is absolutely calibrated, thus is used to constrain the
total number of particles, while the FBCT defines the relative bunch
populations. 
The procedure is described in detail in Ref.~\cite{ref:note-bcnwg}.
All devices have two independent readout systems.
For the DCCT both systems provide reliable information and their average
is used in the analysis, while for the FBCT one of the two systems is
dedicated to tests and cannot be used.

The absolute calibration of the DCCT is determined using a
high-precision current source. 
At low intensity (early data) the noise in the DCCT readings is
relatively important, while at the higher intensities of the data taken
in October 2010 this effect is negligible.
The noise level and its variation is determined by interpolating the
average DCCT readings over long periods of time without beam before and
after the relevant fills. 

In addition to the absolute calibration of the DCCTs, a deviation from the
proportionality of the FBCT readings to the individual bunch charges is
a potential source of systematic uncertainty.
The FBCT charge offsets are cross checked using the ATLAS BPTX (timing)
system~\cite{ref:ATLAS-BPTX}. 
This comparison shows small discrepancies between their offsets.
These deviations are used as an estimate of the uncertainties. 
Since the FBCT equipment is readjusted at regular intervals, the offsets
can vary on a fill-by-fill basis.
Following the discussion in Ref.~\cite{ref:note-bcnwg}, an estimate
of 2.9\% is used for the uncertainty of an individual bunch population
product of a colliding bunch pair.
Owing to the DCCT constraint on the total beam current, the overal
uncertainty is reduced when averaging results of different bunch pairs
within a single fill.
As will be discussed in Sect.~\ref{sec:vdm}, for the analysis of the VDM
data a method can be used which only needs the assumption of the
linearity of the FBCT response.

The LHC radio frequency (RF) system operates at 400 MHz, compared to the
nominal 40 MHz bunch frequency.
If protons circulate in the ring outside the nominal RF buckets, the
readings of the DCCT need to be corrected before they are used to
normalize the sum of the FBCT signals. 
We define ``satellite'' bunches as charges in neighbouring
RF buckets compared to the nominal bucket. 
Satellite bunches inside the nominally filled bunch slots can be
detected by the LHC experiments when there is no (or a very small)
crossing angle between the two beams.  
The satellites would be observed as interactions displaced by a multiple
of 37.5~cm from the nominal intersection point.
For a part of the 2010 run the ATLAS and CMS experiments were operating
with zero crossing angle and displaced interactions were indeed
observed~\cite{ref:note-bcnwg}.  

The ``ghost charge'' is defined as the charge outside the nominally
filled bunch slots. 
The rates of beam-gas  events produced by ``ghost'' and nominal protons
are measured using the beam-gas trigger.
The ghost fraction is determined by comparing the number of beam-gas
interactions during \bx{ee} crossings with the numbers observed in \bx{be} and
\bx{eb} crossings. 
The timing of the LHCb trigger is optimized for interactions in the
nominal RF buckets. 
The trigger efficiency depends on the time of the interaction with
respect to the  phase of the clock (modulo 25~\ns).  
A measurement of the trigger efficiency was performed by shifting the
clock which is usually synchronized with the LHC bunch-crossing time by
5, 10 and 12.5~ns and by comparing the total beam-gas rates in the
nominal crossings.
From these data the average efficiency for ghost charge
is obtained to be 
$\epsilon_{\mathrm{average}} = 0.86 \,\pm\, 0.14$  ($0.84 \,\pm\, 0.16$)
for beam~1 (beam~2).
The ghost charge is measured for each fill during which an absolute luminosity
measurement is performed and is typically 1\% of the total beam charge
or less. 
The contribution of ``ghost'' protons to the total LHC beam current
is subtracted from the DCCT value before the sum of the FBCT
bunch populations is constrained by the DCCT measurement of the total
current.
The uncertainty assigned to the subtraction of ghost charge varies
per fill and is due to the trigger efficiency uncertainty and the
limited statistical accuracy.
These two error components are of comparable size.

\section{The van der Meer scan (VDM) method}
\label{sec:vdm}

The beam position scanning method,  invented by van der Meer, 
provides a direct determination of an effective cross-section \sigeff{\eff} by
measuring the corresponding counting rate as a function of the position
offsets of two colliding beams~\cite{ref:vdm-method}.  
At the ISR only vertical displacements were needed owing to the crossing
angle between the beams in the horizontal plane and to the fact
that the beams were not bunched.
For the LHC the beams have to be scanned in both transverse
directions due to the fact that the beams are bunched~\cite{ref:vdm-LHC}. 
The cross-section \sigeff{\eff} can be measured for two colliding bunches
using the equation~\cite{ref:vdmimaging}
\begin{equation}
 \label{eq:vdm}
  \sigeff{\eff} = \frac{\int\!\mueff{\eff}(\Deltax, \Deltayz)\,d \Deltax\, 
  \int\!\mueff{\eff}(\Deltaxz, \Deltay)\,d\Deltay}{N_1 N_2 
  \mueff{\eff}(\Deltaxz, \Deltayz)\cos{\thetac}} \, ,
\end{equation}
where $\mueff{\eff}(\Deltax, \Deltay)$ is the average number of
interactions per crossing at offset $(\Deltax, \Deltay)$ corresponding
to the cross-section \sigeff{\eff}. 
The interaction rates $R(\Deltax$,$\Deltay)$ are related to 
$\mueff{\eff}(\Deltax, \Deltay)$ by the revolution frequency, 
$R(\Deltax$,$\Deltay) = f\, \mueff{\eff}(\Deltax$,$\Deltay)$.
These rates are measured at offsets $\Deltax$ and $\Deltay$ with
respect to their nominal positions at offsets $(\Deltaxz,\Deltayz)$. 
The scans consist of creating offsets $\Deltax$ and $\Deltay$ such
that practically the full profiles of the beams are explored.
The measured rate integrated over the displacements 
gives the cross-section.

The main assumption is that the  density distributions in the orthogonal
coordinates $x$ and $y$ can be factorized.
In that case, two scans are sufficient to obtain the cross-section:
one along a constant $y$-displacement $\Deltayz$ and 
one along a constant $x$-displacement $\Deltaxz$.
It is also assumed that effects due to bunch evolution during the scans
(shape distortions or transverse kicks  due to beam-beam effects,
emittance growth, bunch current decay), 
effects due to the tails of the bunch density distribution in the
transverse plane and 
effects of the absolute length scale calibration against magnet current
trims are either negligible or can be corrected for.

\subsection{Experimental conditions during the van der Meer scan}
\label{sec:vdm:conditions}

VDM scans were performed in LHCb during dedicated LHC fills at the
beginning and at the end of the 2010 running period, one in April and
one in October.  
The characteristics of the beams are summarized in Table~\ref{tbl1}.  
In both fills there is one scan where both beams moved symmetrically and
one scan where only one beam moved at a time.  
Precise beam positions are calculated from the LHC magnet currents and
cross checked with vertex measurements using the LHCb VELO,
as described below.

\begin{table}[tb]
 \centering
 \caption{Parameters of LHCb van der Meer scans. $N_{1,2}$ is the
 typical number of protons per bunch, 
 $\beta^{\star}$ characterizes the beam optics near the IP,
 \ntot (\ncol) is the total number of (colliding)
 bunches per beam, 
 $\mueff[{\mathrm{max}}]{\eff}$ is the average number of visible interactions
 per crossing at the beam positions with maximal rate.  
 $\tau_{N_1\, N_2}$ is the decay time of the product of the bunch
 populations and $\tau_L$ is the decay time of the luminosity.
 \label{tbl1}
 } 
  \vskip 1mm
 \begin{tabular}{lcccccccc}
  &  \textbf{25 Apr} &  \textbf{15 Oct} \\
  \hline
  LHC fill number              &  1059            & 1422 \\
  $N_{1,2}$ ($10^{10}$ protons)& 1                & 7--8 \\
  $\beta^{\star}$  (m)         &  2               & 3.5\\
  \ncol/\ntot                  & 1/2              & 12/16\\
  $\mueff[{\mathrm{max}}]{\eff}$ & 0.03             & 1\\
  Trigger                      &  minimum bias    & 22.5~kHz random  \\
                               &                  & $\sim$130~Hz minimum bias    \\
                               &                  & beam-gas             \\
  $\tau_{N_1\, N_2}$ (h)       &  950             & 700 \\
  $\tau_L$ (h)                 &   30             &  46 \\
 \end{tabular}
\end{table}

In April the maximal beam movement of $\pm 3\sigma$ was achieved only in the first
scan, as in the second only the first beam was allowed to
move.\footnote{We refer here to $1\sigma$ as the average of the
approximate widths of the beams.}
During the second October scan, both beams moved one after the other,
covering the whole separation range of $\approx 6\sigma$ to both sides.
However, the beam
steering procedure was such that in the middle of the scan the first
beam jumped to an opposite end point and then returned, so that
the beam movement was not continuous. 
This potentially increases hysteresis effects in the LHC magnets. 
In addition, the second scan in October had half the data points, so it
was used only as a cross check to estimate systematic errors.

During the April scans the event rate was low and it was possible to
record all events containing visible interactions. 
A loose minimum bias trigger was used with minimal requirements on the number
of SPD hits ($\ge 3$) and the 
transverse energy deposition in the calorimeters ($\ge 240$~MeV). 
In October the bunch populations were higher by a factor $\sim7.5$,
therefore, in spite of slightly broader beams (the optics defined a
$\beta^\star$ value of 3.5~m instead of 2~m in April), the
rate per colliding bunch pair was higher by a factor of $\sim30$. 
There were twelve colliding bunch pairs instead of one in April. Therefore, 
a selective trigger was used composed of the logical OR of three independent
criteria. 
The first decision accepted random bunch crossings at 22.5~kHz (20~kHz
were devoted to the twelve crossings with collisions, 2~kHz to the
crossings where only one of two beams was present, and 0.5~kHz to the
empty crossings). 
The second  decision
used the same loose minimum bias trigger as the one  used in April but its rate
was limited to 130~Hz. 
The third decision collected events for the beam-gas analysis.

For both the April and October data the systematic error is dominated by
uncertainties in the bunch populations. 
In April this uncertainty is higher (5.6\%) due to a larger contribution from the
offset uncertainty at lower bunch populations~\cite{ref:note-bcnwg}. 
In October the measurement of the bunch populations was more precise, 
but its uncertainty (2.7\%) is still dominant in the
cross-section determination~\cite{ref:bcnwg2}. 
Since the dominant uncertainties are systematic and correlated between
the two scans, we use the less precise April scan only as a cross check.
The scans give consistent results, and in the following we concentrate
on the scan taken in October which gives about a factor two better 
overall precision in the measurement.

The LHC filling scheme
was chosen in such a way that all bunches collided only in one
experiment (except for ATLAS and CMS where the bunches are always
shared), namely twelve bunch pairs in LHCb, three in ATLAS/CMS and one
in ALICE.  
The populations of the bunches colliding in LHCb changed during the two
LHCb scans by less than 0.1\%.  
Therefore, the rates are not normalized by the bunch population
product $N_1\, N_2$ of each colliding bunch pair at every scan point,
but instead only the average of the product over the scan duration is
used.  
This is done to avoid the noise associated with the $N_{1,2}$
measurement. 
The averaged bunch populations are given in Table~\ref{tbl2}. 
The same procedure is applied for the April scan, when the decay time of
${N_1\, N_2}$ was longer, 950 instead of 700 hours in October.
\begin{table}[bt]
 \centering
 \caption{
 Bunch populations (in $10^{10}$ particles) averaged over the two scan periods
 in October separately. The bottom line is 
 the DCCT measurement, all other values are given by the FBCT. 
 The first 12 rows are
 the measurements in bunch crossings (BX) with collisions at LHCb, and
 the last two lines are the sums over all 16 bunches.
 \label{tbl2}
 }
 \vskip 1mm
 \begin{tabular}{ccccccccc}
  & \multicolumn{2}{c}{\bf Scan 1} & \multicolumn{2}{c}{\bf Scan 2} \\
  {BX}     &    $N_1$  &     $N_2$  &
  $N_1$  &    $N_2$  \\
  \hline
 2027  &    8.425  &    7.954  &    8.421  &    7.951  \\
 2077  &    7.949  &    7.959  &    7.944  &    7.957  \\
 2127  &    7.457  &    7.563  &    7.452  &    7.561  \\
 2177  &    6.589  &    7.024  &    6.584  &    7.021  \\
 2237  &    7.315  &    8.257  &    7.311  &    8.255  \\
 2287  &    7.451  &    7.280  &    7.446  &    7.278  \\
 2337  &    7.016  &    7.219  &    7.012  &    7.217  \\
 2387  &    7.803  &    6.808  &    7.798  &    6.805  \\
 2447  &    7.585  &    7.744  &    7.580  &    7.742  \\
 2497  &    7.878  &    7.747  &    7.874  &    7.745  \\
 2547  &    6.960  &    6.244  &    6.955  &    6.243  \\
 2597  &    7.476  &    7.411  &    7.472  &    7.409  \\
  \hline
  All, FBCT  &  120.32  &  119.07  &  120.18  &  118.99  \\
  \hline
  DCCT  &  120.26  &  119.08  &  120.10  &  118.98  \\
 \end{tabular}
\end{table}

\begin{figure}[tbp]
 \centering
 \includegraphics*[width=0.45\textwidth]{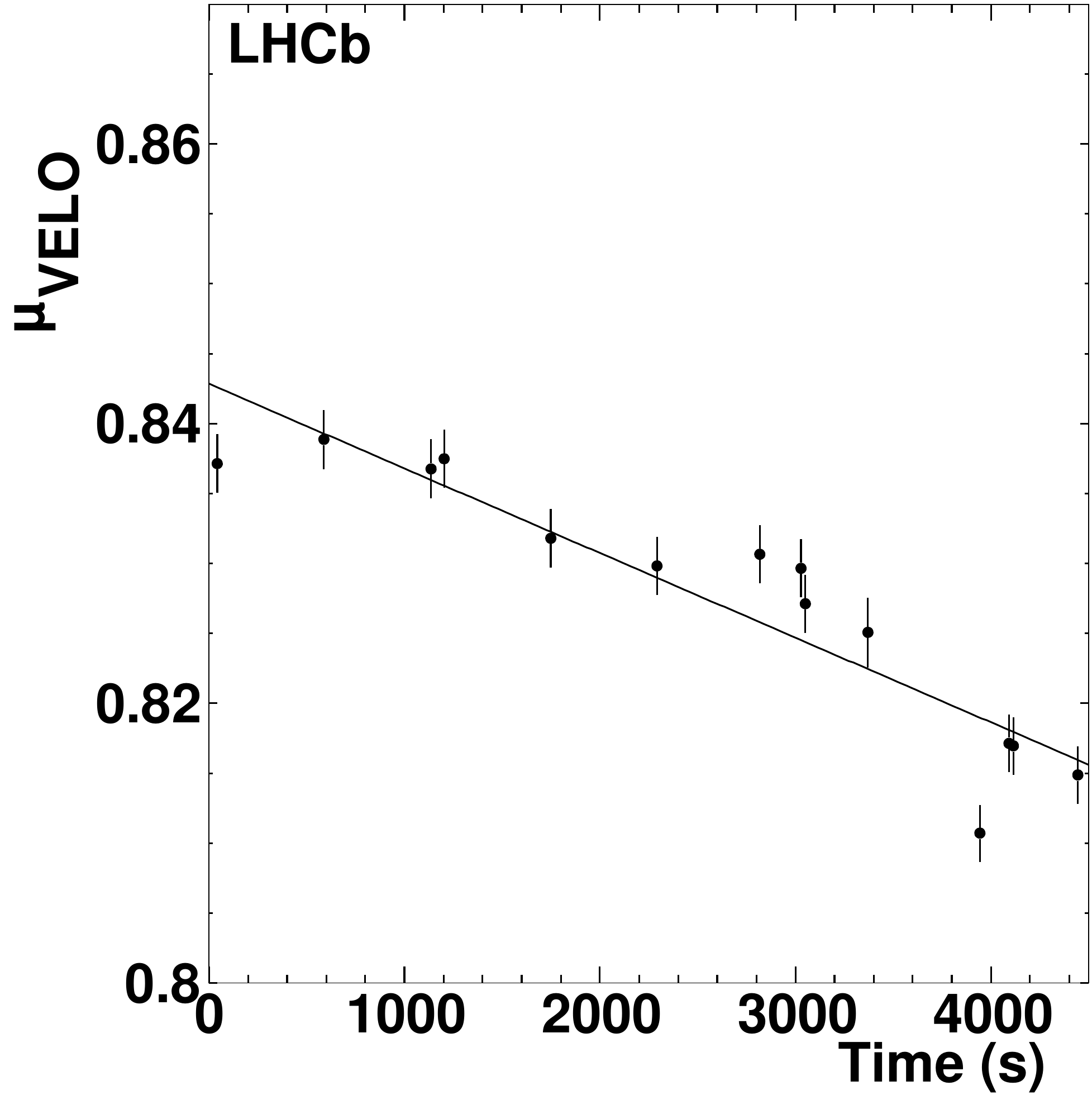}
 \caption{Evolution of the average number of interactions per
 crossing at the nominal beam position during the October scans.  
 In the first (second) scan the  parameters at the nominal beam position
 were  measured three  (four) times both during the \Deltax scan and the  
 \Deltay scan. The straight line is a fit to the data. The
 luminosity decay time is 46~hours.
 \label{fig:evolution}
}
\end{figure}

In addition to the bunch population changes, the luminosity stability may
be limited by the changes in the bunch profiles, {\em e.g.} by emittance
growth.  
The luminosity stability is checked several times during the
scans when the beams were brought back to their nominal position. 
The average number of interactions per crossing is shown in
Fig.~\ref{fig:evolution} for the October scan. 
The luminosity decay time is measured to be 46 hours (30 hours
in April). 
This corresponds to a 0.7\% luminosity drop during the first, longer,
scan along either \Deltax or \Deltay (0.9\% in April).  
The scan points have been taken from lower to higher
\Deltax, \Deltay values, therefore, the luminosity drop
effectively enhances the left part of the integral and
reduces its right part, so that the net effect
cancels to first order since the curve is symmetric. 
The count rate $R(\Deltaxz,\Deltayz)$ at the nominal position entering 
Eq.~\ref{eq:vdm}, is measured in the 
beginning, in the middle and at the end of every scan,
so that the luminosity drop also cancels to first order. 
Therefore, the systematic error due to
the luminosity drop is much less than 0.7\% and is neglected.

The widths of the profiles of the luminous region did
not change within the statistical uncertainties when the beams were
brought to their nominal positions during the first and the second
scans in \Deltax and \Deltay.
In addition, the width of the profiles measured in the two VDM scans did
not change.
These facts also indicate that the effect of the emittance growth on the
cross-section measurement is negligible.

\subsection{Cross-section determination}

\begin{figure}[tbp]
 \centering
 \includegraphics*[width=0.7\textwidth]{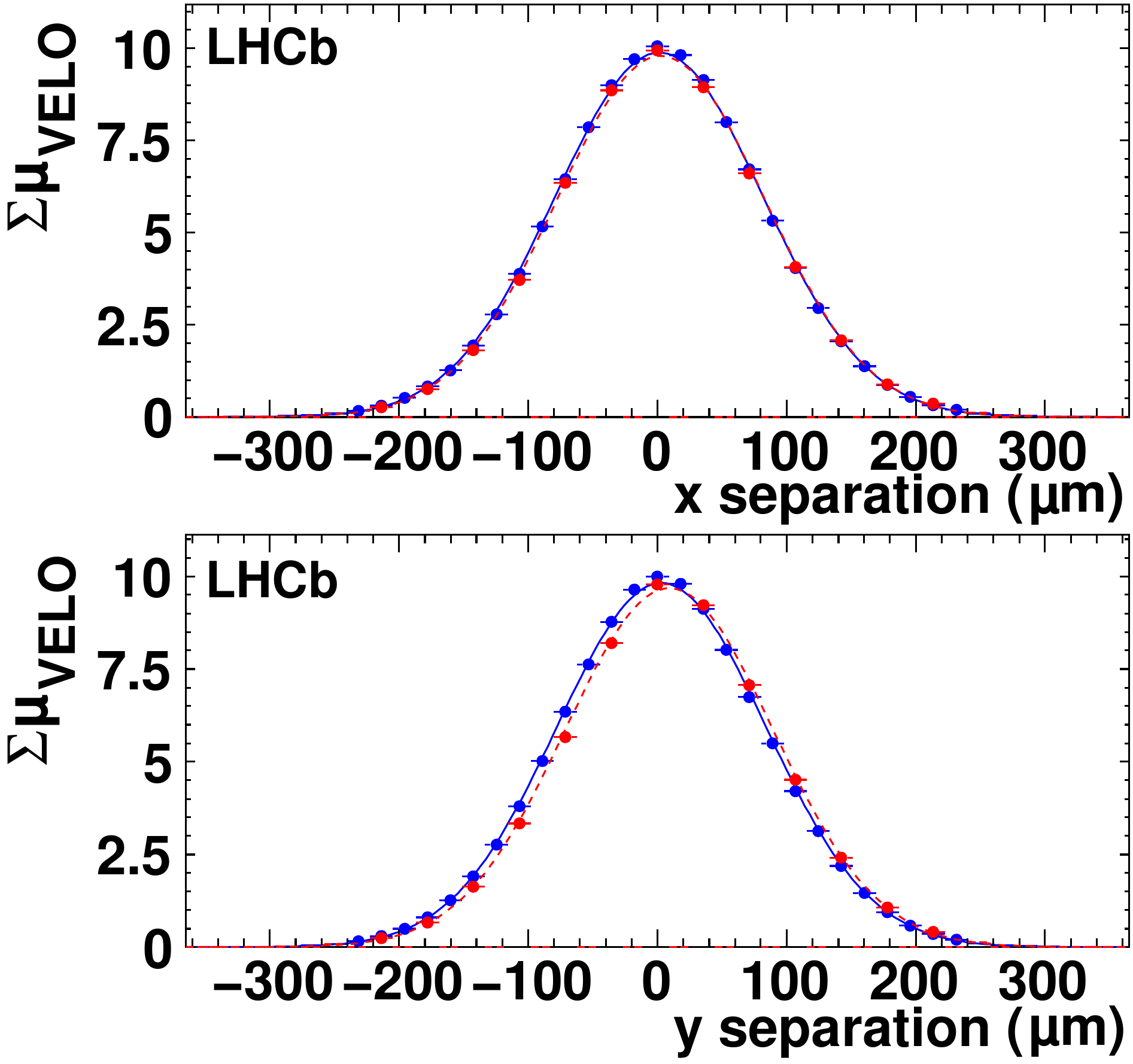}
 \caption{Number of interactions per crossing summed over the twelve
 colliding bunches versus the separations \Deltax (top), \Deltay (bottom) in 
 October. The first (second) scan is represented by the dark/blue (shaded/red)
 points and the solid (dashed) lines.
 The spread of the mean values and widths of the distributions obtained
 individually for each colliding pair are small compared to the widths 
 of the VDM profiles, so that the sum gives a good illustration of the
 shape.
 The curves represent the single Gaussian fits to the data points 
 described in the text.
 \label{fig:vdmprofiles}
 }
\end{figure}

In accordance with the definition of the most stable relative luminosity
counter, a visible event is defined as a $pp$ interaction with at least
two VELO tracks.
The twelve colliding bunch pairs of the VDM scan in October
are analysed individually. 
The dependence on the separation \Deltax and \Deltay of \mueff{\eff}
summed over all bunches is shown in Fig.~\ref{fig:vdmprofiles}. 
Two scans are overlaid, the second is taken at the same values of
\Deltax and \Deltay but with twice as large a step size and different
absolute beam positions. 
One can see that the \Deltay curves are not well reproduced in the two
scans. 
The reason for this apparent non-reproducibility is not understood.  
It may be attributed to hysteresis effects or imperfections in the
description of the optics.\footnote{Imperfections in the description of
the optics can manifest themselves as second order effects in the
translation of magnet settings into beam positions or beam angles.}

\begin{table}[bt]
 \centering
 \caption{
 Mean and RMS of the VDM count-rate profiles summed over the twelve
 colliding bunch pairs obtained from data in the two October scans 
 (scan 1 and scan 2). 
 The statistical errors are 0.05~\micron in the mean position and
 0.04~\micron in the RMS. 
 \label{tbl3}
 }
 \vskip 1mm
 \begin{tabular}{lcccc}
  & {\bf Scan}& \boldmath\Deltax \ {\bf scan} & \boldmath\Deltay \ {\bf scan} \\
  \hline
  Mean (\micron) &$1$& 1.3 & 3.1 \\
                 &$2$& 2.8 & 9.2 \\
  RMS (\micron)  &$1$& 80.6 & 80.8 \\
  &$2$& 80.5 & 80.7 \\
 \end{tabular}
\end{table}

The mean and RMS values of the VDM count-rate profiles shown in
Fig.~\ref{fig:vdmprofiles} are listed in Table~\ref{tbl3}.  
Single Gaussian fits to the individual bunch profiles return 
$\chi^2$ values between 2.7 and 4.3 per degree of freedom.
Double Gaussian fits provide a much better
description of the data and are therefore used in the analysis.  
The single Gaussian fits give cross-section values typically 1.5 to 2\%
larger than the ones obtained with a double Gaussian.
It is found that the fit errors can be reduced by approximately
a factor two if the fits to the \Deltax and \Deltay curves are performed
simultaneously and the value measured at the nominal point 
$\mueff{\eff}(\Deltaxz$,$\Deltayz)$ is constrained to be the same
in both scans.  
The first fit parameter is chosen to be 
$\int\!\mueff{\eff}\,d\Deltax \,
\int\!\mueff{\eff}\,d\Deltay / \mueff{\eff}(\Deltaxz, \Deltayz)$, 
so that a correlation of both
integrals and the value at the nominal point is correctly taken into
account in the resulting fit error. Other fit parameters 
are: the two integrals along \Deltax and \Deltay, and 
$\sigma_1$, $\Delta\sigma$ and a common central position of the Gaussian
function for the \Deltax and similarly for the \Deltay curves. 
Here $\sigma_1$ and  $\sigma_2=\sqrt{\sigma_1^2+\Delta\sigma^2}$ 
are the two Gaussian widths of the fit function.
The relative normalization of the two Gaussian components
and the value at the nominal point are derived from the nine fit
parameters listed above.  The $\chi^2$ value per degree of freedom
of the fit is between 0.7 and 1.8 for all bunch pairs. 

\begin{figure}[tbp]
 \centering
 \includegraphics*[width=0.6\textwidth]{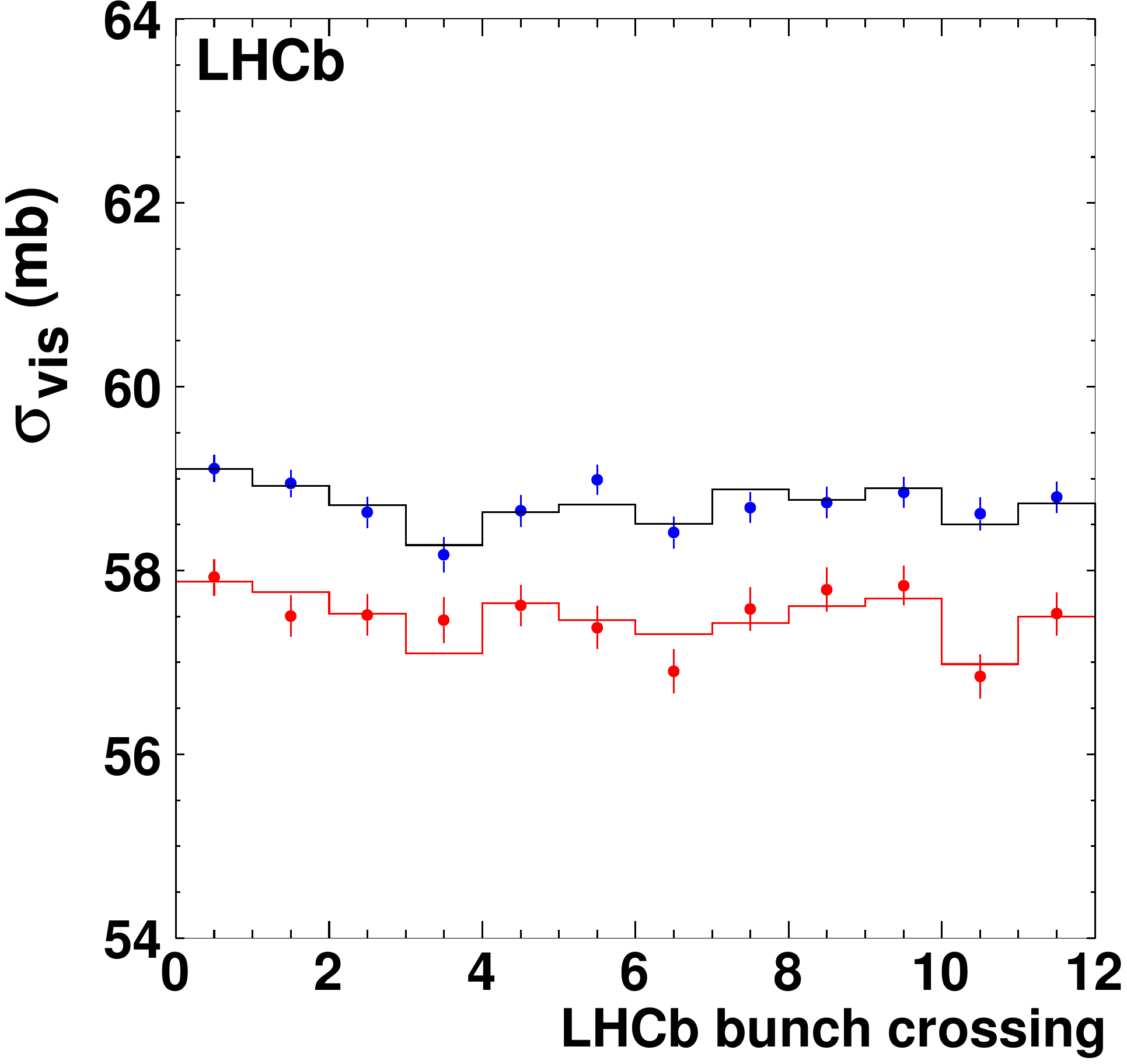}
 \caption{
 Cross-sections without correction for the FBCT offset for the twelve
 bunches of the October VDM fill (data points). 
 The lines indicate the results of the fit as discussed in the
 text. 
 The upper (lower) set of points is obtained in the first
 (second) scan.  
 \label{fig11}
 }
\end{figure}

The product of bunch populations $N_1\, N_2$ of the twelve colliding bunches 
have an RMS spread of 12\%.
The analysis of the individual bunch pairs gives cross-sections
consistent within statistical errors, which typically have values of
0.29\% in the first scan. 
The sensitivity of the method is high enough that it is possible to 
calibrate the {\it relative} bunch populations 
$N_{1,2}^i/\sum_{j=1}^{16} N_{1,2}^j$
measured with the FBCT system by assuming a linear response. 
Here $i$ runs over the twelve bunches colliding in LHCb and $j$ over all
16 bunches circulating in the machine. 
By comparing the FBCT with the  ATLAS BPTX measurements it is observed that
both may have a non-zero offset~\cite{ref:note-bcnwg,ref:bcnwg2}. 
A discrete function $s_\mathrm{\eff}^i$ is fitted to the twelve measurements
$\sigeff[i]{\eff}$ using three free parameters: the common cross-section
\sigeff{\eff} and the two FBCT offsets for the two beams $N_{1,2}^0$
\begin{equation}
\label{eq:offset}
 s_\mathrm{\eff}^i =
 \sigeff{\eff} \prod_{b=1,2}\left[ \frac{(N_b^i-N_b^0)}{N_b^i} 
 \ \frac{\sum_{j=1}^{16}N_b^j}{\sum_{j=1}^{16}(N_b^j-N_b^0)} \right] \, ,
\end{equation}
which corrects the relative populations $N_{1,2}^i$ for
the FBCT offsets $N_{1,2}^0$ and takes into account that the total beam
intensities measured with the DCCT constrain the sums of all bunch
populations obtained from the FBCT values. 
The sum over all bunches of the quantites $N_{1,2}^j$,
$\sum_{j=1}^{16}N_{1,2}^j$, is normalized to the DCCT 
value prior to the fit and the fit using Eq.~\ref{eq:offset}
evaluates the correction due to the FBCT offsets alone. 
The results of this fit are shown in  Fig.~\ref{fig11}, where the data
points $\sigeff[i]{\eff}$ are drawn without offset correction and the
lines represent the fit function of Eq.~\ref{eq:offset}.
The use of two offsets improves the description of the points compared
to the uncorrected simple fit.
The $\chi^2$ per degree of freedom and
other relevant fit results are summarized in Table~\ref{tab:vdm:res}.
In addition, the table also shows results for the case where the 
ATLAS BPTX is used instead of the FBCT system. 

\begin{table}[tbp]
 \begin{center}
 \caption{
 Results for the visible cross-section fitted over the twelve
 bunches colliding in LHCb for the October VDM data together with the
  results of the April scans. 
 $N_{1,2}^0$ are the FBCT or BPTX offsets in units of $10^{10}$
 particles. 
 They should be subtracted from the values measured for individual
 bunches. 
 The first (last) two columns give the results for the first and the second
 scan using the FBCT (BPTX) to measure the relative bunch populations.  
 The cross-section from the first scan obtained with the FBCT  bunch
  populations with offsets determined by the fit is used as final VDM  
 luminosity calibration.
  The results of the April scans are reported on the last row.  Since
  there is only one colliding bunch pair, no fit to the FBCT offsets
  is possible.
 \label{tab:vdm:res}
 }  
 \vskip 1mm
 \begin{tabular}{lcc|cccc}
  &   \multicolumn{4}{c}{\bf October data}  \\	
  \hline
  &   \multicolumn{2}{c|}{\bf FBCT}
  & \multicolumn{2}{c}{\bf ATLAS BPTX}  \\	
  &  {\bf  Scan 1} & {\bf Scan 2}&  {\bf  Scan 1} & {\bf Scan 2}\\
  \hline
  &   \multicolumn{2}{c}{{\bf \it with fitted offsets}} & \multicolumn{2}{c}{{\bf \it with fitted offsets}} \\  
 \sigeff{\eff} (mb) &  ${\bf \underline{58.73\,\pm\,0.05}}$  &  $57.50\,\pm\,0.07$ &  $58.62\,\pm\,0.05$  &  $57.45\,\pm\,0.07$  \\    
  $N_1^0$             &  $0.40\,\pm\,0.10$   &  $0.29\,\pm\,0.15$   		  &  $-0.10\,\pm\,0.12$  &  $-0.23\,\pm\,0.17$  \\  
  $N_2^0$             &  $-0.02\,\pm\,0.10$  &  $0.23\,\pm\,0.13$   		  &  $-0.63\,\pm\,0.12$  &  $-0.34\,\pm\,0.15$  \\  
  $\chi^2$/\ndof  &  5.8 / 9       &  7.6 / 9       				  &  6.9 / 9       &  7.3 / 9       \\	      
  \hline
  &   \multicolumn{2}{c}{{\bf \it with offsets fixed at zero}} & \multicolumn{2}{c}{{\bf \it with offsets fixed at zero}} \\  
  \sigeff{\eff} (mb) &  $58.73\,\pm\,0.05$  &  $57.50\,\pm\,0.07$  		  &  $58.63\,\pm\,0.05$  &  $57.46\,\pm\,0.07$  \\  
  $\chi^2$/\ndof  &  23.5 / 11     &  21.9 / 11
	  &  66.5 / 11     &  23.5 / 11                 \\ 
  \hline
  &   \multicolumn{4}{c}{\bf April data}  \\	
  \hline
  &  {\bf  Scan 1} & {\bf Scan 2}\\
  $\sigeff{\eff}$ (mb)& {\bf $59.6\pm0.5$}   & { $57.0\pm0.5$} \\
 \end{tabular}
 \end{center}
\end{table}

One can see that the offset errors in the first scan are
$(0.10-0.12) \tms 10^{10}$, or 1.5\% relative to the average bunch
population $\langle \, N_{1,2}\, \rangle = 7.5 \tms 10^{10}$. 
The sensitivity of the
method, therefore, is very high, in spite of the fact that the
RMS spread of the bunch population products $N_1\, N_2$ is 12\%. 
The quoted errors are only statistical. 
For the first scan, the relative cross-section error is 0.09\%. 
Since the fits return good $\chi^2$ values, the bunch-crossing dependent 
systematic uncertainties (such as emittance growth and bunch population
product drop) are expected to be lower or comparable.  
An indication of the level of the systematic errors is given by the
difference of about two standard deviations found for $N_{1,2}^0$
between the two scans. 
All principal sources of systematic errors which will be
discussed below (DCCT scale uncertainty, hysteresis, and ghost charges)
cancel when comparing bunches within a single scan.

In spite of the good agreement between the bunches within the same scan,
there is an overall 2.1\% discrepancy between the scans. The reason is
not understood, and may be attributed to a potential hysteresis effect or
similar effects resulting in uncontrollable shifts of the beam as a
whole. 
The results of the first scan
with the FBCT offsets determined by the fit are taken  
as the final VDM luminosity determination (see Sect.~\ref{sec:results-vdm}).  
The 2.1\% uncertainty estimated from the discrepancy is the second
largest systematic error in the cross-section measurement after the
uncertainties in the bunch populations. 
In the April data the situation is similar: the discrepancy between the
cross-sections obtained from the two scans is $(4.4\,\pm\,1.2)$\%, 
the results may be found in Table~\ref{tab:vdm:res}. 
Since the April measurement is performed using corrected trigger rates
proportional to the luminosity instead of VELO tracks, the results
have been corrected for the difference in acceptances.
The correction factor is determined by studying random triggers and is
$\sigeff{VELO}/\sigeff{April~trigger} = 1.066$, where $\sigeff{VELO}$ is
the usual definition of $\sigeff{\eff}$. 

\subsection{Systematic errors}

\subsubsection{Reproducibility of the luminosity at the nominal beam positions}

Figure~\ref{fig:evolution} shows the evolution of the luminosity as a
function of time for the periods where the beams were at their nominal
positions during the VDM scan.
One expects a behaviour which follows the loss of beam particles and the
emittance growth.
Since these effects occur at large time-scales compared to the duration of
the scan, the dependence on these known effects can be approximated by a
linear evolution.
As shown in Fig.~\ref{fig:evolution}, the luminosity did not always
return to the expected value when the beams returned to
their nominal positions. 
The $\chi^2/\ndof$ with respect to the fitted straight line is too large
(40/12), thus, the non-reproducibility cannot be attributed fully to
statistical fluctuations and another systematic effect is present.   
The origin of this effect is not understood but it may be similar to the
one which causes the non-reproducibility of the beam positions observed
in the shift of the two scan curves.
Therefore, a systematic error of 0.4\% is assigned to the absolute scale 
of the \mueff{\eff} measurement to take this observation into
account.
The systematic error is estimated as the amount which should be
added in quadrature to the statistical error of 0.25\% to produce a
$\chi^2/\ndof$ equal to one. 
Since the absolute scale of the \mueff{\eff} measurement enters the
cross-section linearly (Eq.~\ref{eq:vdm}), the same systematic error
of 0.4\% is assigned to the cross-section measurement.

\subsubsection{Length scale calibration}

The beam separation values \Deltax and \Deltay are calculated from
the LHC magnet currents at every scan step. 
There is a small non-reproducibility in the results of two scans, as
shown in Fig.~\ref{fig:vdmprofiles}.
The non-reproducibility may be attributed to a mismatch between the
actual beam positions and the nominal ones. 
Therefore, it is important to check the \Deltax and \Deltay values
as predicted by the magnet currents, and in particular their scales which
enter linearly in the cross-section computation (Eq.~\ref{eq:vdm}).
One distinguishes a possible differential length scale mismatch between
the two beams from a mismatch of their average position calibration. 

A dedicated mini-scan was performed in October where the two beams were  
moved in five equidistant steps both in $x$ and $y$ keeping the
nominal separation between the beams constant.  
During the scan along $x$ the beam separation was 
$80\ \micron$ in $x$ and $0\ \micron$ in $y$.
Here 80~\micron is approximately the width of the luminosity profile of
the VDM scan (see Table~\ref{tbl3}). 
This separation was chosen to maximize the derivative
$dL/d\Delta(x)$, {\em i.e.} the sensitivity of the luminosity to a
possible difference in the length scales for the two beams. 
If {\em e.g.} the first beam moves slightly faster
than the second one compared to the nominal movement, the separation 
$\Delta(x)$ gets smaller and the effect can be visible as an increase of
the luminosity. 
Similarly, the beam separation used in the $y$ scan was 
$0\ \micron$  and $80\ \micron$ in $x$ and $y$, respectively.

\begin{figure}[tbp]
 \centering
 \includegraphics*[width=0.48\textwidth]{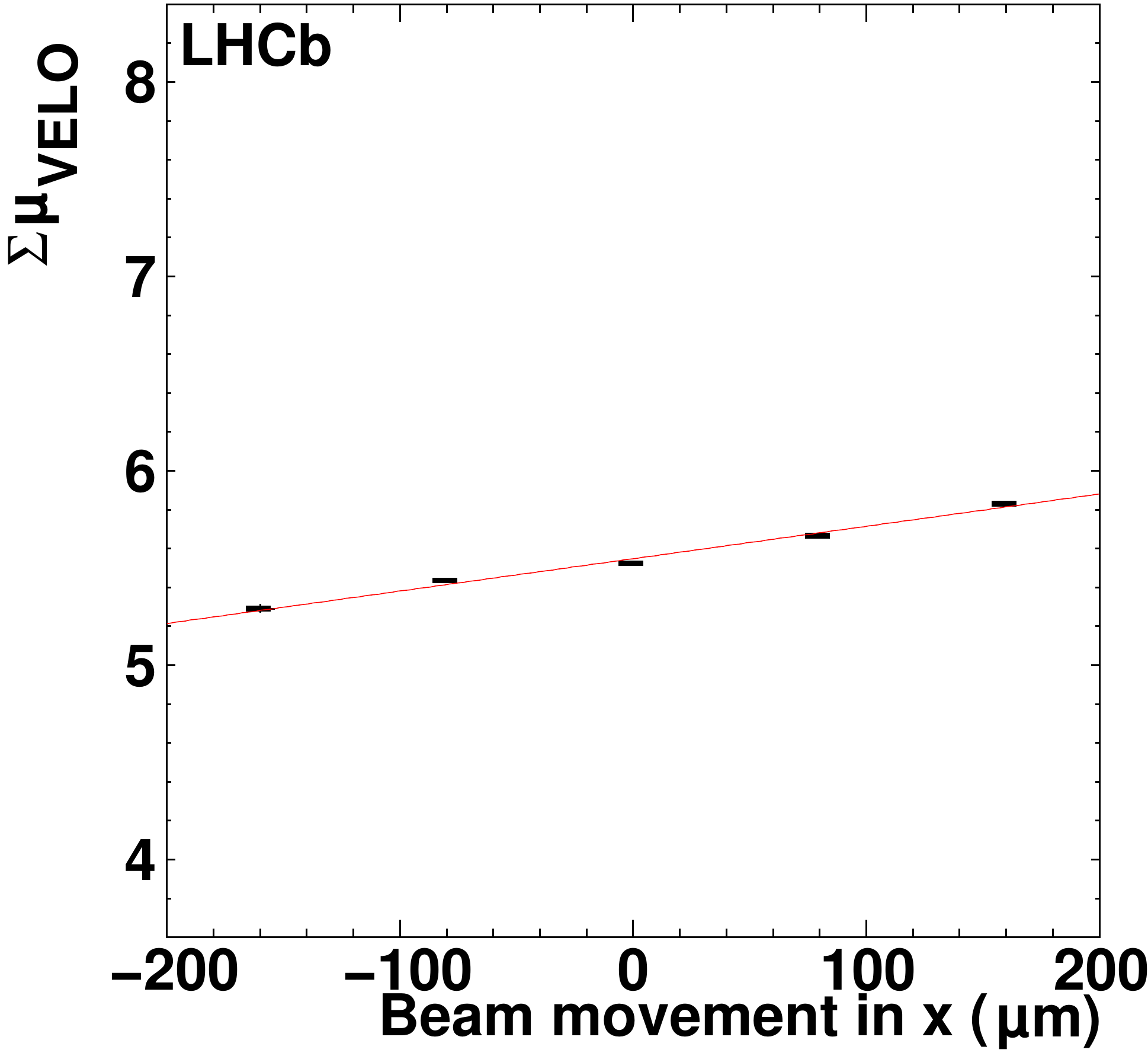}
 \includegraphics*[width=0.48\textwidth]{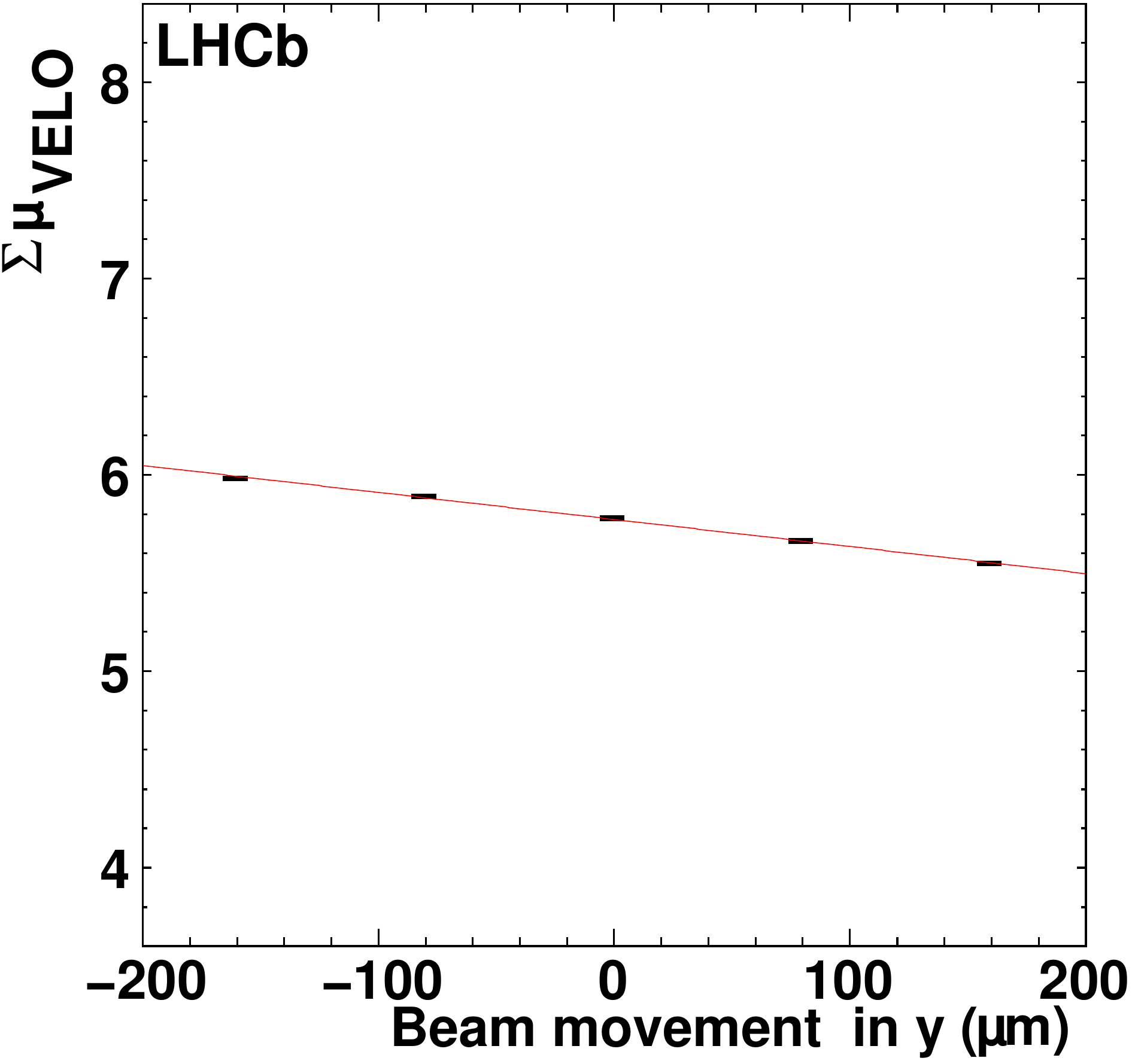}
 \caption{Average number of interactions (\mueff{VELO}) versus the centre of the
 luminous region summed over the twelve colliding bunches and measured
 during the length scale scans in  $x$ (left) and in $y$ (right) taken
 in October.  
 The points are indicated with small horizontal bars, the statistical
 errors are smaller than the symbol size.
 The straight-line fit is overlaid.
 \label{fig:lengthlum}
 }
\end{figure}

The behaviour of the measured luminosity during the length-scale
calibration scans is shown in Fig.~\ref{fig:lengthlum}. 
As one can see, the points show a significant deviation from a
constant. 
This effect may be attributed to different length scales of the two
beams. 
More specifically, we assume that the real positions of the beams
$x_{1,2}$ could be obtained from the values $x_{1,2}^0$ derived from
the LHC magnet currents by applying a correction parametrized by 
$\epsilon_x$ 
\begin{equation}
 \label{eq:vdm:size}
  x_{1,2} = (1\,\pm\,\epsilon_x/2)\, x_{1,2}^0\, ,
\end{equation}
and similarly for $y_{1,2}$.
The $+$ ($-$) sign in front of $\epsilon_x$ holds for beam~1 (beam~2).  
Assuming a Gaussian shape of the luminosity dependence on \Deltax
during the VDM scan, we get 
\begin{equation} 
 \label{eq:vdm:shift}
  \frac{1}{L}\,\frac{dL}{d(x_1+x_2)/2} =
  -\epsilon_x\frac{\Deltax}{\Sigma_x^2} \, .
\end{equation}
Here $\Deltax = 80$~\micron is the fixed nominal beam separation.  
A similar equation holds for the $y$ coordinate.
In the approximation of a single Gaussian shape of the beams,
the width of the VDM profile, $\Sigma_{x}$, is defined as
\begin{equation}
\label{eq:capsigma}
 \Sigma_x = 
 \sqrt{\sigma_{1x}^2+\sigma_{2x}^2+4(\sigxp{z})^2\tan^2{\thetac}} \, ,
\end{equation}
where $\sigxp{z}$ is the width of the luminous region in the $z$ coordinate
and $\sigma_{bx}$ is the width of beam $b$ ($b=1,2$).
A similar equation can be written for $\Sigma_y$. 
From the slopes observed in Fig.~\ref{fig:lengthlum} we obtain 
$\epsilon_x=2.4\%$ and $\epsilon_y=-1.9\%$. 
The same behaviour is observed for all bunches separately.

Since $\Deltax = (x_1^0-x_2^0)+\epsilon\, (x_1^0+x_2^0)/2$, the 
\Deltax correction
depends on the nominal mid-point between the beams $(x_1^0+x_2^0)/2$.  
In the first scan this nominal point was always kept at zero,
therefore, no correction is needed.  
During the second scan this point moved with nominal positions 
$0\to355.9\ \micron\to0$. 
Therefore, a correction to the \Deltax values in
Fig.~\ref{fig:vdmprofiles} is required. 
The central point should be shifted to the right (left) for the  $x$
($y$) scan. 
The left (right) side is thus stretched and the opposite side is
shrunk. 
After correction the shift between the scans is reduced in $y$, but
appears now in $x$, so that the discrepancy cannot be fully explained
by a linear correction alone.
The correction which stretches or shrinks the profiles measured in the
second scan influences the integrals of these profiles and the resulting
cross-sections very little. 
The latter changes on average by only 0.1\%, which we take as an
uncertainty and which we include into the systematic error. 
In Table~\ref{tab:vdm:res} the numbers are given with the correction applied.

\begin{figure}[tbp]
 \centering
 \includegraphics*[width=0.44\textwidth]{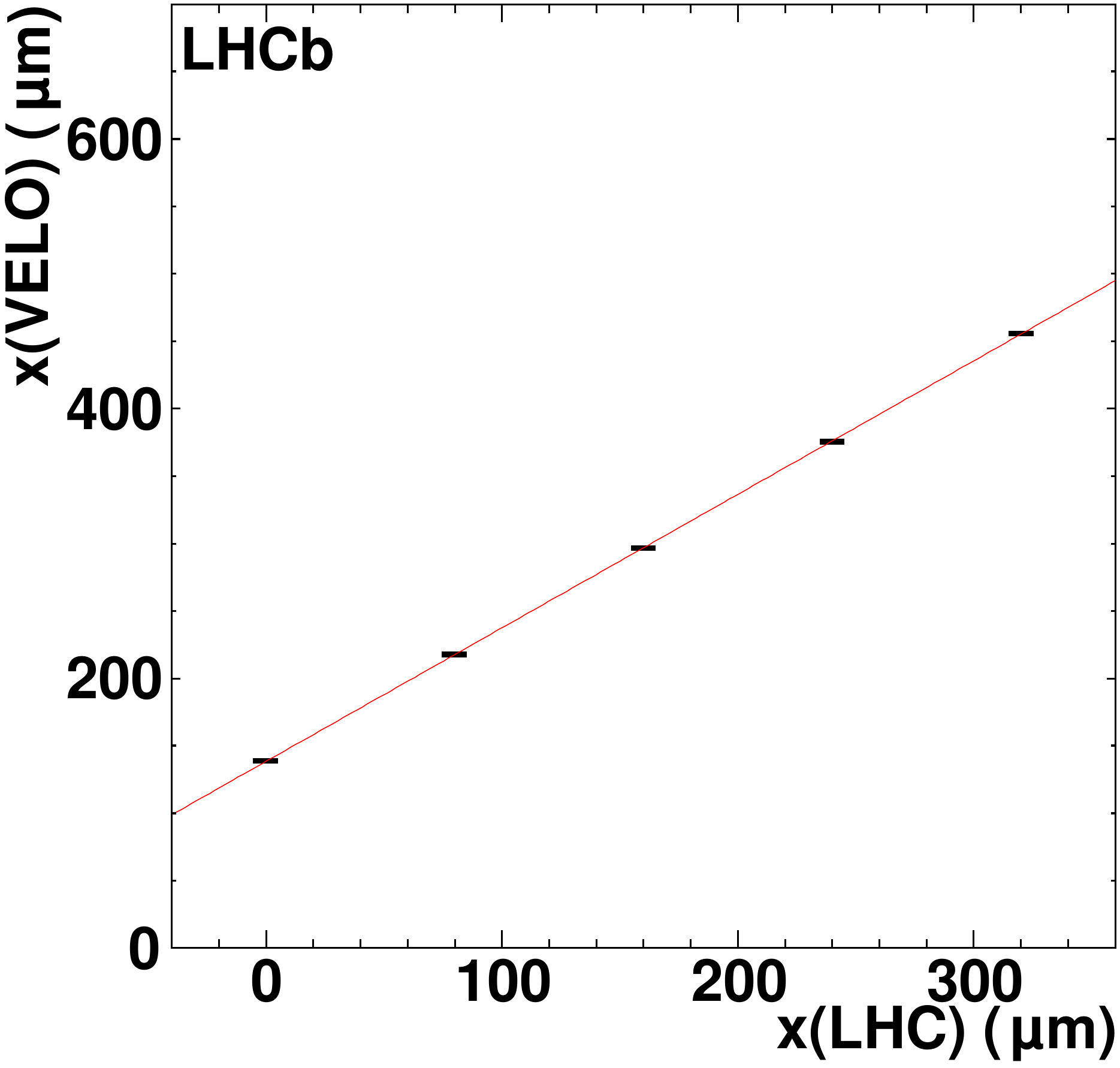}
 \includegraphics*[width=0.44\textwidth]{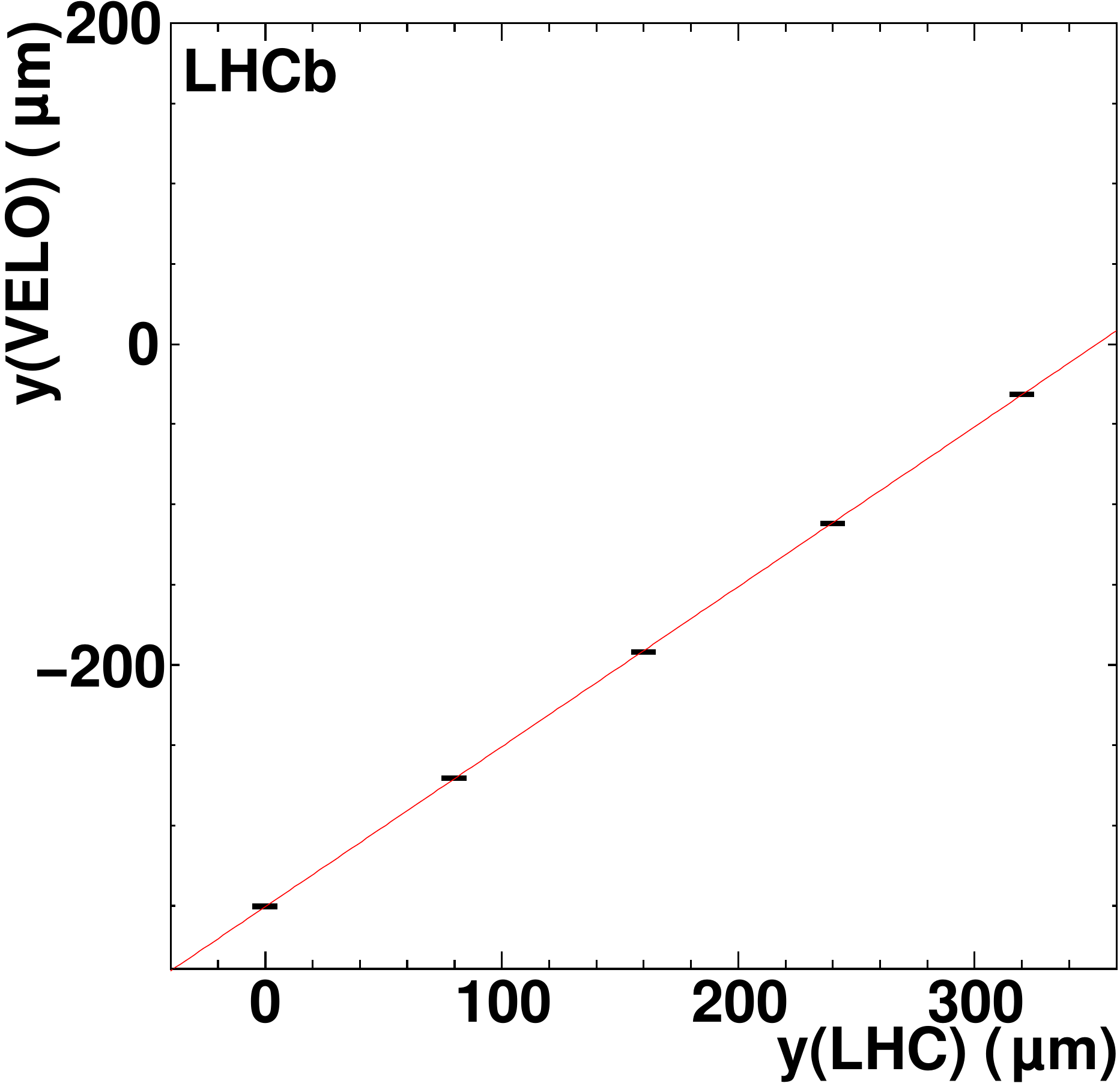}
 \caption{Centre of the luminous region reconstructed with VELO tracks
 versus the position predicted by the LHC magnet currents. 
 The points are indicated with small horizontal bars, the statistical
 errors are smaller than the symbol size.
 The points are fitted to a linear function. 
 The slope calibrates the common length scale.
 \label{fig:parallel}
}
\end{figure}

During a simultaneous parallel translation of both beams, the centre of the
luminous region should follow the beam positions regardless of the bunch
shapes. 
Since it is approximately at $(x_1+x_2)/2=(x_1^0+x_2^0)/2$ and similarly for $y$, the
corrections to the position of the centre due to $\epsilon_{x,y}$ are
negligible. 
The luminous centre can be determined using vertices measured with the VELO. 
This provides a precise cross check of the common
beam length scales $(x_1^0+x_2^0)/2$ and $(y_1^0+y_2^0)/2$.
The result is shown in Fig.~\ref{fig:parallel}. 
The LHC and VELO length scales agree within ($-0.97 \,\pm\, 0.17$)\% and
($-0.33 \,\pm\,0.15$)\% in $x$ and $y$, respectively.  
The scale of the transverse position measurement with the VELO is 
expected to be very precise owing to the fact that it is determined by the
strip positions of the silicon sensors with a well-known geometry.
For the cross-section determination we took the more
precise VELO length scale and multiplied the values 
from
Table~\ref{tab:vdm:res} by \mbox{$(1-0.0097)\times (1-0.0033) = 0.9870$}. 
In addition, we conservatively assigned a 1\% systematic error due to
the common scale uncertainty. 

In April, no dedicated length scale calibration was performed.
However, a cross check is available from the distance between the centre
of the luminous region measured with the VELO and the nominal 
centre position. 
The comparison of these distances between the first and second scan when
either both beams moved symmetrically or only the first beam moved, provides a
cross check which does not depend on the bunch shapes.
From this observation the differences of the length scales between the
nominal beam movements and the VELO reference are found to be
($-1.3\,\pm\,0.9$)\% and ($1.5\,\pm\,0.9$)\% for \Deltax and \Deltay,
respectively.  
Conservatively a 2\% systematic error is assigned to the length
scale calibration for the scans taken in April.

\subsubsection{Coupling between the \boldmath$x$ and \boldmath$y$ coordinates in the LHC beams}
\label{sec:vdm:coupling}

The LHC ring is tilted with respect to the horizontal plane, while the VELO
detector is aligned with respect to a coordinate system where the $x$
axis is in a horizontal plane~\cite{ref:lhcrotation}.
The van der Meer equation (Eq.~\ref{eq:vdm}) is valid only if the particle
distributions in $x$ and in $y$ are independent. 
To check this condition the movement of the centre of the luminous
region along $y$ is measured during the length scale scan in $x$ and
vice versa.
This movement is compatible with the expected tilt of the LHC
ring of 13~mrad at LHCb~\cite{ref:lhcrotation} with respect to the vertical
and the horizontal axes of the VELO. 
The corresponding correction to the cross-section is negligible ($<10^{-3}$).

\begin{figure}[tbp]
 \centering
 \includegraphics*[width=0.5\textwidth]{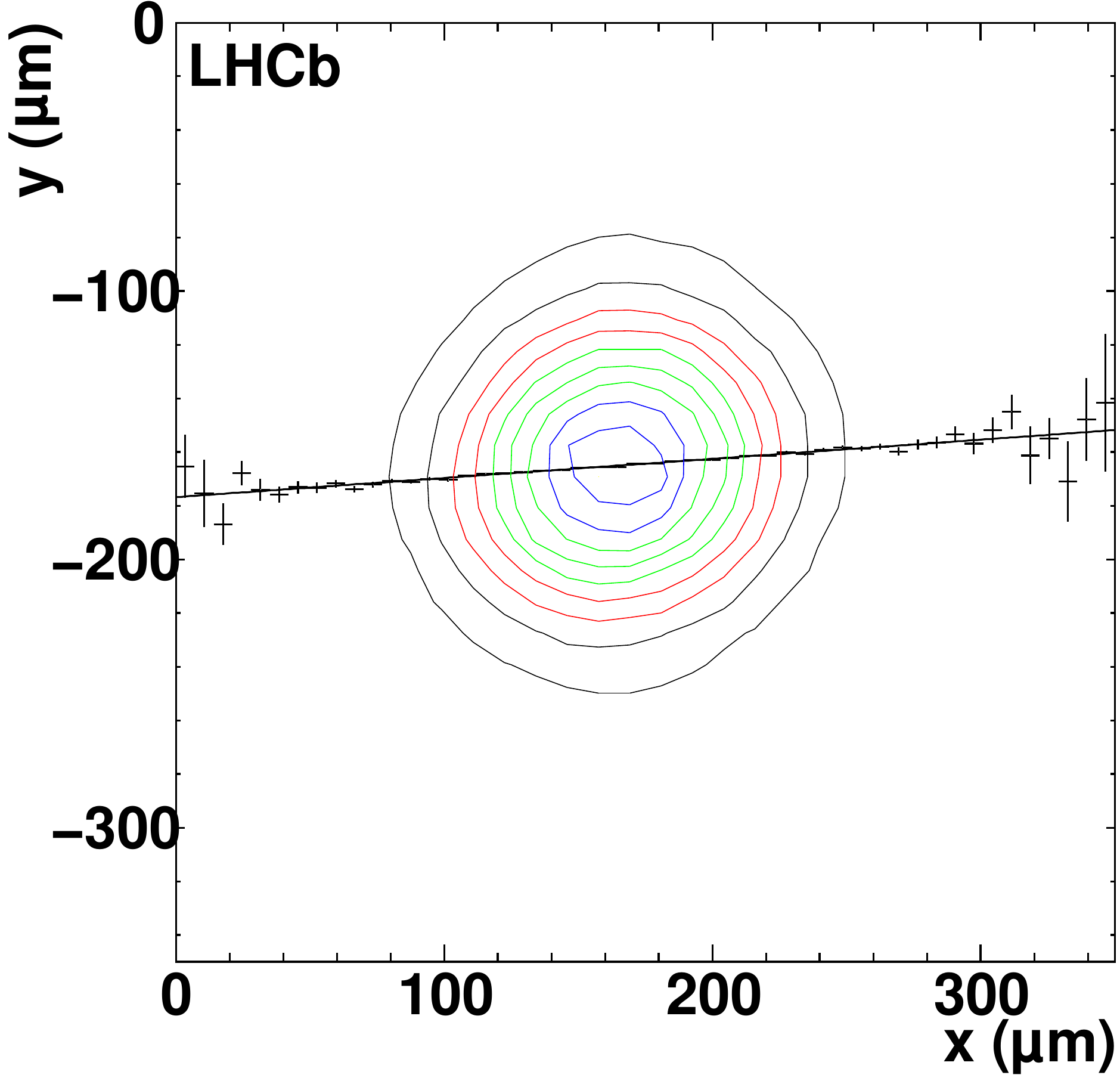}
 \caption{Contours of the distribution of the $x$-$y$ coordinates of the
 luminous region. 
 The contour lines show the values at multiples of 10\% of the maximum. 
 The points represent the $y$-coordinates of the centre of the luminous
 region in different $x$  slices. 
 They are fitted with a linear function.
 \label{fig:xymap}
}
\end{figure}

To measure a possible $x$-$y$ correlation in the machine the
two-dimensional vertex map is studied by determining the centre position
in one coordinate for different values of the other coordinate.  
For the analysis, data were collected with the beams 
colliding head-on at LHCb in Fill 1422, during which also the VDM scan
data were taken.
Figure~\ref{fig:xymap} shows the $x$-$y$ profile of the luminous region.
The centre positions of the $y$ coordinate lie on a straight line with a
slope of 79~mrad. 
The slopes found in the corresponding $x$-$z$ and $y$-$z$ profiles 
are $-92$~\micrad and $44$~\micrad.
These slopes are due to the known fact that the middle line between the two LHC
beams is inclined with respect to the $z$ axis. 
This is observed with beam gas events, the inclination
varies slightly from fill to fill. 
The measurement of the beam directions will be described in detail in
Sect.~\ref{sec:beamgas}. 
Taking into account these known correlations of $x$ and $y$ with $z$
and also the known 13~mrad tilt of the LHC ring, one can calculate the
residual slope of the $x$-$y$ correlation, which is predicted to be
77~mrad. 

If the beam profiles are two-dimensional Gaussian functions 
with a non-zero correlation between the $x$ and $y$ coordinates, the
cross-section relation (Eq.~\ref{eq:vdm}) should be corrected. 
We assume that the $x$-$y$ correlation coefficients of the two beams,
$\zeta$, are similar and, therefore, close to the measured correlation
in the distribution of the vertex coordinates of the luminous region, 
$\zeta=0.077$.  
In this case the correction to the cross-section is $\zeta^2/2=0.3$\%.  
We do not apply a corresponding correction, but instead include 0.3\%
uncertainty as an additional systematic error. 

\subsubsection{Cross check with the \boldmath$z$ position of the
   luminous region}
\label{sec:vdmimaging}

\begin{figure}[btp]
 \centering
 \includegraphics*[width=0.5\textwidth]{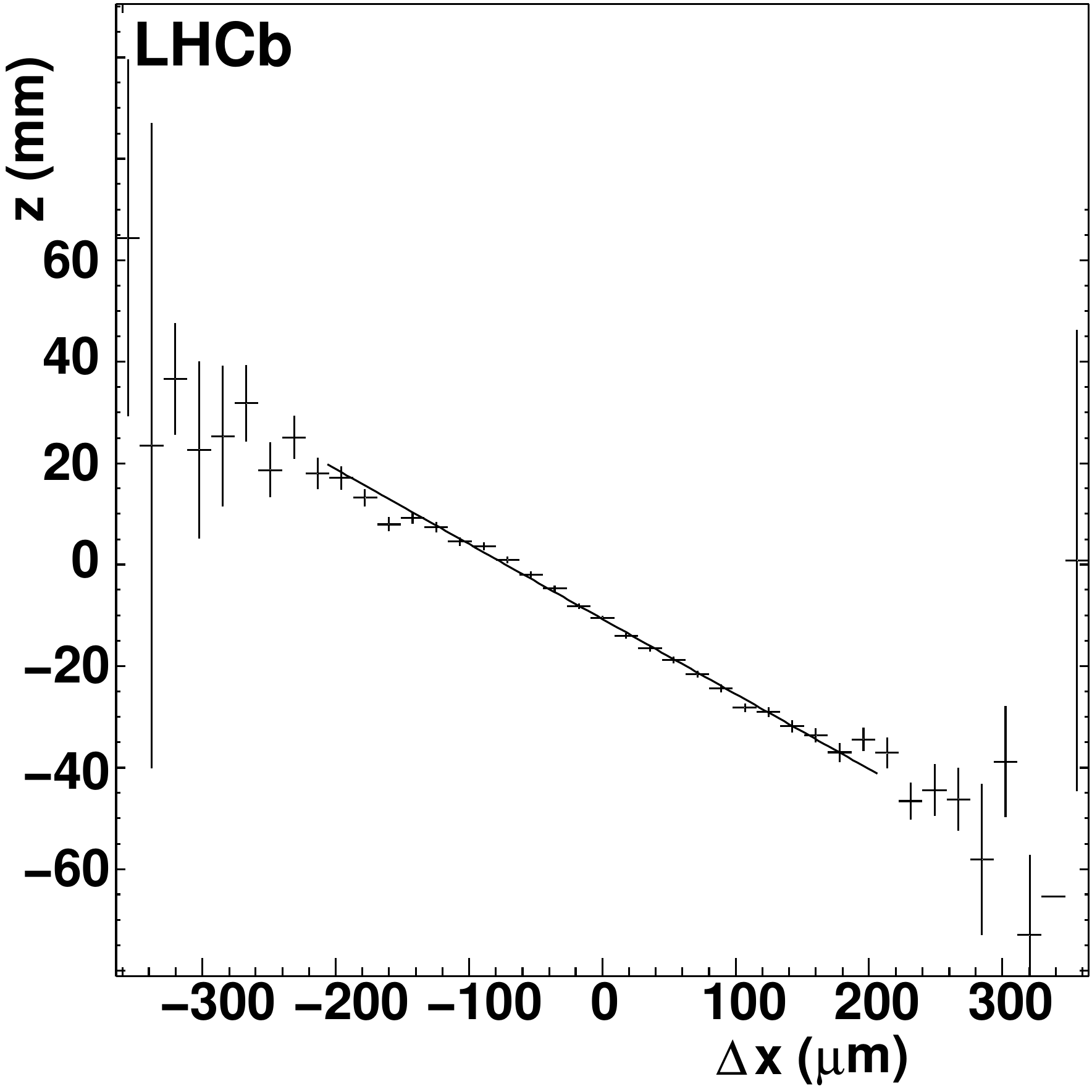}
 \caption{Movement of the centre of the luminous region in $z$ during
 the first scan in $x$ taken in October.
 \label{fig:zmove}
}
\end{figure}

A cross check of the width of the luminosity profile as a function of
\Deltax  is made by measuring the movement of the $z$ position of the
centre of the luminous region during the first VDM scan in the $x$
coordinate in October (see Fig.~\ref{fig:zmove}). 
Assuming Gaussian bunch density distributions and identical widths of
the two colliding beams, the slope is equal to
\begin{equation}
\label{eq:dzlum}
\frac{\mathrm{d} z_{\xp}}{\mathrm{d} (\Deltax)} =
-\frac{\sin2\thetac}{4}\frac{\sigma_z^2-\sigma_x^2}{\sigma_x^2\cos^2\thetac+\sigma_z^2\sin^2\thetac}
\, ,
\end{equation}
where $\sigma_{x,z}$ are the beam widths in the corresponding
directions and $\mathrm{d} z_{\xp}$ is the induced shift in the
$z$-coordinate of the centre of the luminous region. 
We approximate $\sigma_z$ as $\sqrt{2}\sigxp{z}$.  
Using the slope observed in Fig.~\ref{fig:zmove} one gets the expected
width of the luminosity profile versus \Deltax, 
$\sigma_x^\mathrm{VDM} = \sqrt{2}\sigma_x = 78$~\micron, 
in agreement with the measured value of 80~\micron.
A further cross check is described in Sect.~\ref{sec:vdmimagingsigma}.

\subsection{Results of the van der Meer scans}
\label{sec:results-vdm}

The cross-section results obtained with the VDM scans are given in
Table~\ref{tbl8}. 
We recall that the effective cross-section defined by interactions with
at least two VELO tracks is used in the analysis.
The results of the two measurement periods are consistent.
During the second scan in October the beam movements were not
continuous, so the results might suffer from hysteresis
effects. 
In the second scan in April only one beam moved limiting the separation
range.
Therefore, only the first scans are shown in Table~\ref{tbl8} for April
and October.
Both in the April and in the October measurements the difference
observed in the results of the two scans is included as a
systematic error. 
The complete list of errors taken into account is given in
Table~\ref{tbl10}. 
The uncertainties are uncorrelated and therefore added in quadrature.
As already anticipated in Sect.~\ref{sec:vdm:conditions} the
first scan taken in October is retained as the final result of the VDM
method since it has much reduced systematic errors compared
to the April scans.  
\begin{table}[tb]
 \centering
 \caption{Cross-section of $pp$ interactions producing at least two VELO tracks, 
 measured in the two van der Meer scans in April and in October.
 \label{tbl8}
 }
 \vskip 1mm
 \begin{tabular}{lcc}
  & \boldmath$\sigeff{\mathbf\eff}\,\mathbf{(mb)}$ & {\bf Relative uncertainty (\%)} \\
  \hline
  April & 59.7 & 7.5\\
  October & {\bf \underline{58.35}} & 3.64\\
 \end{tabular}
\end{table}
\begin{table}[bt]
 \centering
 \caption{Summary of relative cross-section uncertainties for the van
 der Meer scans in October and April.
 Due to the lower precision in the April data some systematic errors
 could not be evaluated and are indicated with ``$-$''.
 \label{tbl10}
 }  
 \vskip 1mm
 \begin{tabular}{lcc}
  {\bf Source} & \multicolumn{2}{c}{\bf relative uncertainty (\%)} \\
               & {\bf October} & {\bf April} \\
  \hline
  Relative normalization stability     &  0.5       & 0.5  \\
  \hline
  DCCT scale                   	       &  2.7       & 2.7  \\
  DCCT baseline noise          	       &  negligible& 3.9  \\
  FBCT systematics           	       &  0.2       & 2.9  \\
  Ghost charge                         &  0.2       & 0.1  \\
  \hline
  Statistical error                    &  0.1       & 0.9  \\
  \hline
  $N_1\, N_2$ drop             	       &  negligible& negligible  \\ 
  Emittance growth             	       &  negligible& negligible  \\ 
  Reproducibility at nominal position  &  0.4       & $-$  \\ 
  Difference between two scans	       &  2.1       & 4.4  \\ 
  Average beam length scale            &  1.0       & 2.0  \\ 
  Beam length scale difference 	       &  0.1       & $-$  \\ 
  Coupling of $x$ and $y$ coordinates  &  0.3       & $-$  \\ 
  \hline
  Total                        	       &  3.6      & 7.5  \\
 \end{tabular}
\end{table}

\section{The beam-gas imaging (BGI) method}
\label{sec:beamgas}

Tracks measured in the LHCb detector allow vertices from the
interactions of beam particles with the residual gas in the machine
(beam-gas interactions) and from the beam-beam collisions to be
reconstructed. 
The beam-gas imaging method is based on the measurement of the
distributions of these vertices to obtain an image of the 
transverse bunch profile along the beam trajectory.
This allows the beam angles, profiles and relative positions to be
determined. 
The residual gas in the beam vacuum pipe consists mainly of relatively
light elements such as hydrogen, carbon and oxygen.

An important prerequisite for the proper reconstruction of the bunch
profiles is the  transverse homogeneity of the visualizing gas (see
Sect.~\ref{sec:gradient}).  
A dedicated test performed in October 2010 measured the beam-gas
interaction rates as function of beam displacement in a plane
perpendicular to the beam axis. 
The correction to the beam overlap integral due to a possible non-uniform
transverse distribution of the residual gas is found to be smaller than
0.05\% and is neglected. 

Compared to the VDM method, the disadvantage of a small rate is balanced 
by the advantage that the method is non-disruptive, without beam
movements. 
This means that possible beam-beam effects are constant and 
effects which depend on the beam displacement, like hysteresis, are 
avoided.  
Furthermore, the beam-gas imaging method is applicable during physics fills.

The half crossing-angle \thetac is small enough to justify setting
$\cos^2 \thetac =1$ in Eq.~\ref{eq:luminosity}.
In the approximation of a vanishing correlation between the transverse
coordinates the $x$ and $y$ projections can be factorized.
At the level of precision required, the bunch shapes are well described
by Gaussian distributions.
Thus, their shapes are characterized in the $x$-$y$ plane at the 
time of crossing by their widths $\sigma_{bi}$, their mean positions 
$\mpos_{bi}$ ($i=x, y$), and by their bunch length
$\sigma_{bz}$.
The index $b$ takes the values 1 and 2 according to the two beams. 
With these approximations, Eq.~(\ref{eq:luminosity}) 
for a single pair of colliding bunches reduces to~\cite{ref:simonwhite}
\begin{equation}
 L=
 \frac{N_1\, N_2\,f}{2\pi\sqrt{1+\tan^2{\thetac}(\sigma_{1z}^2 +
 \sigma_{2z}^2) /(\sigma_{1x}^2 + \sigma_{2x}^2)}} ~ 
 \prod_{i=x,y} \frac{1}{\sqrt{\sigma_{1i}^2 + \sigma_{2i}^2}} 
 \exp{\left(-\frac{1}{2}
       \frac{(\mpos_{1i}-\mpos_{2i})^2}{\sigma_{1i}^2+\sigma_{2i}^2}\right)}
 \, ,
\label{eq:obslumi}
\end{equation}
where the denominator of the first factor in the product corrects for
the crossing angle.  
The analysis is applied for each individual colliding bunch pair, {\em
i.e.} bunch populations, event rates and beam profiles are considered
per bunch pair.
Thus, each colliding bunch pair provides an internally consistent
measurement of the same visible cross-section.
The observables $\sigma_{bi}$ and $\mpos_{bi}$ are extracted 
from the transverse distributions of the beam-gas vertices 
reconstructed in the \bx{bb} crossings of the colliding bunch pairs 
(see Sect.~\ref{sec:bganalysis}).

The beam overlap-integral is then calculated from the two individual
bunch profiles. 
The simultaneous imaging of the $pp$ luminous region further constrains
the beam parameters.  
The distribution of $pp$-collision vertices, produced by the 
colliding bunch pair and identified by requiring $-150<z<150$~mm, 
is used to measure the parameters of the luminous region. 
Its positions  \posxp{i} and transverse widths \sigxp{i},  
 \begin{equation}
   \posxp{i} = \frac{\mpos_{1i}\sigma_{2i}^2 +
   \mpos_{2i}\sigma_{1i}^2}{\sigma_{1i}^2 + \sigma_{2i}^2} 
   ~~~ \mbox{and} ~~~ 
   \sigxp[2]{i} = \frac{\sigma_{1i}^2 \sigma_{2i}^2}%
   {\sigma_{1i}^2 + \sigma_{2i}^2}  \,,
   \label{eq:lumi-gaussian}
 \end{equation}
constrain the bunch observables. 
Owing to the higher statistics of $pp$ interactions
compared to beam-gas interactions,
the constraints of Eq.~(\ref{eq:lumi-gaussian})
provide the most significant input to the overlap integral.
Equation~(\ref{eq:lumi-gaussian}) is valid only for a zero crossing
angle.
It will be shown in Sect.~\ref{sec:crossing-effect} that the
approximation is justified for this analysis. 

The bunch lengths $\sigma_{bz}$ are extracted
from the longitudinal distribution of the $pp$-collision
vertices.\footnote{In fact, only the combination 
$(\sigma_{1z}^2 + \sigma_{2z}^2)$ can be obtained.}
Because the sizes $\sigma_{bz}$ are approximately 1000 times larger than
$\sigma_{bx}$, the crossing angle reduces the luminosity by a non-negligible
factor equal to the first square root factor in Eq.~(\ref{eq:obslumi}).
The case of non-collinear beams is
described in more detail in Sect.~\ref{sec:crossing-effect}. 

The BGI method requires a vertex resolution comparable to or smaller than 
the transverse beam sizes.  
The knowledge of the vertex resolution is necessary to unfold the
resolution from the measured beam profiles.
The uncertainty in the resolution also plays an essential role in
determining the systematic error.

The beam-gas interaction rate determines the time needed to take a
{\em snapshot} of the beam profiles and the associated statistical
uncertainty. 
When the time required to collect enough statistics is large compared to
the time during which the beam stays stable, it becomes necessary to
make additional corrections.  
This introduces systematic effects.

\subsection{Data-taking conditions}
\label{sec:bg:conditions}

The data used for the results described in the BGI analysis were taken
in May 2010.
In the data taken after this time the 
event rate was too high to select beam-gas events at the trigger level.
In October a more selective trigger was in place and sufficient data
could be collected.
However, in this period difficulties were observed with the DCCT data
for LHC filling schemes using bunch trains.
One should observe that the  VDM data taken in October used a dedicated
fill with individually injected bunches so that these problems were not
present. 

In the selected fills, there were 2 to 13 bunches per beam in the machine.
The number of colliding pairs at LHCb varied between 1 and 8. 
The trigger included a dedicated selection for events
containing beam-gas interactions (see Sect.~\ref{sec:lhcb}). 

The HLT runs a number of algorithms designed to select beam-gas
interactions with high efficiency and low background.
The same vertex algorithm is used for the \bx{be}, \bx{eb} and \bx{bb}
crossings, but different $z$-selection cuts are applied. 
For \bx{bb} crossings the region $-0.35 < z < 0.25$~m is excluded to reject
the overwhelming amount of $pp$ interactions.\footnote{Due to the
asymmetric VELO geometry, the background from $pp$ vertices near the
upstream end of the VELO is more difficult to reject, hence the
asymmetric $z$ selection.}

\subsection{Analysis and data selection procedure}
\label{sec:bganalysis}

The standard vertex reconstruction algorithms in LHCb are optimized to find
$pp$ interaction vertices close to $z=0$.
This preference is removed for this particular analysis such
that no explicit bias is present in the track and vertex selection as a
function of $z$. 
The resolution of the vertex measurement has to be known with high
precision.
Details of the resolution study are given in
Sect.~\ref{sec:vxresolution}. 

\begin{table}
 \small
 \begin{center}
  \caption{LHC fills used in the BGI analysis. 
  The third and fourth columns show the total number of (colliding) 
  bunches \ntot (\ncol), the fifth the typical number of particles per
  bunch, the sixth the period of time used for the analysis, and for the
  fills used in the BGI analysis the seventh and eighth the measured
  angles in $x$ (in mrad) of the individual beams with respect to the LHCb reference
  frame (the uncertainties in the angles range from 1 to 5 \micrad).
  The last two columns give the typical number of events per bunch used in the BGI
  vertex fits for each of the two beams.
  \label{tab:fills}
  }
  \vskip 1mm
   \begin{tabular}{lcrrccrrlrr}
    {Fill} & {part} & ntot&
    \ncol& $N$ & {time (h)} &
     $\thetab_\mathrm{beam~1}$&  $\thetab_\mathrm{beam~2}$& {
    analysis}&{events 1}&{events 2}\\ \hline
\boldmath    1089 &   &  2 & 1 & $2 \tms 10^{10}$ & 15 & $0.209$   & -0.371& BGI 1  &1270&720\\
    1090 &   &  2 & 1 & $2 \tms 10^{10}$ & 4  & $0.215$   & -0.355& BGI 2  &400&300\\
    1101 &   &  4 & 2 & $2 \tms 10^{10}$ & 6  & $-0.329$  & 0.189 & BGI 3  &730&400\\
    1104 & A &  6 & 3 & $2 \tms 10^{10}$ & 5  & $0.211$   & -0.364& BGI 4  &510&350\\
    1104 & B &  6 & 3 & $2 \tms 10^{10}$ & 5  & $0.211$   & -0.364& BGI 5  &520&350\\
    1117 &   &  6 & 3 & $2 \tms 10^{10}$ & 6  & $-0.327$  & 0.185 & BGI 6  &700&500\\
    1118 &   &  6 & 3 & $2 \tms 10^{10}$ & 5  & $-0.332$  & 0.181 & BGI 7  &500&400\\
    1122 &   & 13 & 8 & $2 \tms 10^{10}$ & 3  & $-0.329$  & 0.182 & BGI 8  &300&250\\
   \end{tabular}
 \end{center}
\end{table}

The BGI method relies on the unambiguous selection of beam-gas
interactions, also during \bx{bb} crossings where an overwhelming majority of
$pp$ collisions is present.
The beam-gas fraction can be as low as $10^{-5}$ depending on the beam
conditions. 
The criteria to distinguish beam-gas vertices from $pp$ interactions 
exploit the small longitudinal size of the beam spot (luminous region). 
As an additional requirement only  vertices formed with exclusively
forward (backward) tracks are accepted as beam~1(2)-gas interactions and
vertices are required to be made with more than ten tracks.
A further selection on the transverse distance from the measured beam-axis is
applied to reject spurious vertices\footnote{Interactions in material
and random associations of tracks.} ($\pm 2$~mm). 
Due to the worsening of the resolutions for large distances from 
$z=0$ and due to the presence of $pp$ interactions near $z=0$, to
determine the width of the beams the analysis regions are limited to 
$-700<z<-250$~mm for beam~1 and $250 < z < 800$~mm for beam~2.\footnote{The
vertex resolution for beam~2 has a weaker $z$ dependence, so the
sensitivity is improved by enlarging the region.} 
The selection of $pp$ events requires $-150 < z < 150$~mm and only
accepts vertices with more than 20 tracks.
The background of beam-gas interactions in the $pp$ interaction sample
is negligible owing to the high $pp$ event rate.

The transverse profiles of the two beams are measured for each
individual colliding bunch by projecting the vertex position 
on a plane perpendicular to the beam direction. 
The direction of the beam is determined on a fill-by-fill basis using
the beam-gas interactions observed in \bx{be} and \bx{eb} crossings,
which are free of $pp$ interactions.  
The direction of the beam axis can be determined with 1 to 5~\micrad
precision depending on the fill. 

Out of the many LHC fills only seven are selected for the BGI analysis. 
It is required that all necessary data (DCCT, FBCT,
luminosity counters  and vertex measurements) are present during a
sufficiently long period 
and that the bunch populations and emittances are sufficiently stable
during the selected period. 
The list of used fills is given in Table~\ref{tab:fills}.
The table shows the total number of bunches and the number of bunches colliding at
LHCb, the typical number of protons per bunch, the measured beam slopes
with respect to the LHCb reference frame, and the duration of the period
used for the analysis.  

The bunch population and size cannot be assumed to be constant during the
analysis period.
Therefore, the DCCT and FBCT data and the vertex measurements using $pp$
interactions are binned in periods of 900 seconds.
This binning choice is not critical. 
The chosen value maintains sufficient statistical precision while
remaining sensitive to variations of the beams.   
The distributions of beam-gas interactions do not have sufficient
statistics and are accumulated over the full periods shown in 
Table~\ref{tab:fills}.
The analysis proceeds by determining a time-weighted average for the
bunch-pair population product and the width and position of the $pp$
beam spot.
The weighting procedure solves the difficulty introduced by short
periods of missing data by a logarithmic interpolation for the bunch
populations and a linear interpolation for the bunch profiles. 
The averages defined by the latter procedure can be directly compared to
the single measurement of the profiles of the single beams accumulated
over the full period of multiple hours. 
A systematic error is assigned to account for the approximations
introduced by this averaging procedure.

\subsection{Vertex resolution}
\label{sec:vxresolution}

The measured vertex distribution is a convolution of the true 
width of each beam with the resolution function of the detector. 
Since the resolution is comparable
to the actual beam size (approximately 35~\micron in the selected fills),
it is crucial to understand this effect, and to be able to unfold it
from the reconstructed values. 

The vertex resolution is parametrized as a function of the multiplicity,
or number of tracks used to reconstruct the vertex, and as a function of
the $z$ position of the interaction. Beam-gas vertices alone are used to
measure the positions and spatial extent of each beam; however, these
events are rare in comparison to beam-beam vertices. To avoid binning
the beam-gas vertices in both number of tracks and $z$ position,
beam-beam events are initially used to measure the dependence of the
resolution on the number of tracks.
Once this
dependence is known, the beam-gas vertices are used to find the $z$
dependence of the resolution, taking into account the contribution to
the resolution given by the number of tracks found in each vertex. 

The resolution as a function of the number of tracks in a vertex is
determined using $pp$ interactions which occur around $z = 0$. 
The reconstructed tracks from each event are randomly split into two
independent sets of equal size. 
The vertex reconstruction is applied to each set of tracks, and if
exactly one vertex is found from each track collection it is assumed to
be from the same original interaction. 
Then, if the number of tracks making each of these two vertices is the
same, the residuals in $x$ and $y$ are taken as an estimate of the vertex
resolution.

\begin{figure}[tbp]
 \begin{center}
  \includegraphics*[width=0.8\textwidth,clip,trim=0mm 0mm 0mm 0mm]{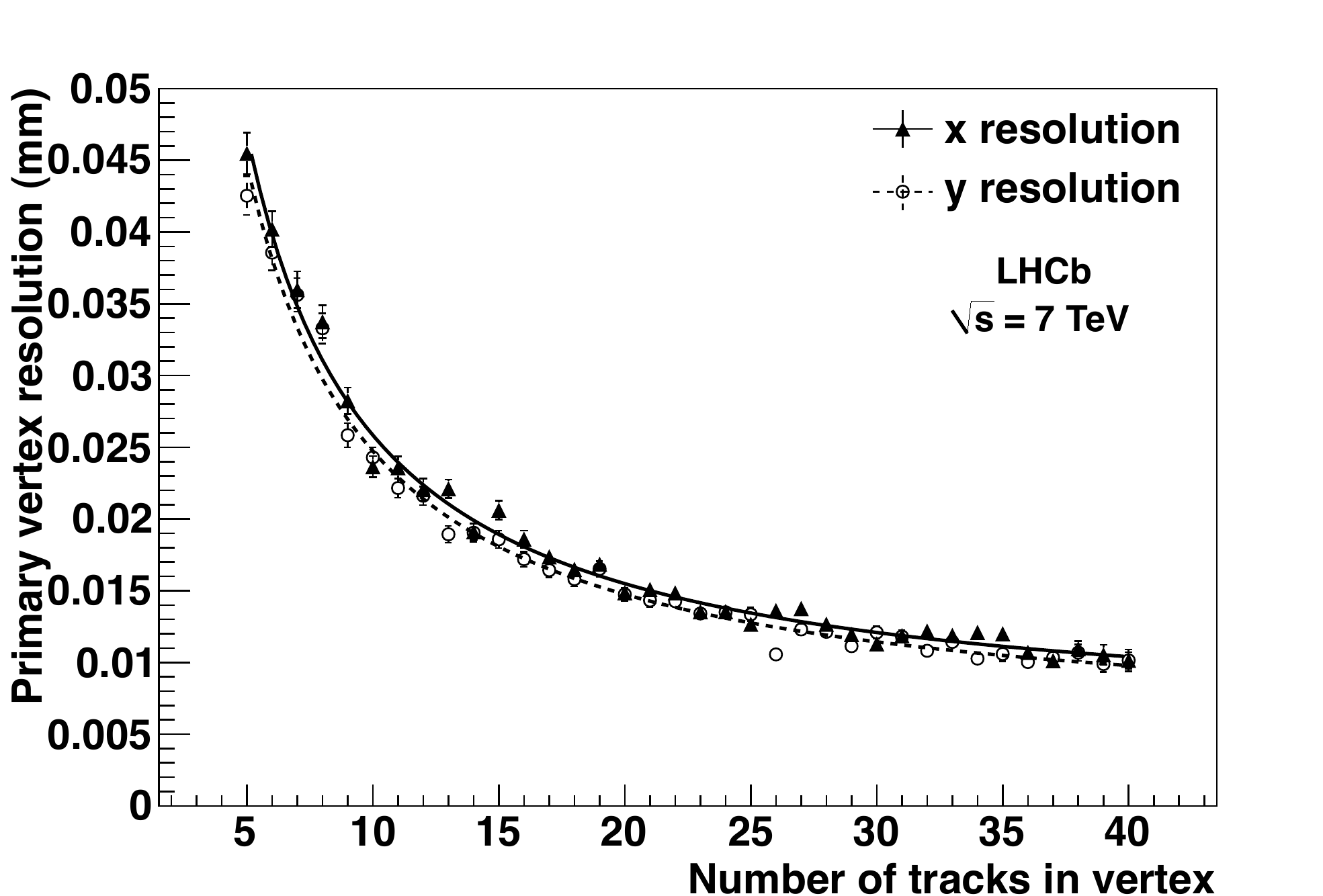}
  \caption{Primary vertex resolution $\sigma_{\mathrm{res}}$ in the
  transverse directions $x$ (full circles) and $y$ (open circles)  
  for beam-beam interactions as a function of the number of tracks in the
  vertex, \ntrk.
  The curves are explained in the text.
  \label{fig:vtx-resol} 
  }
 \end{center}
\end{figure}

\begin{table}[tbp]
\centering
 \caption{
 Fit parameters for the resolution of the transverse positions $x$ and $y$
 of reconstructed beam-beam interactions as a function of the number of
 tracks. 
 The errors in the fit parameters are correlated. 
 \label{table:res_fit_params}
 }
 \vskip 1mm
\begin{tabular}{ lcc }
                   	    & \boldmath$x$     & \boldmath$y$ \\\hline   
Factor $A$ (mm)    	    & $0.215 \,\pm\, 0.020$    & $0.202 \,\pm\, 0.018$ \\
Power $B$          	    & $1.023 \,\pm\, 0.054$    & $1.008 \,\pm\, 0.053$ \\
Constant $C$ ($10^{-3}$~mm)  & $5.463 \,\pm\, 0.675$    & $4.875 \,\pm\, 0.645$ \\
\end{tabular}
\end{table}

The resolution is calculated as the
width of the Gaussian function fitted to the residual distributions divided by
$\sqrt 2$, as there are two resolution contributions in each residual
measurement.  
The resolutions are shown as a function of the number of tracks  in
Fig.~\ref{fig:vtx-resol}.  
Since the VELO is not fully symmetric in the $x$ and $y$ coordinates,
the analysis is performed in both coordinates separately.
Indeed, one observes a small, but significant difference in the resolution.
The points of Fig.~\ref{fig:vtx-resol} are fitted with a function which
parametrizes the resolution in terms of a factor $A$, a power $B$ and a
constant $C$, as a function of the track multiplicity \ntrk
\begin{equation}
\label{eq:track:dep} 
\sigma_{\mathrm{res}} = \frac{A}{\ntrk^{B}} + C \, .
\end{equation}
The values of the fit parameters are given in
Table~\ref{table:res_fit_params}.  

 \begin{figure}[btp]
 \begin{center}
  \includegraphics*[width=0.495\textwidth,clip,trim=3mm 0mm 10mm 0mm]{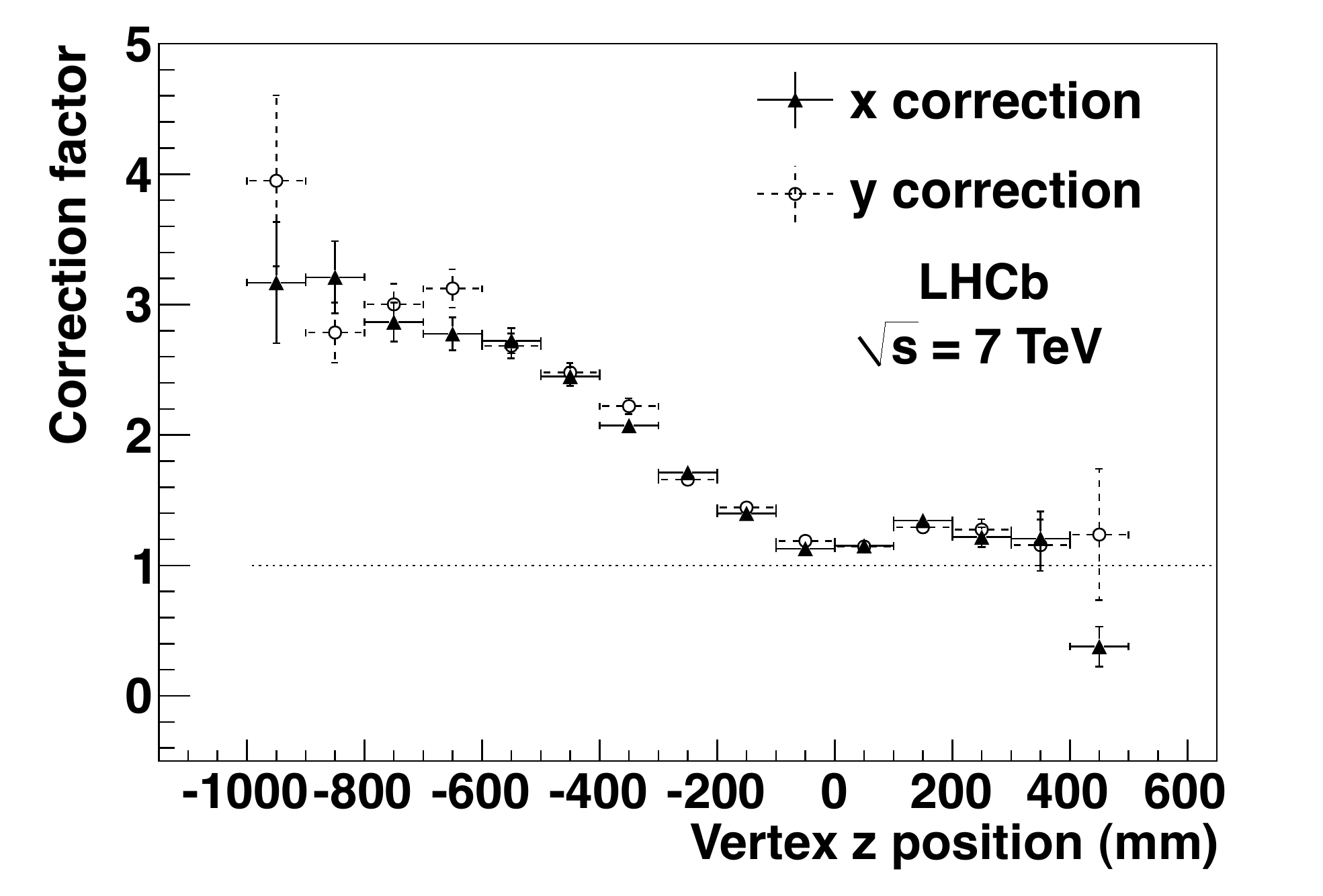}
  \includegraphics*[width=0.495\textwidth,clip,trim=3mm 0mm 10mm 0mm]{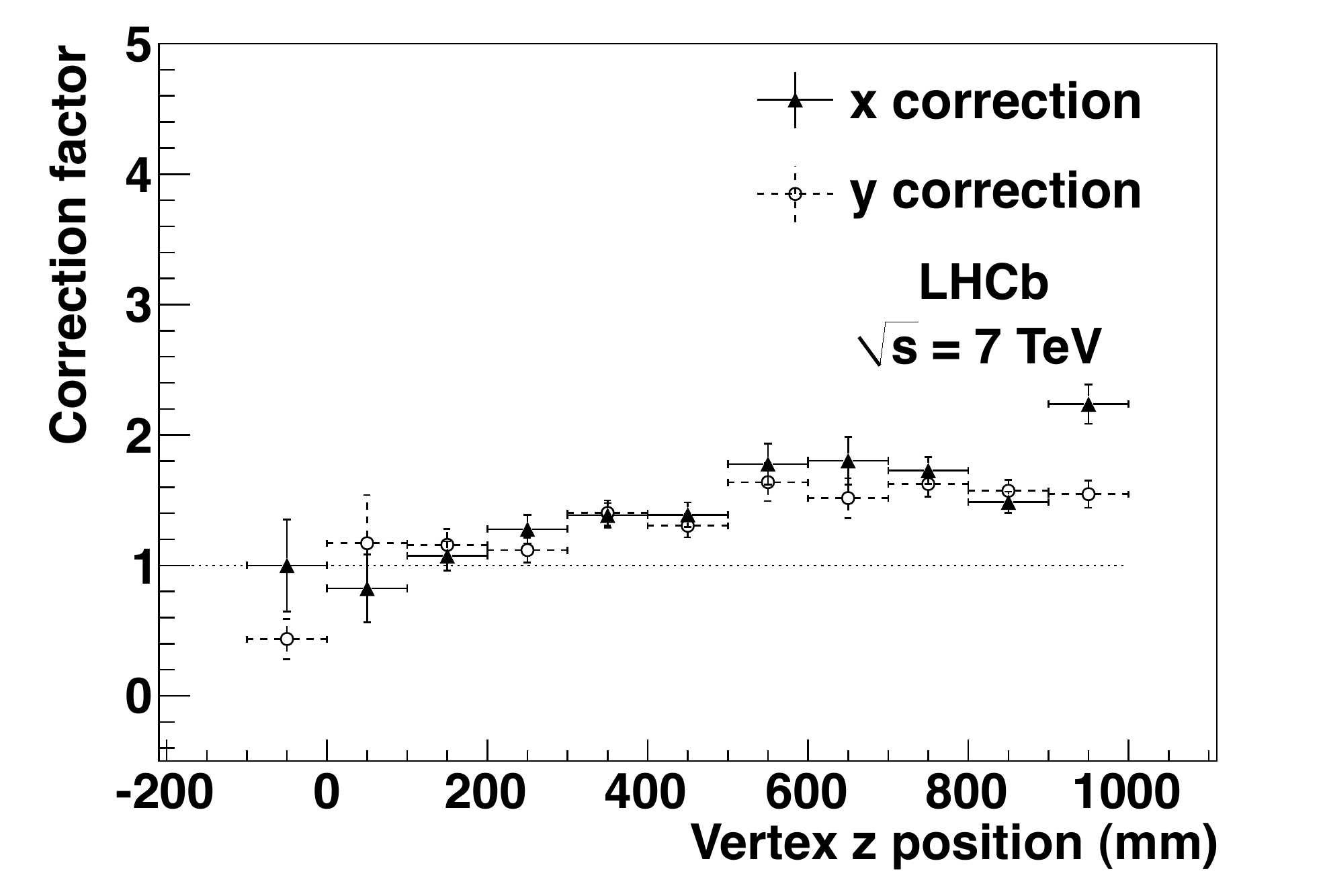}
  \caption{The $z$ dependence of the resolution correction factor \cres
  in $x$ (full circles) and $y$ (open circles) for
  beam~1-gas (left)   and beam~2-gas (right) interactions.
  \label{fig:vtx-resol2}
  }
 \end{center}
\end{figure}

Beam-gas vertices, selected in \bx{be} and \bx{eb} bunch crossings, 
are reconstructed in the same manner as the $pp$ vertices.
Every beam-gas event which yields two vertices after splitting the track
sample into two is used in the analysis, without requiring that the two
vertices are reconstructed with an equal number of tracks.  
Accounting for the resolution coming from the track multiplicity means
that a correction factor \cres is calculated as a function of $z$
position. 
This is the factor by which the beam-beam resolution at $z = 0$ 
must be multiplied to find the true resolution for an event with a
particular number of tracks, at a certain position in $z$. 
The contribution to the resolution from the track multiplicity is taken
into account for coordinate $v$ ($v=x,y$) according to
\begin{equation}
\label{eq:zdep} 
\cres = \frac{v_{1} - v_{2}}{\sqrt{\sigma_{\ntrks{1}}^{2} + \sigma_{\ntrks{2}}^{2}}} \, ,
\end{equation}
where the index ${1,2}$ signifies the two vertices with measured position $v_1$
and $v_2$, and ${\ntrks{1}, \ntrks{2}}$ the number of tracks in each. 
The quantities $\sigma_{\ntrks{1,2}}$ are the resolutions $\sigma_{\mathrm{res}}$
expected for vertices at $z=0$ made of \ntrk tracks as defined in
Eq.~\ref{eq:track:dep}. 

The correction factors \cres are plotted in Fig.~\ref{fig:vtx-resol2}. 
In order to better understand the resolution as a function of $z$, it is
instructive to consider the geometry of the VELO, shown in 
Fig.~\ref{fig:velo-sketch}. 
Figure~\ref{fig:vtx-resol2} shows that around the interaction point of $z = 0$ the
correction factor is close to one, which signifies that the resolution
is nearly independent of the type of event, whether beam-beam or
beam-gas. 
This is expected, since the VELO is optimized to reconstruct vertices near
$z=0$. 
The correction factor increases as the vertices
move away from the interaction region.

\subsection{Measurement of the beam profiles using the BGI method}
\label{sec:profiles}

\begin{figure}[tbp]
 \centering
 \includegraphics[width=0.95\textwidth]{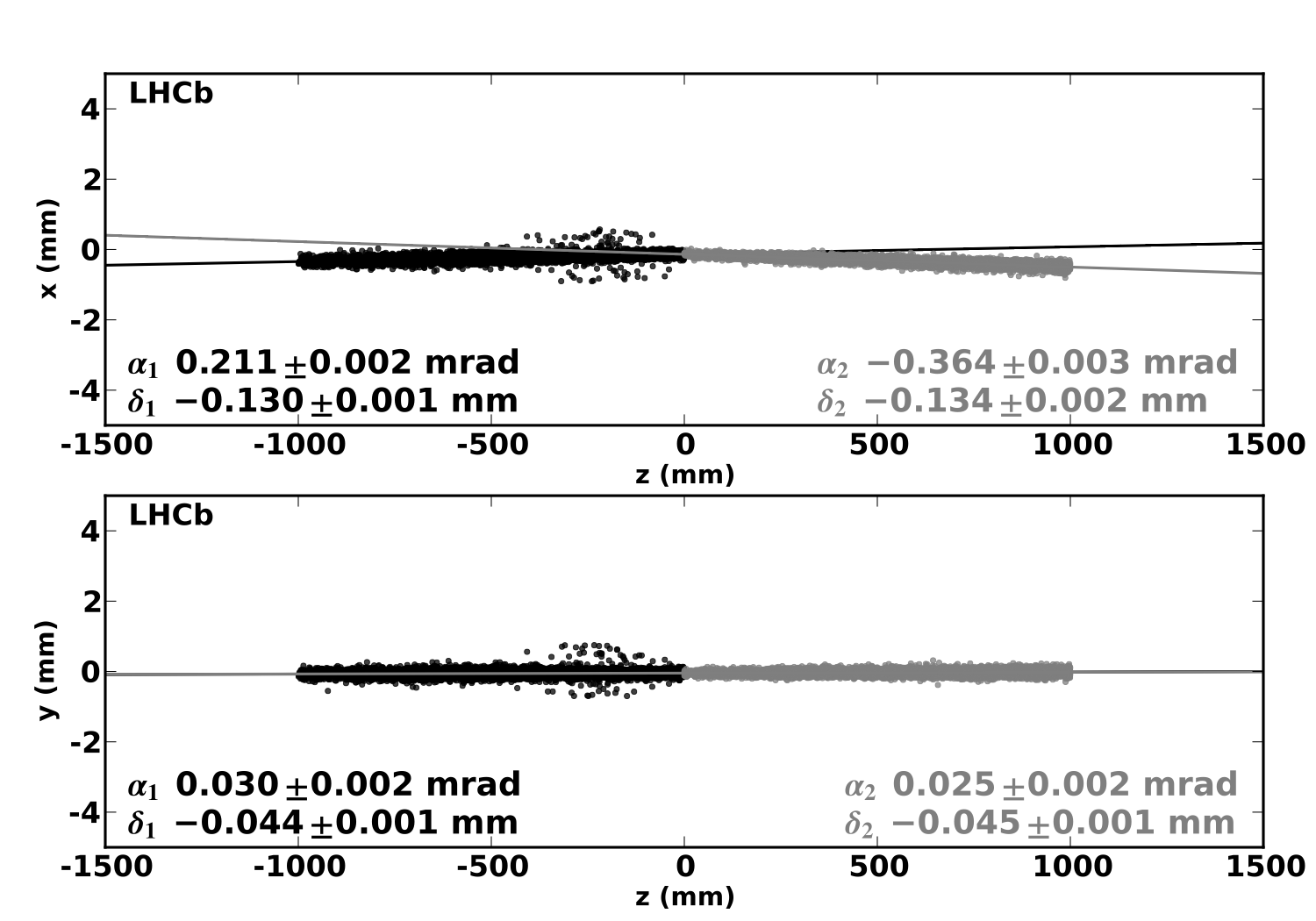}
 \caption{Positions of reconstructed beam-gas interaction vertices
 for \mbox{\bx{be}} (black points) and \mbox{\bx{eb}} (grey points)
 crossings during Fill 1104.  
 The measured beam angles $\alpha_{1,2}$ and offsets $\delta_{1,2}$
 at $z=0$  in the horizontal (top) and  vertical (bottom)
 planes are shown in the figure. 
 \label{fig:beamAngles}
 }
\end{figure}

In Fig.~\ref{fig:beamAngles} the positions of the vertices of 
beam-gas interactions of the single beams in {\bx{be}} 
and {\bx{eb}} crossings are shown in the $x$-$z$ and $y$-$z$ planes. 
The straight line fits provide the beam angles in the corresponding planes.
Whereas we can use the non-colliding bunches to determine the beam directions,
the colliding bunches are the only relevant ones for luminosity 
measurements.

\begin{figure}[tbp]
 \begin{center}
  \includegraphics[width=0.45\textwidth]{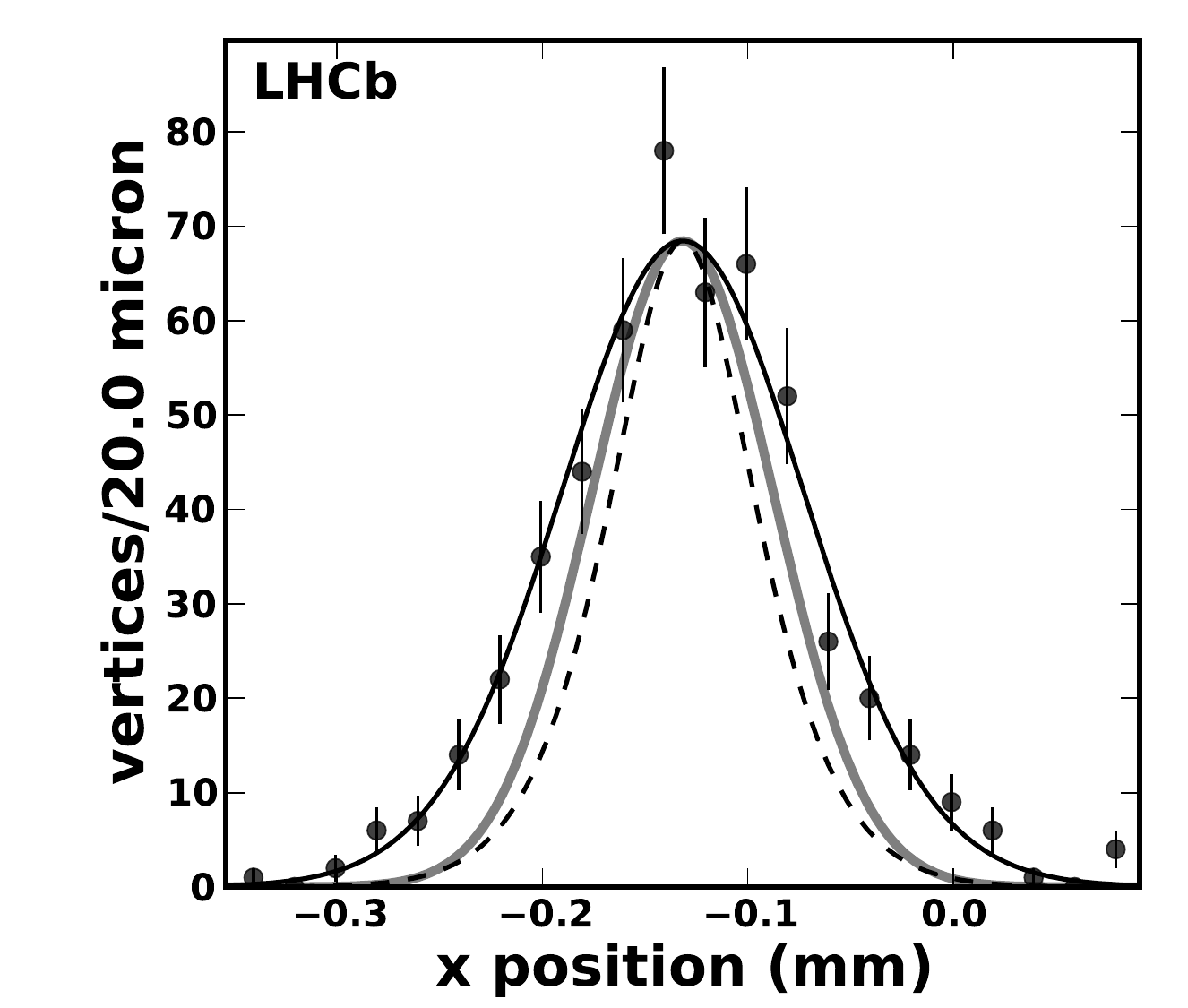}
  \includegraphics[width=0.45\textwidth]{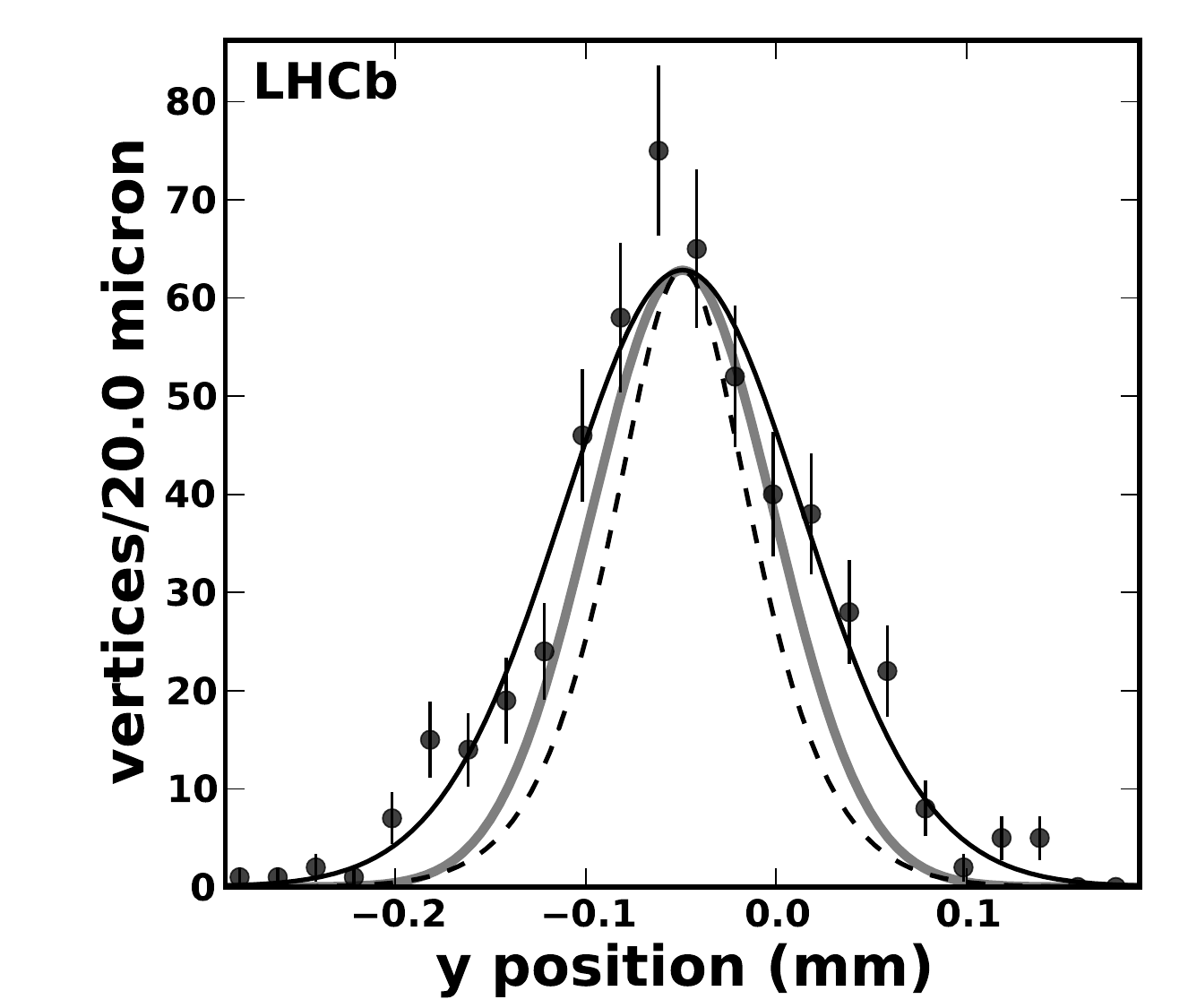}
  \includegraphics[width=0.45\textwidth]{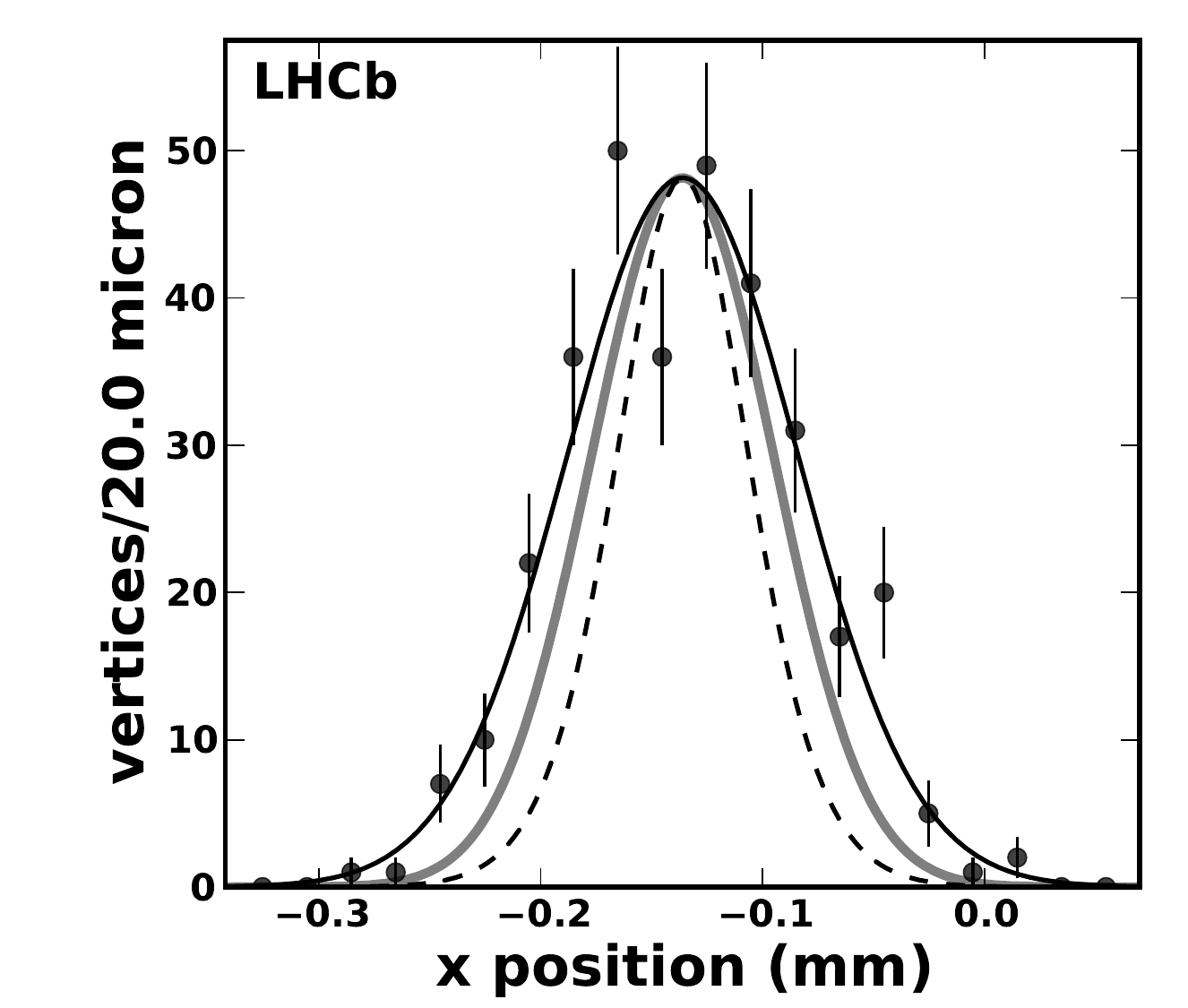}
  \includegraphics[width=0.45\textwidth]{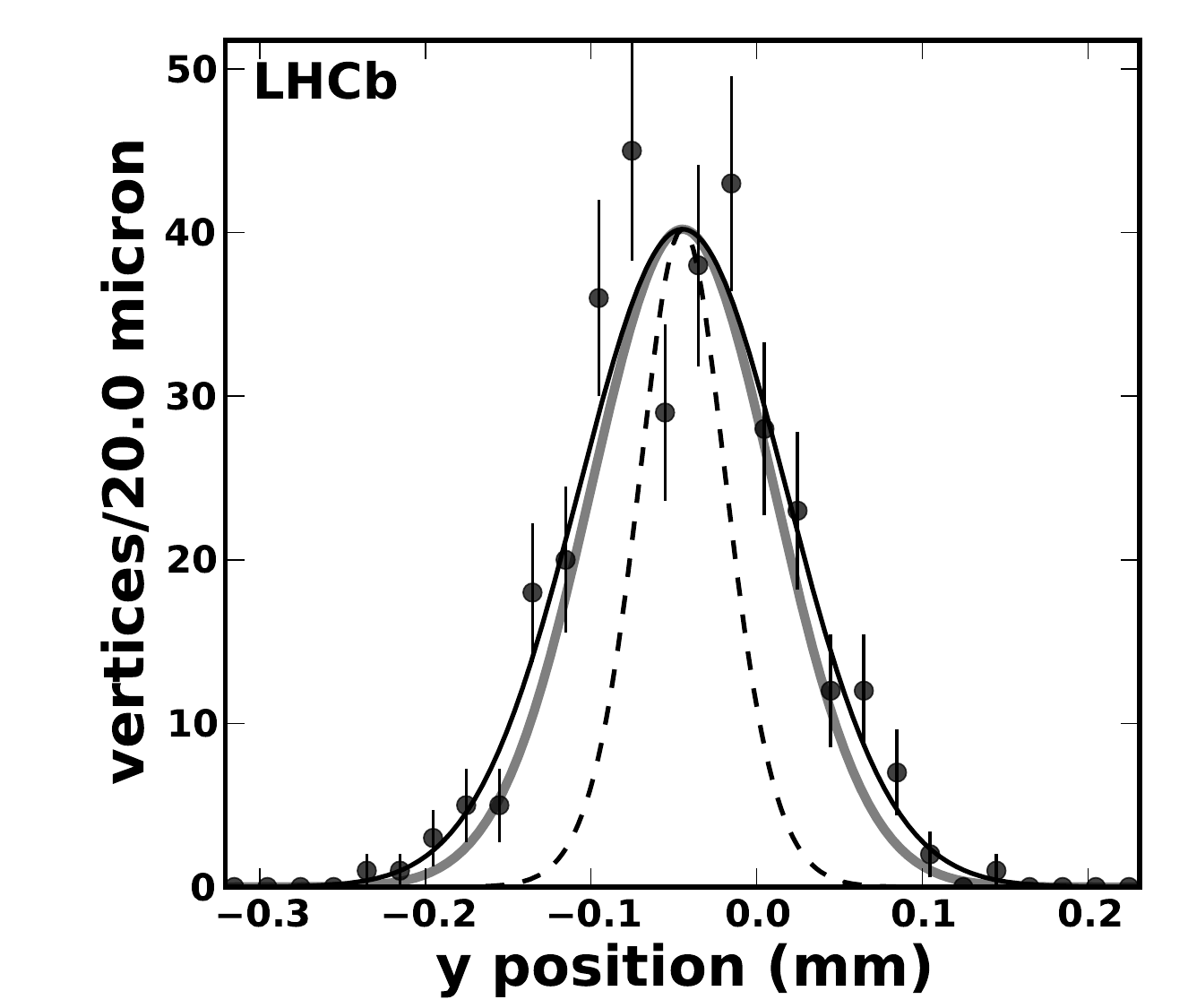}
 \end{center}
 \caption{Distributions of the vertex positions of beam-gas events for beam~1 (top) and
 beam~2 (bottom) for one single bunch pair (ID 2186) in Fill 1104.  
 The left (right) panel shows the  distribution in $x$ ($y$). 
 The Gaussian fit to the measured vertex positions is shown as a solid
 black curve together with  the resolution function (dashed) and the
 unfolded beam profile (shaded).
 Note the variable scale of the horizontal axis.
 \label{fig:beamprofiles-12}
 }
\end{figure}

\begin{figure}[tbp]
 \begin{center}
  \includegraphics[width=0.45\textwidth]{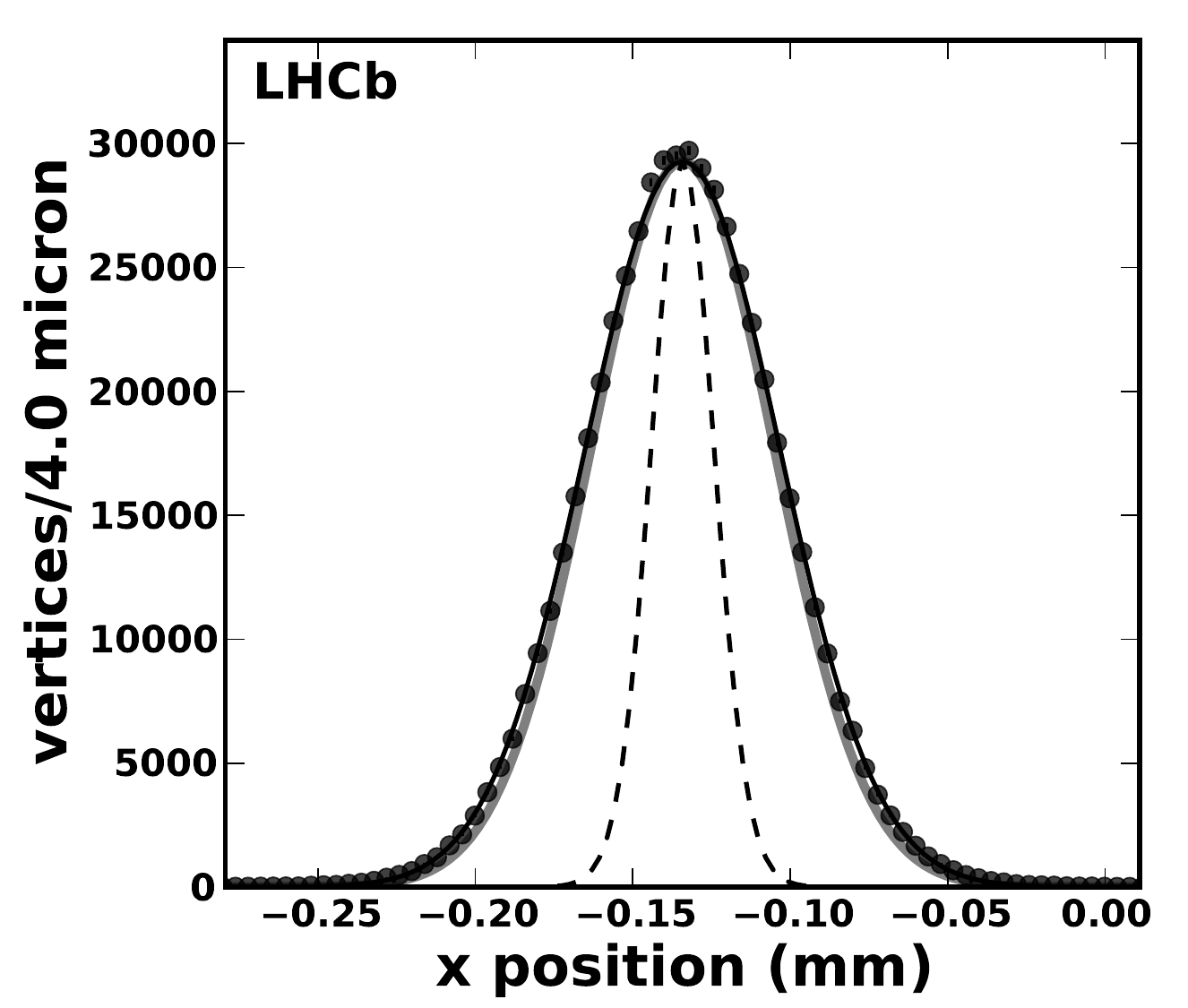}
  \includegraphics[width=0.45\textwidth]{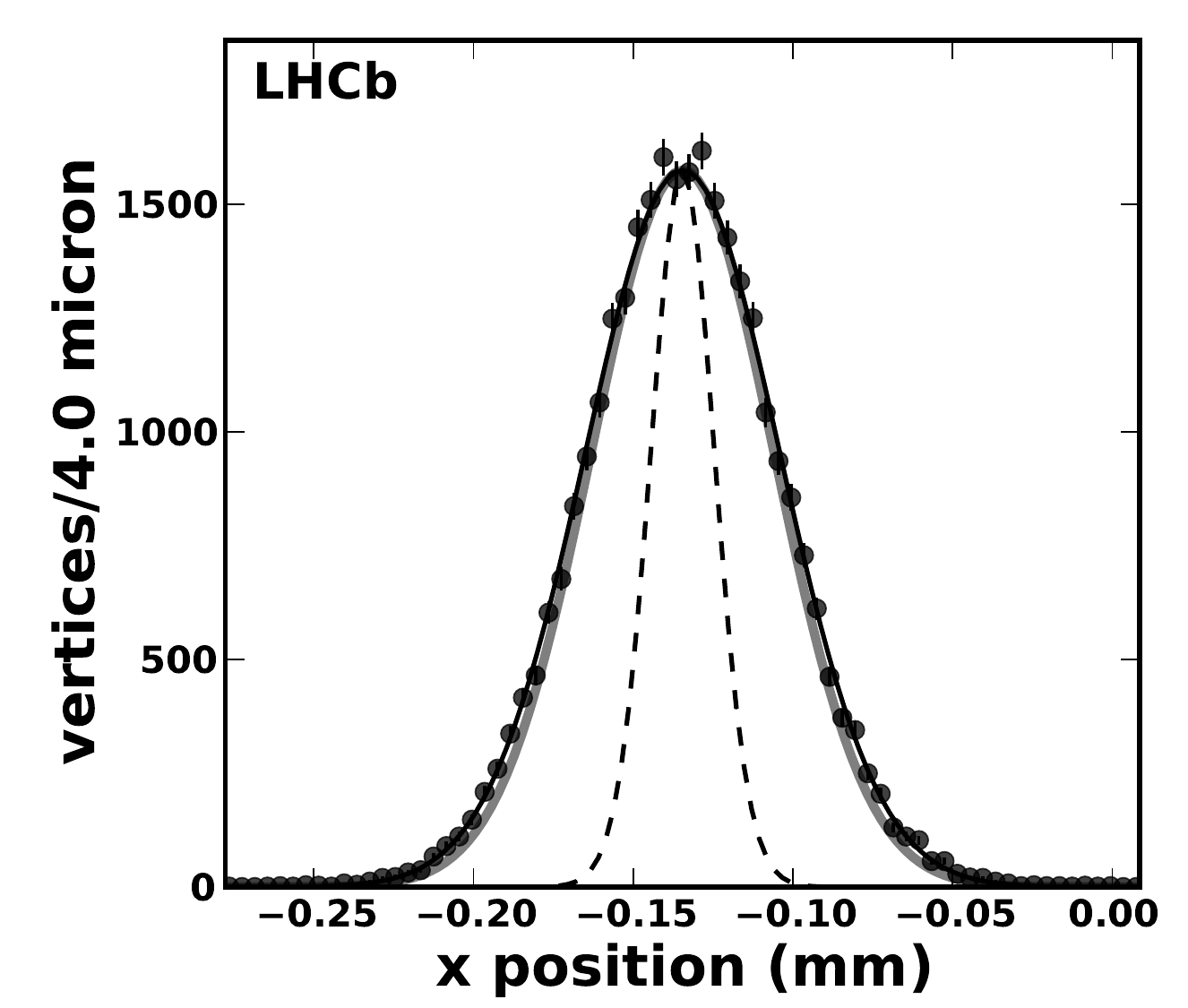}
  \includegraphics[width=0.45\textwidth]{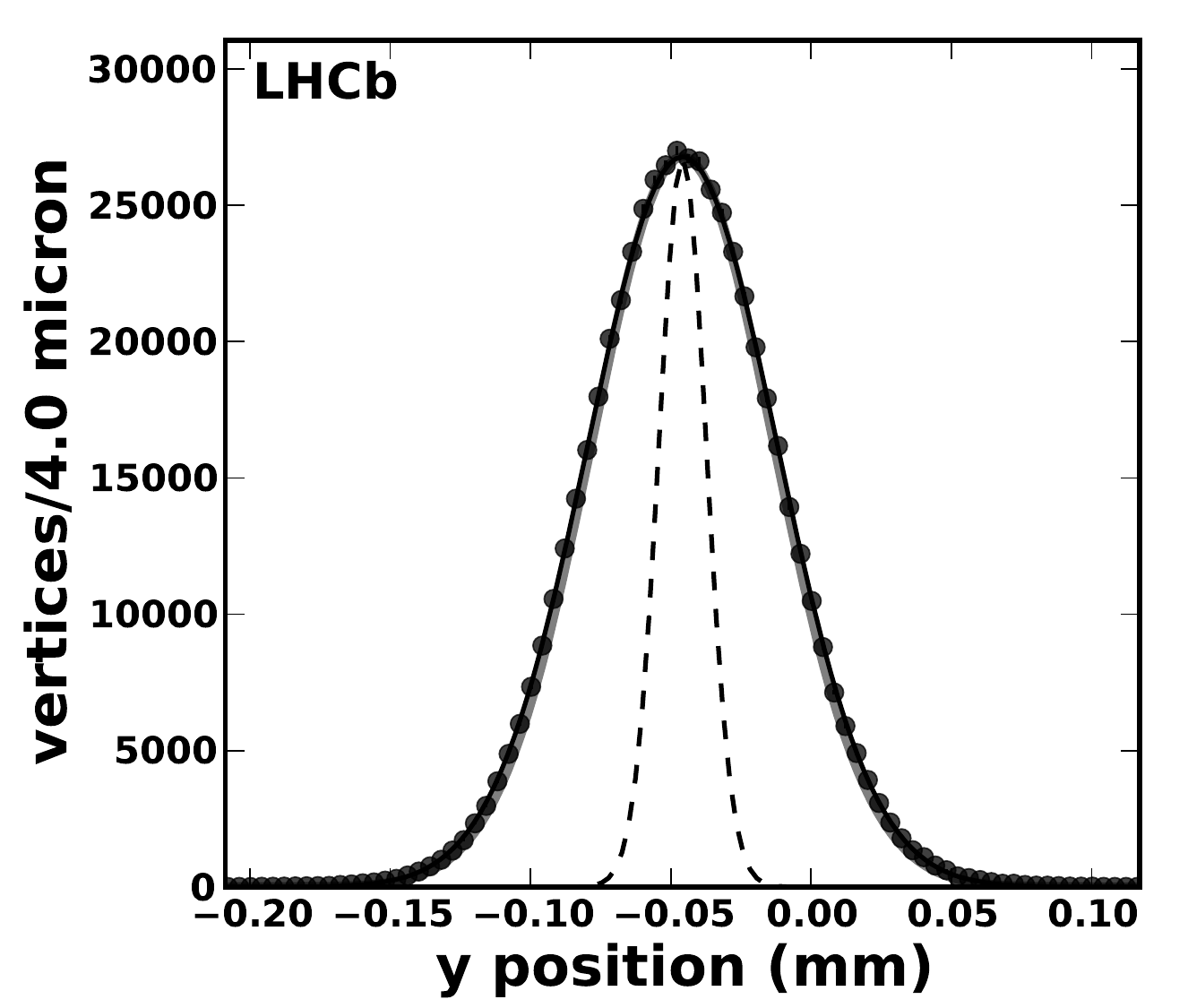}
  \includegraphics[width=0.45\textwidth]{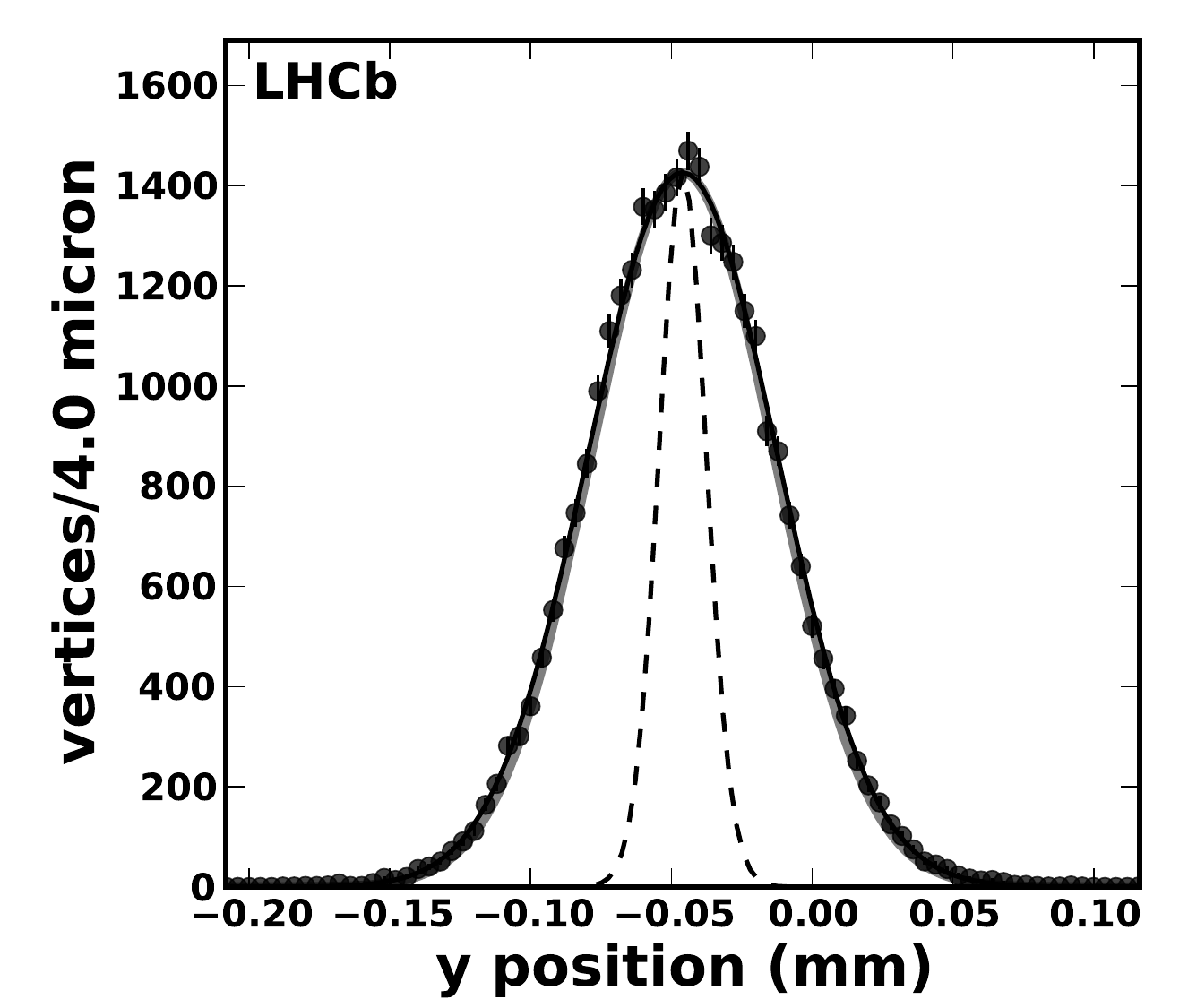}
  \includegraphics[width=0.45\textwidth]{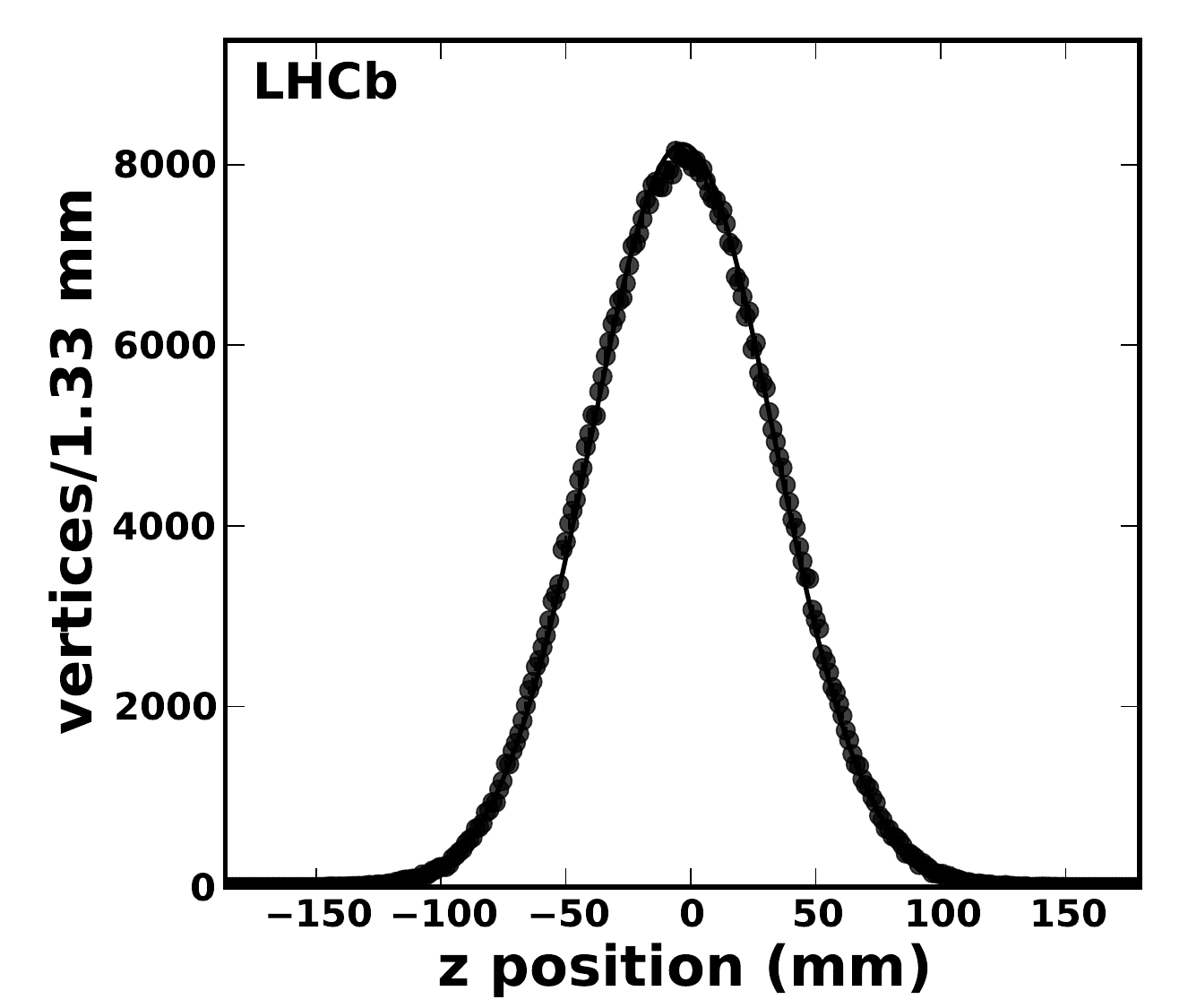}
  \includegraphics[width=0.45\textwidth]{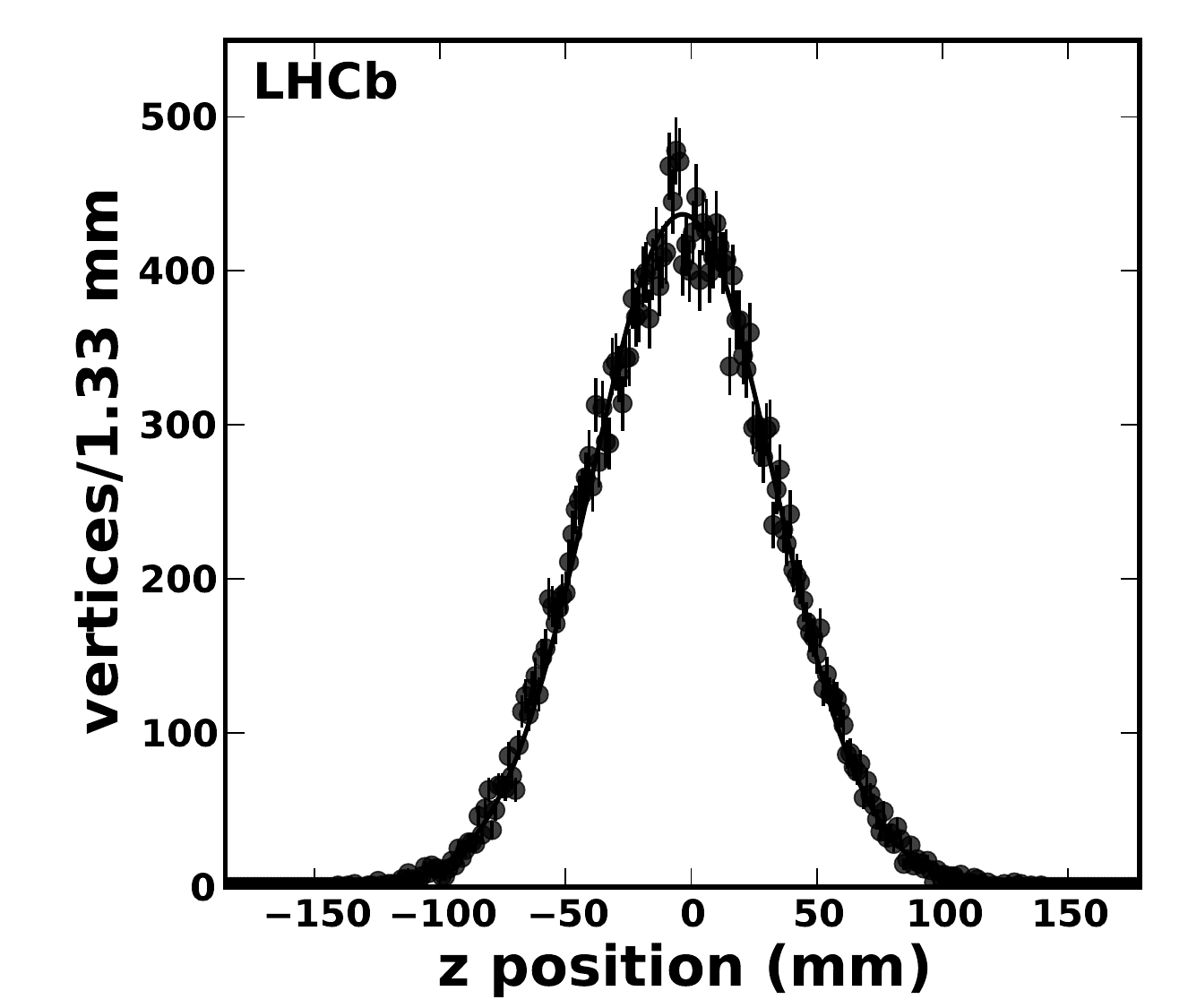}
 \end{center}
 \caption{Distributions of vertex positions of $pp$
 interactions for the full fill duration (left) and
 for a 900 s period in the middle of the fill (right) for one colliding
 bunch pair (ID 2186) in Fill 1104.  
 The top, middle and bottom panels show the
 distributions in $x$, $y$ and $z$, respectively. 
 The Gaussian fit to the measured vertex positions is shown as a solid
 black curve together with  the resolution function (dashed) and the
 unfolded luminous region  (shaded).
 Owing to the good resolution the shaded curves are close to the solid
 curves and are therefore not clearly visible in the figures.
 The fit to the $z$ coordinate neglects the vertex resolution.
 Note the variable scale of the horizontal axis.
 \label{fig:beamprofiles-pp}
 }
\end{figure}

As an example the $x$ and $y$ profiles of one colliding bunch pair
are shown in Fig.~\ref{fig:beamprofiles-12}. 
The physical bunch size is obtained after deconvolving the vertex
resolution.
The resolution function and physical beam profile are drawn separately
to show the importance of the knowledge of the resolution.
In Fig.~\ref{fig:beamprofiles-pp} the corresponding fits to the luminous
region of the same bunch pair are shown, both for the full fill duration
and for a short period of 900 s.
The fits to the distributions in $x$ and $y$ of the full fill have a
$\chi^2/{\ndof} \approx 10$, probably due to the
emittance growth of the beam.
The corresponding fits to data taken during the shorter period of 900~s
give satisfactory values, $\chi^2/{\ndof} \approx 1$.
The fact that the resolution at $z = 0$ is small compared to the size of
the luminous region makes it possible to reach small systematic uncertainties in
the luminosity determination as shown in Sect.~\ref{sec:bg:results}.

\begin{figure}[tb]
 \begin{center}
  \includegraphics[width=0.7\textwidth]{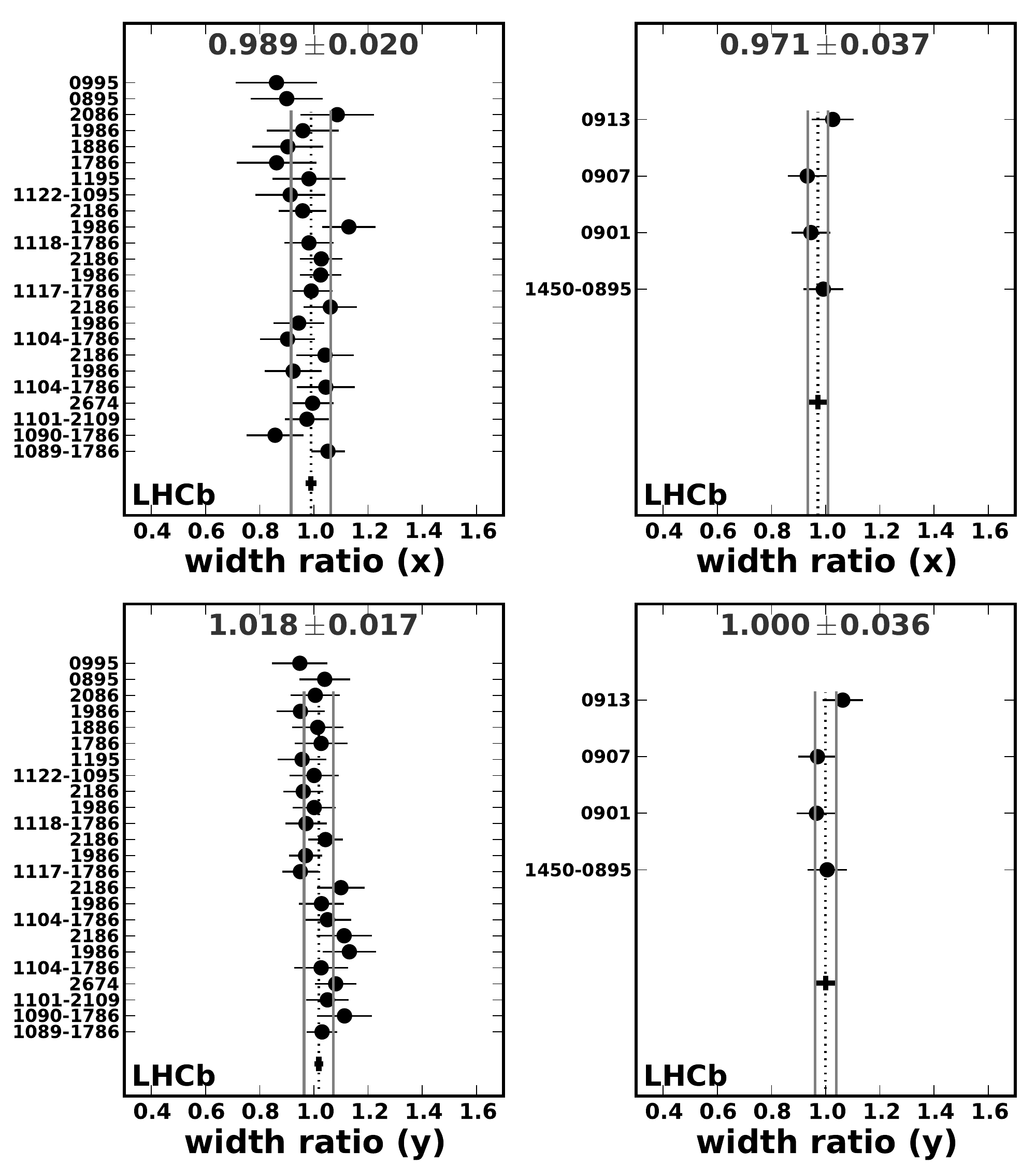}
 \end{center}
 \caption{Comparison of the prediction for the luminous region width from
 measurements based on beam-gas events of individual bunches which are
 part of a colliding bunch pair  with the direct measurement of 
 the luminous region width for these colliding  bunches.
 The panels on the left show the results for bunches in the fills with
 $\beta^* = 2$~m
 optics used in this analysis, the right panels show four colliding bunches in a
 fill taken with $\beta^* = 3.5$~m optics.
 The fill and bunch numbers are shown on the vertical axis.
 The vertical dotted line indicates the average and the solid lines the
 standard deviation of the data points.
 The lowest point indicates the weighted average of the individual
 measurements; its error bar represents the corresponding uncertainty in
 the average.
 The same information is given above the data points.
 The fills with the $\beta^* = 3.5$~m optics are not used for the
 analysis due to the fact that larger uncertainties in the DCCT
 calibration were observed.
 \label{fig:constraint-pp}
 }
\end{figure}

For non-colliding bunches it is possible to measure the width of the
beam in the region of the interaction point at $z=0$ since there is
no background from $pp$ collisions. 
One can compare the measurement at the IP with the measurement in the
region outside the IP which needs to be used for the colliding bunches. 
After correcting for the resolution no difference is observed, as expected
from the values of $\beta^*$ of the beam optics used during the data
taking.  
One can also compare the width measurements of the colliding bunches far
from the IP using the beam-gas events from beam~1 and beam~2 to predict
the width of the luminous region using Eq.~\ref{eq:lumi-gaussian}.
Figure~\ref{fig:constraint-pp}  shows that there is overall
consistency. 
In addition to the data used in the BGI analysis described here, also
higher statistics data from later fills are used for this comparison.
The cross check reaches a precision of 1--1.6\% for the consistency of
the width measurements at large $z$ compared to the measurement at $z=0$,
providing good evidence for the correctness of the parametrization of the
$z$ dependence of the vertex resolution.

The relations of Eq.~\ref{eq:lumi-gaussian} are used to constrain the
 width and position measurements of the single beams
and of the luminous region in both coordinates separately. 
Given the high statistics of vertices of the luminous region the $pp$
events have the largest weight in the luminosity calculation.
Effectively, the beam-gas measurements determine the relative 
offsets of the two beams and their width ratio  $\rho$
\begin{equation}
 \label{eq:rho}
  \rho_i = \sigma_{2i}/\sigma_{1i} \ \ (i = x,y)\, .
\end{equation}

According to Eq.~\ref{eq:obslumi}, neglecting crossing angle effects and
beam offsets, the luminosity is proportional to $A_{\mathrm{eff}}^{-1}$,
 \begin{equation}
\label{eq:invarea}
 A_{\mathrm{eff}}^{-1} = 
 \frac{1}{\sqrt{ \left( \rule{0pt}{10pt} {{\sigma_{1x}}^2 + {\sigma_{2x}}^2 } \right)
 \left( \rule{0pt}{10pt} {{\sigma_{1y}}^2 + {\sigma_{2y}}^2} \right) }}
 \, .
 \end{equation}
This quantity can be rewritten using the definition
of $\rho_i$ and Eq.~\ref{eq:lumi-gaussian}
\begin{equation}
 \label{eq:rhofactor}
  A_{\mathrm{eff}}^{-1} = 
  \prod_{i=x,y} \frac{\rho_i}{(1+\rho_i^2)\sigxp{i}} \, ,
\end{equation}
which shows, especially for nearly equal beam sizes, the weight of the
measurement of the width of the luminous region $\sigxp{i}$ in the luminosity
determination.  
The expression has its maximum for $\rho_i=1$ which minimizes the
importance of the measurement errors of $\rho_i$.

The luminosity changes if there is an offset between the two
colliding bunches in one of the transverse coordinates.
An extra factor appeared already in Eq.~\ref{eq:obslumi} to take this
into account
 \begin{equation}
  C_{\mathrm{offset}} = 
  \prod_{i=x,y} 
  \exp{\left(-\frac{1}{2}
  \frac{(\mpos_{1i}-\mpos_{2i})^2}{\sigma_{1i}^2+\sigma_{2i}^2}\right)}
  \, ,
  \label{eq:offsetfactor}
 \end{equation}
which is unity for head-on beams.

By examining both relations of Eq.~\ref{eq:lumi-gaussian} a
system of two constraint equations and six measurable quantities
emerges which can be used to improve the precision for each transverse
coordinate separately.  
This fact is exploited in a combined fit where the individual beam
widths 
$\sigma_{1i}$, $\sigma_{2i}$, and the luminous region width
\sigxp{i} together with the corresponding position values $\mpos_{bi}$
and \posxp{i} are used as input measurements.
Several choices are possible for the set of four fit-parameters,
trivially the set $\sigma_{1i}$, $\sigma_{2i}$, $\mpos_{1i}$,  $\mpos_{2i}$
can be used.
The set $\Sigma_{i}$ (Eq.~\ref{eq:capsigma}), 
$\rho_{i}$ (Eq.~\ref{eq:rho}), 
$\Delta_{\mpos i} = \mpos_{1i} - \mpos_{2i}$ and $\posxp{i}$ 
is used which makes it easier to evaluate the corresponding luminosity
error propagation. 
The results for the central values are identical, independently of the
set used. 

\subsection{Corrections and systematic errors}
\label{sec:bgsyst}

In the following, corrections and systematic error sources
affecting the BGI analysis will be described.
The uncertainties related to the bunch population normalization have
already been discussed in Sect.~\ref{sec:bct}.

\subsubsection{Vertex resolution}

As mentioned above, the uncertainty of the resolution potentially
induces a significant systematic error in the luminosity measurement.  
To quantify its effect the fits to the beam profiles have been made with
different choices for the parameters of the resolution functions within
the limits of their uncertainty.
One way to estimate the uncertainty is by comparing the resolution of
simulated $pp$ collision events determined using the method described in
Sect.~\ref{sec:vxresolution}, {\em i.e.} by dividing the tracks into
two groups, with the true resolution which is only accessible in the
simulation owing to the prior knowledge of the vertex position. 
The uncertainty in the number of tracks (\ntrk) dependence at $z=0$ is
estimated in this way to be no larger than 5\%.

The uncertainty in the $z$ dependence is estimated by analysing events
 in \bx{eb} and \bx{be} crossings.
For these crossings all events, including the ones near $z=0$, can be assumed to
 originate from beam-gas interactions.
A cross check of the resolution is obtained by comparing the measurement
 of the transverse beam profiles at the  $z$ range used in \bx{bb}
 events with the ones obtained near $z=0$.
Together with the comparison of the width of the luminous region and its
 prediction from the beam-gas events we estimate a 10\% uncertainty
 in the $z$ dependence of the resolution.
It should be noted that the focusing effect of the beams towards the
 interaction point is negligible compared to the precision needed to
 determine the resolution.

The effect of the uncertainties in the \ntrk dependence and $z$
dependence on the final results are estimated by repeating the analysis
varying the resolution within its uncertainty.
The conservative approach is taken to vary the different dependencies
coherently in both $x$  and $y$ and for the dependence on \ntrk and $z$
simulataneously. 
The resulting uncertainties in the cross-section depend on the widths of the
distributions and are therefore different for each analysed bunch pair.

\subsubsection{Time dependence and stability}

The beam and data-taking stability are taken into account when selecting
suitable fills to perform the beam-gas imaging analysis.
This is an essential requirement given the long integration times needed
to collect sufficient statistics for the beam-gas interactions. 
A clear decay of the bunch-populations and emittance growth is observed
over these long time periods.
It is checked that these variations are smooth and that the time-average is a
good approximation to be used to compare with the average beam profiles
measured with vertices of beam-gas interactions.
No significant movement of the luminous region is observed during the
fills selected for the BGI analysis.
The systematics introduced by these variations are minimized by
the interpolation procedure described in Sect.~\ref{sec:bganalysis} and
are estimated to amount to less than 1\%.

\subsubsection{Bias due to unequal beam sizes and beam offsets}

When the colliding bunches in a pair have similar widths ($\rho_i=1$),
the $\rho$-dependence in Eq.~\ref{eq:rhofactor}, 
$\rho_i/(1+\rho_i^2)$, is close to its maximum.
Thus, when the precision of measuring $\rho$ is similar to its
difference from unity, the experimental estimate of the $\rho$-factor is
biased towards smaller values.  
In the present case the deviation from unity is compatible with the
statistical error of the measurement for each colliding bunch pair.
These values are typically 15\% in the $x$ coordinate and 10\% in the
$y$ coordinate.
The size of the ``$\rho$ bias'' effect is of the order of 1\% in $x$ and 
0.5\% in $y$. 

A similar situation occurs for the offset factor $C_{\mathrm{offset}}$
for bunches colliding with non-zero relative transverse offset.
The offsets are also in this case compatible with zero within their
statistical errors and the correction can only take values smaller than
one. 
The average expected ``offset bias'' is typically 0.5\% per
transverse coordinate.

Since these four sources of bias (unequal beam sizes and offsets in both transverse
coordinates) act in the same direction, their overall effect is
no longer negligible and is corrected for on a bunch-by-bunch basis.
We assume a systematic error equal to half of the correction, {\em i.e.} 
typically $1.5\%$.
The correction and associated uncertainty depends on the
measured central value and its statistical precision and therefore
varies per fill.

\subsubsection{Gas pressure gradient}
\label{sec:gradient}

The basic assumption of the BGI method is the fact that the residual gas
pressure is uniform in the plane transverse to the beam direction and
hence the interactions of the beams with the gas produce an image of the
beam profile. 
An experimental verification of this assumption is performed by
displacing the beams and recording the rate of beam-gas interactions at
these different beam positions.
In Fill 1422 the beam was displaced in the $x$ coordinate by a maximum
of 0.3~mm.
Assuming a linear behaviour, the upper limit on the gradient of the
interaction rate is 0.62~Hz/mm at
95\%~CL compared to a rate of $2.14 \,\pm\, 0.05$~Hz observed with the
beam at its nominal position.
When the profiles of beam~1-gas and beam~2-gas interactions are used
directly to determine the overlap integral $A_{\mathrm{eff}}$, the 
relative error $\delta_A$ on the overlap integral is given by
\begin{equation}
\label{eq:gradient}
\delta_A=\frac{A_{\mathrm{eff}}(a=0)}{A_{\mathrm{eff}}(a\ne0)}-1 =
 \frac{{a}^{2}\,{\sigma_{x}^{2}}}{2\,{b}^{2}} \, ,
\end{equation}
where $a$ is the gradient, $b$ the rate when the beam is at its nominal
position, and $\sigma_{x}$ is the true 
beam width, using the approximation of equal beam sizes.
This result has been derived by comparing the overlap integral for 
beam images distorted by a linear pressure gradient with the one
obtained with ideal beam images.
With the measured limit on the gradient, the maximum relative effect on
the overlap is then estimated to be less than $4.2  \tms 10^{-4}$.
However, the BGI method uses the width of the luminous region measured
using $pp$ interactions as a constraint.
This measurement does not depend on the gas pressure gradient.
The gas pressure gradient enters through the measurements of the
individual widths which are mainly used to determine the ratio between 
the two beam widths.
These are equally affected, thus, the overall effect of an eventual gas
pressure gradient is much smaller that the estimate from
Eq.~\ref{eq:gradient} and can safely be neglected in the analysis. 

\subsubsection{Crossing angle effects}
\label{sec:crossing-effect}

\begin{figure}[tbp]
 \begin{center}
  \includegraphics[width=0.48\textwidth, clip, trim=0.2cm 0cm 0.1cm 0cm]{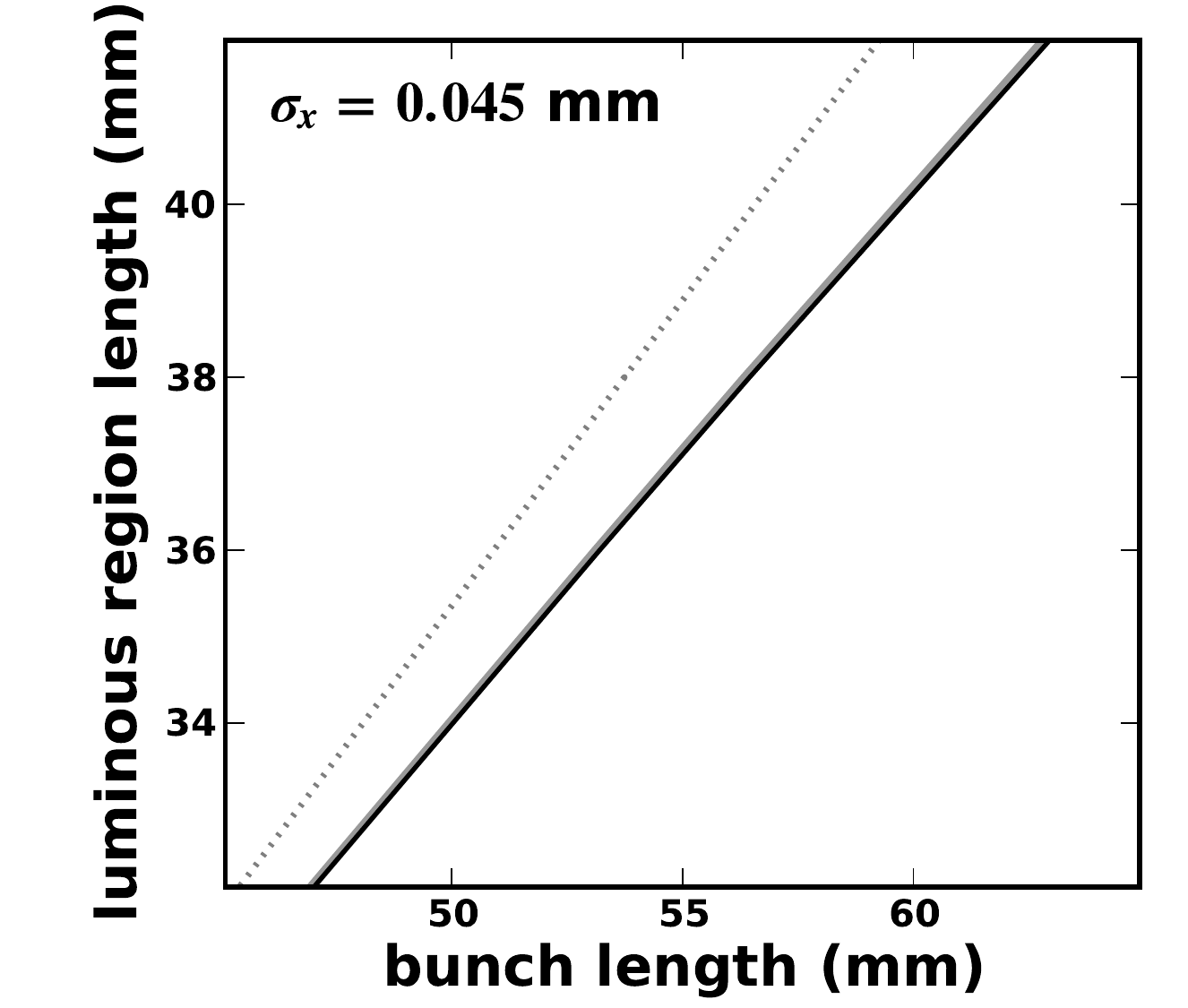}
  \includegraphics[width=0.48\textwidth, clip, trim=0.2cm 0cm 0.1cm 0cm]{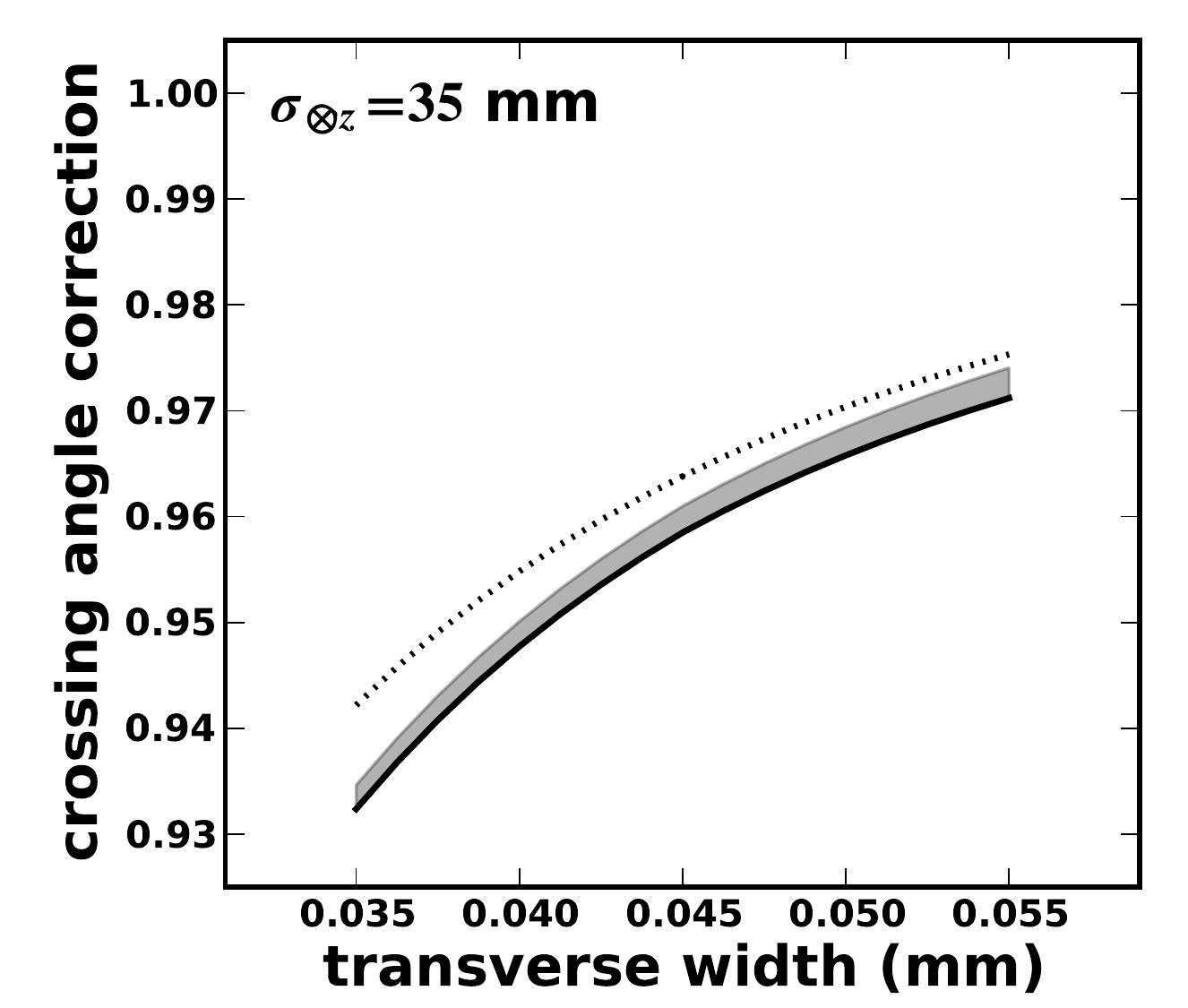}
 \end{center}
 \caption{
 Left: the dependence of the length of the luminous region
 \sigxp{z} on the single bunch length $\sigma_z$ under the
 assumption that both beams have equal length bunches.
 The dotted line shows the $\sqrt{2}$ behaviour expected in the absence
 of a crossing angle.
 The solid black line shows the dependence for equal transverse beam
 sizes $\sigma_x = 0.045$~mm, the shaded region shows the change for
 $\rho = 1.2$ keeping the average size constant.
 Right: the dependence of the luminosity reduction factor
 $C_{\mathrm{\thetac}}$ on the transverse width of the beam $\sigma_x$
 for a value of $\sigxp{z} = 35$~mm. 
 The solid line shows the full calculation for $\rho = 1$ (equal beam
 widths) with the shaded area the change of the value up to $\rho =
 1.2$, keeping the transverse luminous region size constant.
 The dotted line shows the result of the naive calculation assuming a
 simple $\sqrt{2}$ relation for the length of the
 individual beams.
 All graphs are calculated for a half crossing-angle $\thetac = 0.2515$~mrad.
 \label{fig:crossingangle}
 }
\end{figure}

The expression for the luminosity (Eq.~\ref{eq:obslumi}) contains a
correction factor for the crossing angle $C_{\mathrm{\thetac}}$ of the
form~\cite{ref:simonwhite} 
\begin{equation}
 \label{eq:anglecor}
  C_{\mathrm{\thetac}} = 
  \left[
  {1+\tan^2{\thetac} (\sigma_{1z}^2 + \sigma_{2z}^2)/(\sigma_{1x}^2 +
  \sigma_{2x}^2)}
  \right]^{-\frac{1}{2}} \, .
\end{equation}
For a vanishing crossing angle and equal bunch lengths, the bunch length 
$\sigma_z$ is obtained from the beam spot measurement assuming
that the two beams have equal size, by $\sigma_z = \sqrt{2} \, \sigxp{z}$.  
In the presence of a crossing angle the measured length of the luminous
region depends on the lengths of the bunches, on the crossing angle and
on the transverse widths of the two beams in the plane of the crossing
angle.  
The bunch lengths need not necessarily be equal.
Evaluating the overlap integral of the two colliding bunches over the
duration of the bunch crossing, one finds for
the width of the luminous region in the $z$ coordinate
\begin{equation}
 \label{eq:lumz}
  \sigxp{z} = \left[ 
  \frac{\tan^2{\thetac}}{\sigxp[2]{x}}  + 
  \frac{4 \, \cos^2{\thetac}}{\sigma_{1z}^2 + \sigma_{2z}^2}
  \right]^{-\frac{1}{2}}\, .
\end{equation}
Solving for ${\sigma_{1z}^2 + \sigma_{2z}^2}$, the right-hand side  of
Eq.~\ref{eq:anglecor} can be written in terms of the measured quantities
\thetac, \sigxp{z}, \sigxp{x}, $\sigma_{1x}$, and $\sigma_{2x}$
\begin{equation}
 \label{eq:lumzanglecor}
  C_{\mathrm{\thetac}} = \left[ 
  1+\frac{4\sin^2{\thetac} \sigxp[2]{z}} 
  {\left( 1-(\tan{\thetac}{\sigxp{z}}/{\sigxp{x}})^2 \right)(\sigma_{1x}^2+\sigma_{2x}^2)}
  \right]^{-\frac{1}{2}}\, .
\end{equation}
The dependence of the estimate of  \sigxp{z} on $\sigma_z$ and of the
overall correction on $\sigma_x$ is shown in Fig.~\ref{fig:crossingangle}
for typical values of the parameters. 
The difference with respect to a naive calculation assuming equal beam
sizes and using the simple $\sqrt{2}$ factor to obtain the bunch lengths
from the luminous region length is in all relevant cases smaller than $1\%$. 
For the beam conditions in May 2010 the value of the crossing angle 
correction factor $C_{\mathrm{\thetac}}$ is about 0.95. 
To take into account the accuracy of the calculation 
a 1\% systematic error is conservatively assigned to this factor.

There are other small effects introduced by the beam angles.  
The average angle of the beams is different from 0 in the LHCb
coordinate system.
This small difference introduces a broadening of the
measured transverse widths of the luminous region, since in this case the projection 
is taken along the nominal LHCb axis.  
Another effect is more subtle.
The expression (Eq.~\ref{eq:lumi-gaussian}) for the width of the
luminous region assumes a vanishing crossing angle.
It is still valid for any crossing angle if one considers the width for
a fixed value of $z$.
When applying Eq.~\ref{eq:lumi-gaussian} as a function of $z$ one can
show that the centre of the luminous region is offset if in the presence
of a non-vanishing crossing angle the widths of the two beams are not
equal. 
Thus, when these two conditions are met the luminous region is rotated.
The rotation angle $\phi_i$ ($i=x,y$) is given by
\begin{equation} 
\label{eq:phi-rotation}
 \tan{\phi_i} = \tan{\thetac_i}\frac{1-\rho_i^2}{1+\rho_i^2} \, ,
\end{equation}
where $\rho$ is defined in Eq.~\ref{eq:rho}. 
With the parameters observed in this analysis the effect of the
rotation is smaller than $10^{-3}$.

\subsection{Results of the beam-gas imaging method}
\label{sec:bg:results}

\newcommand{\un}{\ensuremath{{{u}}}}
\newcommand{\co}{\ensuremath{{{c}}}}
\newcommand{\fl}{\ensuremath{{{f}}}}
\begin{table}
 \begin{center}
  \caption{
  Measurements of the cross-section \sigeff{\eff} with the BGI method  per fill and
  overall average (third column). 
  All errors are quoted as percent of the cross-section values.
  {\em DCCT scale}, 
  {\em DCCT baseline noise}, 
  {\em FBCT systematics} 
  and {\em Ghost charge}
  are combined into the overall {\em Beam normalization} error.
  The {\em Width syst} row is the combination of 
  {\em Resolution syst} (the systematic error in the vertex resolution
  for $pp$ and beam-gas events), 
  {\em Time dep. syst} (treatment of time-dependence) 
  and {\em Bias syst} (unequal beam sizes and beam offset biases), 
  and is combined with 
  {\em Crossing angle} (uncertainties in the crossing angle correction)
  into {\em Overlap syst}.
  The {\em Total error} is the combination of 
  {\em Relative normalization stability}, 
  {\em Beam normalization}, 
  {\em Statistical error}, and
  {\em Overlap syst}. 
  {\em Total systematics} is the combination of the latter three only and can be broken
  down into {\em Uncorrelated syst} and {\em Correlated syst}, where ``uncorrelated''
  applies to the avergaging of different fills.
  Finally, {\em Excluding norm} is the uncertainty excluding the overall 
  {\em DCCT scale} uncertainty.
  The grouping of the systematic errors into (partial) sums is expressed as an indentation
  in the first column of the table.
  The error components are labelled in the second column by \un, \co\ or \fl\ dependending
  on whether they are uncorrelated, fully correlated or correlated within one fill, 
  respectively.
  \label{tab:bgfinal}
  }
  \vskip 1mm
  \begin{tabular}{lcrrrrrrrrrrrr}
   \small
   &&{\bf average} &{\bf 1089}&{\bf 1090}&{\bf 1101}&{\bf 1104}&{\bf 1117}&{\bf 1118}&{\bf 1122}  \\
   \hline
Cross-section \sigeff{\eff} (mb) && {\bf \underline{59.94}} & 61.49& 59.97& 57.67& 56.33& 61.63& 61.84& 61.04\\
   \hline	                                                     
Relative normalization  &\co& 0.50& 0.50& 0.50& 0.50& 0.50& 0.50& 0.50& 0.50\\
 stability & &&&&&&&\\	 
   \hline		 
~~~DCCT scale           &\co& 2.70& 2.70& 2.70& 2.70& 2.70& 2.70& 2.70& 2.70\\
~~~DCCT baseline noise  &\fl& 0.36& 0.97& 1.01& 0.43& 0.29& 0.29& 0.29& 0.14\\
~~~FBCT systematics     &\fl& 0.91& 3.00& 3.00& 2.61& 2.10& 2.41& 2.41& 1.98\\
~~~Ghost charge         &\fl& 0.19& 0.70& 0.65& 1.00& 0.60& 0.38& 0.55& 0.35\\
Beam normalization      && 2.88& 4.21& 4.21& 3.91& 3.48& 3.65& 3.67& 3.37\\
   \hline		 
Statistical error       &\un& 0.96& 4.06& 4.73& 3.09& 2.56& 1.89& 2.66& 1.82\\
   \hline		 
~~~~~~Resolution syst   &\co& 2.56& 2.79& 2.74& 2.54& 2.86& 2.37& 2.47& 2.44\\
~~~~~~Time dep. syst    &\co& 1.00& 1.00& 1.00& 1.00& 1.00& 1.00& 1.00& 1.00\\
~~~~~~Bias syst         &\co& 1.61& 1.14& 1.81& 1.35& 1.89& 1.19& 1.35& 2.05\\
~~~~~~Gas homogeneity   &&negl.&negl.&negl.&negl.&negl.&negl.&negl.&negl.\\  
~~~Width syst           && 3.20& 3.18& 3.43& 3.05& 3.56& 2.83& 2.99& 3.34\\
~~~Crossing angle       &\co& 1.00& 1.00& 1.00& 1.00& 1.00& 1.00& 1.00& 1.00\\
Overlap syst            && 3.35& 3.33& 3.58& 3.21& 3.70& 3.00& 3.15& 3.49\\
   \hline	                                                    
~~~Uncorrelated syst    &\fl& 0.93& 3.08& 3.07& 2.80& 2.18& 2.44& 2.47& 2.01\\
~~~Correlated syst      &\co& 4.35& 4.43& 4.62& 4.25& 4.62& 4.08& 4.19& 4.44\\
Total systematics       && 4.45& 5.39& 5.55& 5.08& 5.11& 4.75& 4.87& 4.88\\
   \hline		 
Total error             && 4.55& 6.75& 7.29& 5.95& 5.71& 5.11& 5.55& 5.20\\
Excluding norm          && 3.63& 6.17& 6.75& 5.28& 5.01& 4.31& 4.82& 4.42\\
  \end{tabular}
 \end{center}
\end{table}

\begin{figure}[tbp]
 \centering
 \includegraphics[width=0.6\textwidth]{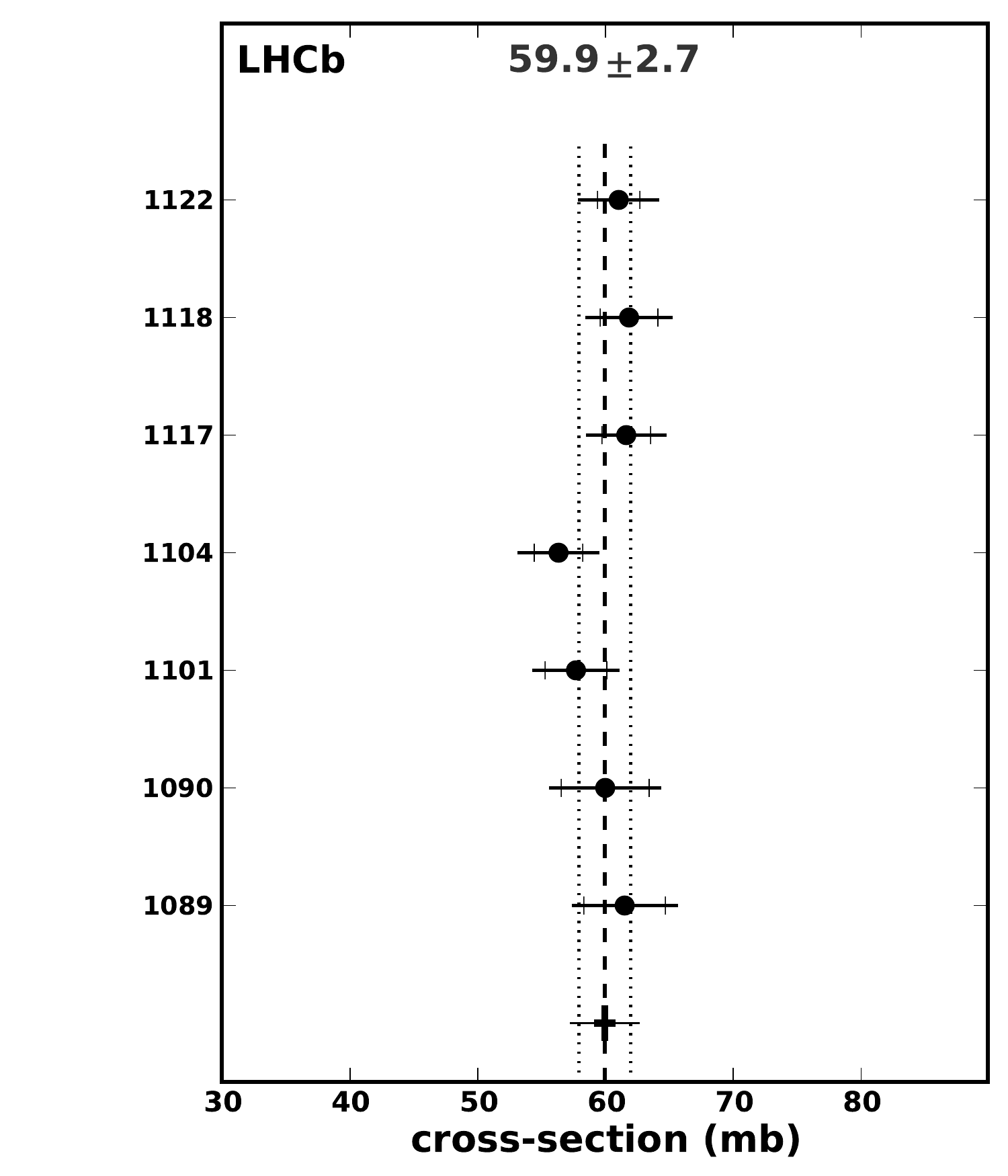}
 \caption{
 Results of the beam-gas imaging method for the visible
 cross-section of the $pp$ interactions producing at least two VELO
 tracks, \sigeff{\eff}.  
 The results for each fill (indicated on the vertical axis) are obtained
 by averaging over all colliding bunch pairs.
 The small vertical lines on the error bars indicate the size of the
 uncorrelated errors, while the length of the error bars show the total
 error. 
 The dashed vertical line indicates the average of the data points and
 the dotted vertical lines show the one standard-deviation interval.
 The weighted average is represented by the lowest data point, where the
 error bar corresponds to its total error.
 \label{fig:bgresult-fill}
 }
\end{figure}

With the use of the beam-gas imaging method 
seven independent measurements of an effective reference cross-section
were performed.
The main uncertainties contributing to
the overall precision of the cross-section measurement come from the
overlap integral and from the measurement of the product of the bunch
populations.  
The systematic error in the overlap integral is composed of the effect
of the resolution uncertainties, the treatment of the time dependence,
the treatment of the bias due to the non-linear dependencies in $\rho$ 
and $\Delta(\mpos)$ and the crossing angle corrections.
It also takes into account small deviations of the beam shape from a
single Gaussian observed in the VDM scans. 
The normalization error has components from the DCCT scale, and its 
baseline fluctuations, FBCT systematics, and systematics in the relative
normalization procedure (Sect.~\ref{sec:relative}).
For multi-bunch fills the results obtained for each colliding pair are
first averaged taking the correlations into account.
The results of the averaging procedure applied on a per-fill basis are 
shown in Table~\ref{tab:bgfinal}. 
For fills with multiple bunches the numbers are a result of an average 
over individual bunches. 
Errors are divided into two types: correlated and uncorrelated errors.
On a fill-by-fill basis the statistical errors, ghost charge and DCCT baseline 
corrections are treated as uncorrelated errors.
The latter two sources are of course correlated when bunches within one 
fill are combined.
The FBCT systematic uncertainty, which is dominated by the uncertainty
in its offset is treated taking into account the fact that 
the sum is constrained.
Owing to the constraint on the total beam current provided by 
the DCCT, averaging results for different colliding bunch pairs within
one fill reduces the error introduced by the FBCT offset uncertainty.
A usual error propagation is applied taking the inverse square of the 
uncorrelated errors as the weights.
The difference with respect to the procedure applied for the VDM method is
due to the fact that a fit using the FBCT offsets as free parameters
cannot be applied here.
Not all fills have the required number of crossing bunch pairs,
and the uncorrelated bunch-to-bunch errors are too large to
obtain a meaningful result for the FBCT offset.

Each colliding bunch pair provides a self-consistent effective cross-section 
determination. 
The analysis proceeds by first averaging over all individual colliding
bunch pairs  within a fill and then by averaging over fills taking all
correlations into account. 
Thus, an effective cross-section result can also be quoted per fill.
These are shown in Fig.~\ref{fig:bgresult-fill}.
The spread in the results is in good agreement with the expectations from 
the uncorrelated errors.

The final beam-gas result for the effective cross-section is:
$\sigeff{\eff} = 59.9 \,\pm\, 2.7 \ \mbox{mb}$. 
The uncertainties in the DCCT scale error and the systematics of the
relative normalization procedure are in common with the VDM method.
The uncertainty in $\sigeff{\eff}$ from the BGI method without these
common errors is 2.2 mb.  

\section{Cross checks with the beam-beam imaging method}
\label{sec:vdmimagingsigma}

During the VDM scan the transverse beam images can be reconstructed with
a method described in Ref.~\cite{ref:vdmimaging}. 
When one beam ({\em e.g.} beam~1) scans across the other (beam~2), the
differences of the measured coordinates of $pp$
vertices with respect to the nominal position of beam~2 are accumulated.
These differences are projected onto a plane transverse to beam~2 and
summed over all scan points. 
The resulting distribution  represents the density profile of beam~2 when
the number of scan steps is large enough and the step size is small
enough.  
By inverting the procedure, beam~1 can be imaged using the relative
movement of beam~2.
Since the distributions are obtained using measured vertex positions,
they are convolved with the corresponding vertex resolution.
After deconvolving the vertex distributions with the transverse vertex
resolution a measurement of the transverse beam image is obtained. 
This approach is complementary to the BGI and VDM methods. 

\begin{figure}[tbp]
 \centering
 \includegraphics*[width=0.6\textwidth]{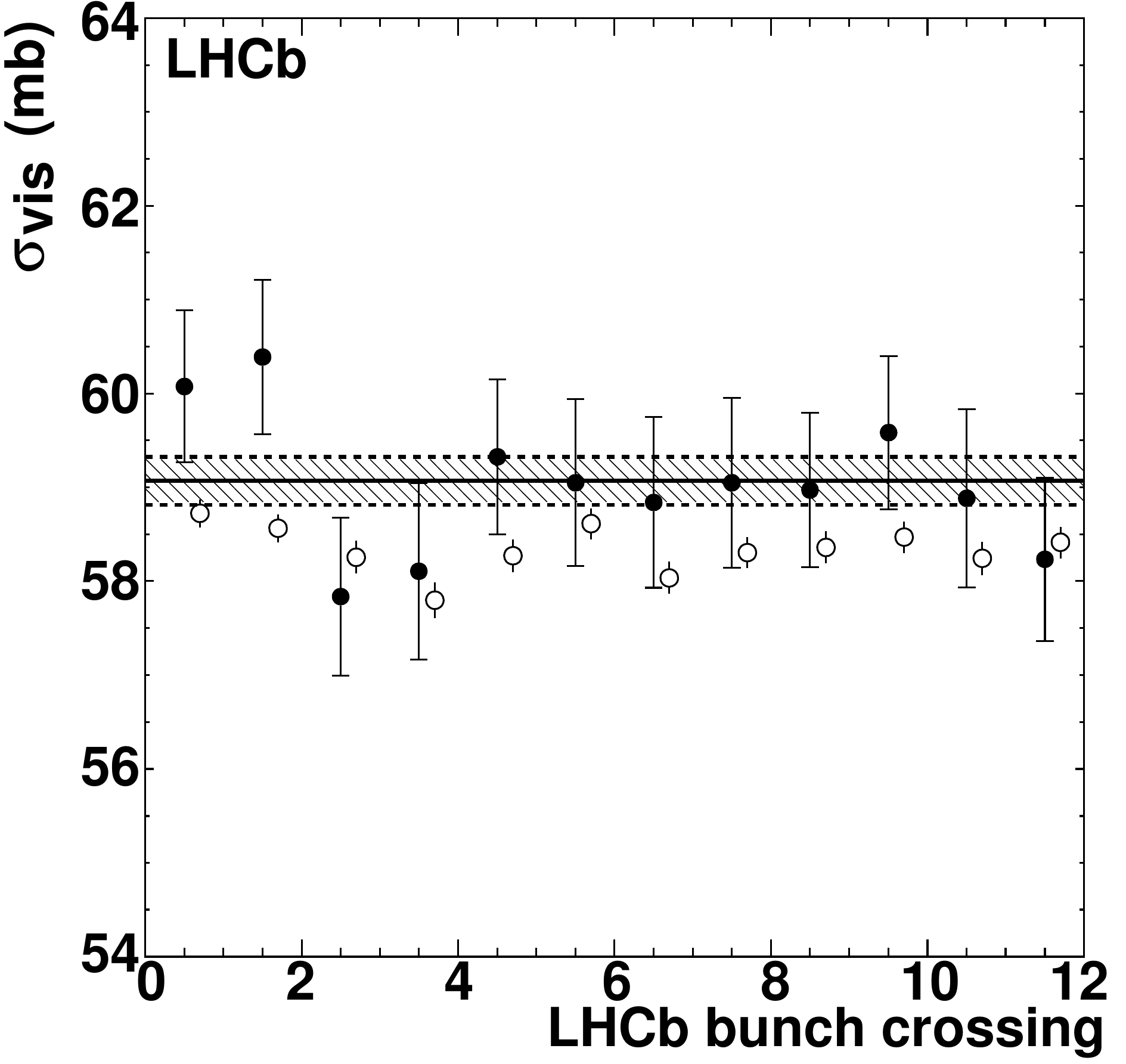}
 \caption{
 Visible cross-section measurement using the beam-beam imaging method
 for twelve different bunch pairs (filled circles) compared to 
 the cross-section measurements using the VDM method (open circles).  
 The horizontal line represents the average of the twelve beam-beam imaging points.
 The error bars are statistical only and neglect the correlations
 between the measurements of the profiles of two beams.
 The band corresponds to a one sigma variation of the
 vertex resolution parameters. 
 \label{fig:vdm-bbimaging}
 }
\end{figure}

The beam-beam imaging method is applied to the first October VDM scan
since the number of scanning steps of that scan is twice as large as
that of the second scan.   
The events are selected by the minimum bias trigger. 
Contrary to the 22.5~kHz random trigger events with only luminosity
data, they contain complete detector information needed for the
tracking. 
The minimum-bias trigger rate was limited to about 130~Hz on average. 
The bias due to the rate limitation is corrected by normalizing the vertex
distributions at every scan point to the measured average number of
interactions per crossing.

The transverse planes with respect to the beams are defined using the
known half crossing-angle of 170~\micrad and the measured inclination
of the luminous ellipsoid with respect to the $z$ axis as discussed in
Sect.~\ref{sec:vdm:coupling}.
The measured
common length scale correction and the difference in the length scales of
the two beams described by the asymmetry parameters $\epsilon_{x,y}$ in
Eq.~\ref{eq:vdm:size},
are also taken into account.

The luminosity overlap integral is calculated numerically
from the reconstructed normalized beam profiles. 
The effect of the VELO smearing is measured and subtracted by comparing
with the case when the smearing is doubled. 
The extra smearing is
performed on an event-by-event basis using the description of the
resolution given in Eq.~\ref{eq:track:dep} with the parameters $A$,
$B$ and $C$ taken from Table~\ref{table:res_fit_params}. 
To improve the vertex resolution, only vertices made with more than 25 tracks
are considered. 
This reduces the average cross-section correction due to the VELO
resolution to 3.7\%.

Similar to the BGI method, the beam-beam imaging method measures the
beam profiles perpendicular to the beam directions. 
For the luminosity determination in the presence of a crossing angle
their overlap should be corrected by the factor $C_{\mathrm{\thetac}}$
(see Eq.~\ref{eq:anglecor}) due to a contribution from the length of
the bunches. 
The average correction for the conditions during the VDM fill in October
is 2.6\%. 
A bunch-by-bunch comparison of the cross-section measurement
with the beam-beam imaging method and the VDM
method is shown in Fig.~\ref{fig:vdm-bbimaging}. 
The cross-section is measured at the nominal beam positions. 
The FBCT bunch populations with zero offsets normalized to the DCCT
values and corrected for the ghost charge are used for the cross-section 
determination.
The band indicates the variation obtained by changing
the vertex resolution by one standard deviation in either direction.
The obtained cross-section of 59.1~mb is in good agreement with the
value of 58.4~mb reported in Table~\ref{tab:vdm:res}.
The comparison is very sensitive since the overall bunch population
normalization and the length scale uncertainty are in common.
Uncorrelated errors amount to about 1\%. 
The main uncorrelated errors in the beam-beam imaging method are the
VELO systematics and the statistical error which are each at the level
of 0.4\%.
The main uncorrelated errors in the VDM method are the stability of the
working point (0.4\%) and the statistics (0.1\%).
The difference between the two methods is smaller than 
but similar to the difference between the two October scan results
observed with the VDM method. 

The width of the VDM rate profile and the widths of the individual
beams are related following Eq.~\ref{eq:capsigma}.
Thus the widths $\Sigma_{x,y}$ are directly {\em measured} with the VDM
rate profile and {\em predicted} using the measured widths of the beams.
The widths can be compared directly using the RMS of the distributions,
the widths (RMS) of single Gaussian fits or the RMS of double Gaussian
fits. 
The variation among these different values are of the order of 1\% and
limit the sensitivity. 
However, it should be noted that these are just numerical differences; 
Eq.~\ref{eq:capsigma} holds for arbitrary beam shapes.
The ratio of the measured and predicted width is
0.994 and 0.996 in the $x$ and $y$ coordinate, respectively.
The statistical uncertainties are 0.3\% and the uncertainties due to the
knowledge of the vertex resolution are 0.2\%.
Considering the sensitivity of the comparison, we note good agreement.

\section{Results and conclusions}
\label{sec:results-conclusions}

The beam-gas imaging method is applied to data collected by LHCb in May
2010 using the residual gas pressure and provides an absolute luminosity
normalization with a relative uncertainty of 4.6\%, dominated by the knowledge of
the bunch populations. 
The measured effective cross-section is in agreement with the
measurement performed with the van der Meer scan method using dedicated
fills in April 2010 and October 2010. 
The VDM method has an overall relative uncertainty of 3.6\%.
The final VDM result is based on the October data alone which give 
significantly lower systematic uncertainties.
The common DCCT scale error represents a large part of the overall uncertainty 
for the results of both methods and is equal to 2.7\%.
To determine the average of the two results the common scale should be 
removed before calculating the relative weights.
Table~\ref{tab:final} shows the ingredients and results of the averaging
procedure.
The combined result has a 3.4\% relative error.

Since the data-sets used for physics analysis contain only a subset of all
available information (see Sect.~\ref{sec:relative}), a small additional
error is introduced {\em e.g.}\ by using \mueff{\eff} information
averaged over bunch crossings.
Together with the uncertainty introduced by the long term stability of
the relative normalization this results in a final uncertainty in
the integrated  luminosity determination of 3.5\%.
We have taken the conservative approach to assign a 0.5\% uncertainty
representing the relative normalization variation to all data-sets and
not to single out one specific period as reference. 
The results of the absolute luminosity measurements are expressed as a
calibration of the visible cross-section \sigeff{\eff}.
This calibration has been used to determine the 
inclusive $\phi$ cross-section in $pp$ collisions at $\sqrt{s} =
7$~TeV~\cite{ref:lhcb-phi}.\footnote{In fact, for the early data-taking 
period on which this measurement is based, the hit count in the SPD is
used to define the visible cross-section.  This cross-section differs
from \sigeff{\eff} defined in this paper by 0.5\%.} 

The relative normalization and its stability have been studied for the
data taken with LHCb in 2010 (see Sect.~\ref{sec:relative}).
Before the normalization can be used for other data-sets an appropriate
study of the relative normalization stability needs to be performed.
\begin{table}
 \begin{center}
  \caption{Averaging of the VDM and BGI results and additonal
  uncertainties when applied to data-sets used in physics analyses.
  \label{tab:final}}
  \vskip 1mm
  \begin{tabular}{lrrrrrrrrrrrr}
                                                &{\bf Average}&{\bf VDM}&{\bf BGI}\\
   {Cross-section (mb)}                         & 58.8 & 58.4 & 59.9\\
   \hline
   {~~~~~~DCCT scale uncertainty (\%)}                & 2.7 & 2.7 & 2.7\\
   {~~~~~~Uncorrelated uncertainty (\%)}              & 2.0 & 2.4 & 3.7\\
   {~~~Cross-section uncertainty (\%)}                & 3.4 & 3.6 & 4.6\\
   \hline
   {~~~~~~Relative normalization stability (\%)}      & 0.5 & & \\
   {~~~~~~Use of average value of \mueff{\eff} (\%)}  & 0.5 & & \\
   {~~~Additional uncertainty for other data-sets (\%)}  & 0.7 & & \\
   {Total uncertainty for large data sets (\%)} & 3.5 & & \\
  \end{tabular}
 \end{center}
\end{table}

While the VDM data have been taken during dedicated fills, no dedicated data
taking periods have yet been set aside for the BGI method.
It is, therefore, remarkable that this method can reach a comparable precision.
A significantly improved precision in the DCCT scale can be expected in the 
near future.
In addition, a controlled pressure bump in the LHCb interaction region
would allow us to apply the beam-gas imaging method in a shorter period, at the
same time decreasing the effects from non-reproducibility of beam
conditions and increasing the statistical precision.
The main uncertainty in the VDM result, apart from the scale error, is
due to the lack of reproducibility found between different scanning
strategies. 
Dedicated tests will have to be designed to understand these differences
better.  
Finally, it is also very advantageous to perform beam-gas measurements
in the same fill as the  van der Meer scans.
This would allow cross checks to be made with a precision which does not
suffer from scale uncertainties in the bunch population measurement.
Furthermore, a number of parameters which limit the precision of the BGI
method can be constrained independently using the VDM scan data, such as
the relative beam positions.
To improve the result of the BGI method a relatively large $\beta^*$
value should be chosen, such as 10~m.
The precision of the VDM method does not, in principle, depend on
$\beta^*$.

\section{Acknowledgements}

We express our gratitude to our colleagues in the CERN accelerator
 departments for their support and for the excellent performance of
 the LHC. In particular, we thank S. White and H. Burkhardt for their
 help on the van der Meer scans.
We thank the
technical and administrative staff at CERN and at the LHCb institutes,
and acknowledge support from the National Agencies: CAPES, CNPq, FAPERJ
and FINEP (Brazil); CERN; NSFC (China); CNRS/IN2P3 (France); BMBF, DFG,
HGF and MPG (Germany); SFI (Ireland); INFN (Italy); FOM and NWO
(Netherlands); SCSR (Poland); ANCS (Romania); MinES of Russia and
Rosatom (Russia); MICINN, XuntaGAL and GENCAT (Spain); SNSF and SER
(Switzerland); NAS Ukraine (Ukraine); STFC (United Kingdom); NSF
(USA). We also acknowledge the support received from the ERC under FP7
and the R\'{e}gion Auvergne.

\end{document}